\documentclass[aps,prd,twocolumn,showpacs,amsmath,amssymb]{revtex4}
\usepackage{amsfonts}
\usepackage{amsmath}
\usepackage{graphicx}
\usepackage{subfigure}
\usepackage{dcolumn}
\usepackage{bm}
\usepackage{booktabs}
\usepackage{graphicx,graphics,dcolumn,booktabs,bm}
\usepackage{longtable,lscape}
\pdfoutput=1
\usepackage{txfonts}
\usepackage{overpic}
\usepackage{amssymb}
\usepackage{indentfirst}
\usepackage{epsfig}
\usepackage{feynmf}
\usepackage{epstopdf}   
\usepackage{slashed}  
\usepackage{multirow}

\usepackage{ulem}
\usepackage[usenames,dvipsnames]{color}

\usepackage[colorlinks,
            citecolor=blue,
            anchorcolor=red,
            menucolor=red,
            linkcolor=red,
            filecolor=red,
            runcolor=red,
            urlcolor=blue,
            frenchlinks=red]{hyperref}

\begin{document}

\title{Surveying exotic pentaquarks with the typical $QQqq\bar{q}$ configuration}
\author{Qin-Song Zhou$^{1,2}$}
\author{Kan Chen$^{1,2}$}
\author{Xiang Liu$^{1,2}$}
\email{xiangliu@lzu.edu.cn} \affiliation{
$^1$School of Physical Science and Technology, Lanzhou University, Lanzhou 730000, China\\
$^2$Research Center for Hadron and CSR Physics, Lanzhou University
and Institute of Modern Physics of CAS, Lanzhou 730000, China }
\author{Yan-Rui Liu$^3$}
\email{yrliu@sdu.edu.cn} \affiliation{ $^3$School of Physics and Key
Laboratory of Particle Physics and Particle Irradiation (MOE),
Shandong University, Jinan 250100, China}

\author{Shi-Lin Zhu$^{4,5,6}$}
\email{zhusl@pku.edu.cn} \affiliation{
$^4$School of Physics and State Key Laboratory of Nuclear Physics and Technology, Peking University, Beijing 100871, China\\
$^5$Collaborative Innovation Center of Quantum Matter, Beijing 100871, China\\
$^6$Center of High Energy Physics, Peking University, Beijing
100871, China }

\date{\today}
\begin{abstract}
As a hot issue, exploring exotic pentaquarks is full of challenges and opportunities for both theorist and experimentalist. In this work, we focus on a type of pentaquark with the $QQqq\bar{q}$ ($Q=b,c$; $q=u,d,s$) configuration, where their mass spectrum is estimated systematically. Especially, our result indicates that there may exist some stable or narrow exotic pentaquark states.  Obviously, our study may provides valuable information for further experimental search for the $QQqq\bar{q}$ pentaquarks. With the running of LHCb and forthcoming Belle II, we have a reason to believe that these predictions present here can be tested.

\end{abstract}
\maketitle

\section{Introduction}\label{sec1}

Nowadays it is still a hot topic to identify multiquark states from
both theoretical side and experimental side since the proposal of
the quark model \cite{GellMann:1964nj,Zweig:1981pd}. More and more
exotic $XYZ$ states observed by experiments in recent years
\cite{Choi:2007wga,Mizuk:2008me,Mizuk:2009da,Liu:2013dau,Xiao:2013iha,Ablikim:2013mio,Ablikim:2013wzq,Ablikim:2013xfr,Chilikin:2014bkk,Aaij:2014jqa,Ablikim:2015gda,Ablikim:2015swa,Ablikim:2015tbp}
are considered as possible tetraquark candidates
\cite{Swanson:2006st,Zhu:2007wz,Voloshin:2007dx,Drenska:2010kg,Chen:2016qju,Hosaka:2016pey,Richard:2016eis,Lebed:2016hpi,Esposito:2016noz}.
With one more quark component, the intriguing pentaquark states were
also studied in various colliders. Although the subsequent
experiments \cite{Hicks:2012zz} did not confirm the light $\Theta^+$
pentaquark with component $uudd\bar{s}$ claimed by the LEPS
Collaboration \cite{Nakano:2003qx}, the LHCb experiment brought us
new findings in the heavy quark realm in 2015 \cite{Aaij:2015tga}.
Two hidden-charm pentaquark-like resonances $P_{c}(4380)$ and
$P_{c}(4450)$ are extracted in the $J/\psi p$ invariant mass
distribution of the $\Lambda_{b}^{0}$ decay into $J/\Psi K^{-}p$.
This observation stimulated further studies on pentaquark states
\cite{Chen:2016qju,Chen:2016heh,Aaij:2017jgf}. In this paper, we pay
attention to the $QQqq\bar{q}$ systems, where $Q=b,c$ and $q=u,d,s$,
and estimate the masses of such pentaquark states roughly.

In the quark model, the doubly charmed baryon $\Xi_{cc}$
($J^P=\frac12^+$ or $\frac32^+$) is in a 20-plet representation of
the flavor SU(4) classification \cite{Patrignani:2016xqp}.
Although its study started 40 years ago \cite{DeRujula:1975qlm}, its
existence is confirmed very recently
\cite{Mattson:2002vu,Ocherashvili:2004hi,Aaij:2017ueg}. The
confirmation from LHCb motivates further theoretical studies on the
possible stable $T_{QQ}$ ($QQ\bar{q}\bar{q}$) states, which had been
predicted in various models. Both the $\Xi_{cc}$ baryon and the
$T_{QQ}$ meson contain a heavy diquark. Now we would like to add one
more light quark component and discuss the spectra of the doubly
heavy pentaquarks within a simple model. The so-called heavy
diquark-antiquark symmetry was used to relate the mass splittings of
$QQq$ and $QQ\bar{q}\bar{q}$ in Ref. \cite{Cohen:2006jg}. Hopefully,
the present investigation can also be helpful to further study on
such a symmetry in multiquark systems.

Compared to the $QQq$ baryon, the $QQqq\bar{q}$ pentaquark state
should be heavier. However, the complicated interactions within
multiquark systems may lower the mass, which probably makes it
difficult to distinguish experimentally a conventional baryon from a
pentaquark baryon just from the mass consideration. One example for
this feature is the five newly observed $\Omega_c$ states
\cite{Aaij:2017nav,Yelton:2017qxg}. They can be accommodated in both
$3q$ configuration
\cite{Chen:2017sci,Karliner:2017kfm,Wang:2017hej,Wang:2017vnc,Padmanath:2017lng,Cheng:2017ove,Wang:2017zjw,Chen:2017gnu,Zhao:2017fov}
and $5q$ configuration
\cite{Yang:2017rpg,An:2017lwg,Montana:2017kjw,Debastiani:2017ewu,Wang:2017smo,Nieves:2017jjx,Huang:2018wgr}
and much more measurements are needed to resolve their nature. As a
theoretical prediction, the basic features for the pentaquark
spectra may be useful for us to understand possible structures of
heavy quark hadrons.

For the doubly heavy five-quark systems, we have a compact
$QQqq\bar{q}$ configuration and two baryon-meson moleucle
configurations, $(QQq)(q\bar{q})$ and $(Qqq)(Q\bar{q})$. As for the
latter molecule configuration, there are theoretical studies in the
meson exchange methods
\cite{Xu:2010fc,Chen:2017vai,Shimizu:2017xrg}. Here, we discuss the
mass splittings of the compact $QQqq\bar{q}$ pentaquark states by
considering the color-magnetic interactions between quarks and
estimate their rough positions. It is still an open question how to
distinguish the two configurations. For example, if we compare the
prediction for the $\Lambda$-type hidden charm state in the molecule
picture \cite{Wu:2010jy} and the estimation for the mass of the
lowest $c\bar{c}uds$ compact pentaquark \cite{Wu:2017weo}, one gets
consistent results. However, the numbers of possible states in these
two pictures are different. The present study should be useful in
looking for genuine pentaquark states rather than molecules.

This paper is organised as follows. In Sec. \ref{sec2}, we construct
the $flavor\otimes color \otimes spin$ wave functions for the
$QQqq\bar{q}$ pentaquark states. In Sec. \ref{sec3}, the relevant
Hamiltonians for various systems are presented. In Sec. \ref{sec4},
we give numerical results and discuss the mass spectra of the
pentaquark states and their strong decay channels. Finally, we
present a summary in Sec. \ref{sec5}.

\section{Color-magnetic interaction and wave functions}\label{sec2}

{Few-body problem is difficult to deal with and there are scarce dynamical studies on pentaquark systems without substructure assumptions \cite{Hiyama:2018ukv,Hiyama:2005cf}. To understand systematically the basic features for the properties of multiquark states, as the first step, we here adopt a color-magnetic model and mainly focus on the mass splittings of the $S$-wave pentaquark states. For the pentaquark masses, we just present some estimations. Their accurate values need further dynamical calculations.} For the ground state hadrons with the same quark content, e.g. $\Delta$ and $N$, their mass splitting is mainly determined by the color-magnetic interaction (CMI). The Hamiltonian in this model reads
\begin{eqnarray}\label{Eq1}
H&=&\sum_im_i+H_{CM},\nonumber\\
H_{CM}&=&-\sum_{i<j}C_{ij} \vec\lambda_i\cdot \vec\lambda_j
\vec\sigma_i\cdot\vec\sigma_j =-\sum_{i<j}C_{ij}
\lambda_i^a\lambda_j^a \sigma_i^b\sigma_j^b,
\end{eqnarray}
where $\lambda_i^a$ ($a=1,\cdots,8$) are the Gell-Mann matrices for
the $i$-th quark and $\sigma_{j}^b$ ($b=1,2,3$) are the Pauli
matrices for the $j$-th quark. For antiquarks, the $\vec{\lambda}_i$
is replaced with $-\vec{\lambda}_i^{*}$. The effective mass $m_i$
for the $i$-th quark includes the constituent quark mass and
contributions from color-electric interactions and color
confinements. The effective coupling constants $C_{ij}$ depend on
the quark masses and the ground state spatial wave functions.

{The model is an oversimplified one of the realistic quark interactions. We may check its relation with the leading order Hamiltonian in nonrelativistic approximation in Ref. \cite{DeRujula:1975qlm} (ignore the electromagnetic part),
\begin{eqnarray}\label{H}
\hat{H}=L(\vec{r}_{1},\vec{r}_{2},...)+\sum\limits_{i}(m_{0i}+\frac{\vec{p}_{i}}{2m_{0i}})+\frac{1}{4}\sum\limits_{i>j}\alpha_{s}
\vec{\lambda_{i}}\cdot\vec{\lambda_{j}}S_{ij}.
\end{eqnarray}
Here, $L$ is responsible for quark binding and $\vec{r}_{i}$, $\vec{p}_{i}$,
and $m_{0i}$ are the position, momentum, and mass of the $i$-th quark, respectively. $S_{ij}$ has the form
\begin{eqnarray}
S_{ij}&=&\frac{1}{|\vec{r}|}-\frac{1}{2m_{0i}m_{0j}}\Big(\frac{\vec{p}_{i}\cdot \vec{p}_{j}}{|\vec{r}|}+\frac{\vec{r}\cdot(\vec{r}\cdot\vec{p}_{i}) \vec{p}_{j}}{|\vec{r}|^{3}}\Big)\nonumber\\
&&-\frac{\pi}{2}\delta^{3}(\vec{r})\Big(\frac{1}{m_{0i}^2}+\frac{1}{m_{0j}^2}+\frac{4\vec{\sigma}_{i}\cdot\vec{\sigma}_{j}}{3m_{0i}m_{0j}}\Big)\nonumber\\
&&-\frac{1}{4|\vec{r}|^{3}}\Big\{\frac{\vec{r}\times\vec{p}_{i}\cdot \vec{\sigma}_{i}}{m_{0i}^{2}}-\frac{\vec{r}\times\vec{p}_{j}\cdot \vec{\sigma}_{j}}{m_{0j}^{2}}
+\frac{1}{m_{0i}m_{0j}}\Big[2\vec{r}\times\vec{p}_{i}\cdot \vec{\sigma}_{j}\nonumber\\
&&-2\vec{r}\times\vec{p}_{j}\cdot \vec{\sigma}_{i}
-\vec{\sigma}_{i}\cdot\vec{\sigma}_{j}+3\frac{(\vec{\sigma}_{i}\cdot\vec{r})(\vec{\sigma}_{j}\cdot\vec{r})}{|\vec{r}|^{2}}\Big]     \Big\},
\end{eqnarray}
where $\vec{r}=\vec{r}_i-\vec{r}_j$. For $S$-wave hadrons, the last two lines (spin-orbit and tensor parts) have vanishing contributions. By calculating the average value with the orbital wave function $\Psi_0$ ($L=0$), one may write the Hamiltonian as
\begin{eqnarray}
H&=&\langle\Psi_0|\hat{H}|\Psi_0\rangle\nonumber\\
&=&\Big\{\langle\Psi_{0}|\Big[L(\vec{r}_{1},\vec{r}_{2},...)+\sum\limits_{i}(m_{0i}+\frac{\vec{p}_{i}}{2m_{0i}})\Big]|\Psi_{0}\rangle\nonumber\\
&&+\frac14\sum\limits_{i>j}\vec{\lambda_{i}}\cdot\vec{\lambda_{j}}
\langle\Psi_{0}|\alpha_{s}\Big[\frac{1}{|\vec{r}|}-(\frac{1}{m_{0i}^{2}}+\frac{1}{m_{0j}^{2}})\frac{\pi}{2}\delta^{3}(\vec{r})
\nonumber\\
&&-\frac{1}{2m_{0i}m_{0j}}\Big(\frac{\vec{p}_{i}\cdot \vec{p}_{j}}{|\vec{r}|}+\frac{\vec{r}\cdot(\vec{r}\cdot\vec{p}_{i}) \vec{p}_{j}}{|\vec{r}|^{3}}\Big)\Big]|\Psi_{0}\rangle\Big\}\nonumber\\
&&-\sum\limits_{i>j}\frac{\pi}{6}\langle\Psi_{0}|\alpha_{s}\delta^{3}(\vec{r})|\Psi_{0}\rangle
\frac{\vec{\lambda_{i}}\cdot\vec{\lambda_{j}}\vec{\sigma}_{i}\cdot\vec{\sigma}_{j}}{m_{0i}m_{0j}}\nonumber\\
&\equiv&M_0-\sum\limits_{i>j}C_{ij}\vec{\lambda_{i}}\cdot\vec{\lambda_{j}}\vec{\sigma}_{i}\cdot\vec{\sigma}_{j}.
\end{eqnarray}
For states with the same quark content, $M_{0}$ is a constant and it can be expressed as the summation of effective quark masses $M_0=\sum_im_i$. Then the model Hamiltonian we will use is obtained. In principle, the values of $m_i$ and $C_{ij}$ should be different for various systems. However, it is difficult to exactly calculate these parameters for a given system without knowing the spatial wave function. In the present study, they will be extracted from the masses of conventional hadrons. That is to say, we use the assumption that quark-quark interactions are the same for various systems. This assumption certainly leads to uncertainties on hadron masses. The uncertainty cause by $m_i$ does not allow us to give accurate pentaquark masses while the uncertainty in coupling parameters has smaller effects and the mass splittings should be more reliable. In order to reduce the uncertainties and obtain more appropriate estimations, we will try to use an alternative form of the mass formula. Whether this manipulation gives results close to realistic masses or not can be tested in future measurments.

 }


Obviously, we can calculate the color-magnetic matrix elements and
investigate the mass spectra for the $QQqq\bar{q}$ systems if the
wave functions were constructed. Now we move on to the construction
of the flavor-color-spin wave function of a system, which is a
direct product of SU(3)$_f$ flavor wave function, SU(3)$_c$ color
wave function, and SU(2)$_s$ spin wave function. We construct these
wave functions separately and then combine them together by noticing
the possible constraint from the Pauli principle. We will use the
diquark-diquark-antiquark bases to construct the wave function. In
principle, the selection of wave function bases is irrelevant with
the final results since we will diagonalize the Hamiltonian in this
CMI model. Here, the notation ``diquark'' only means two quarks and
it does not mean a compact substructure.

\begin{figure}[!h]\centering
\begin{tabular}{c}
\includegraphics[width=180pt]{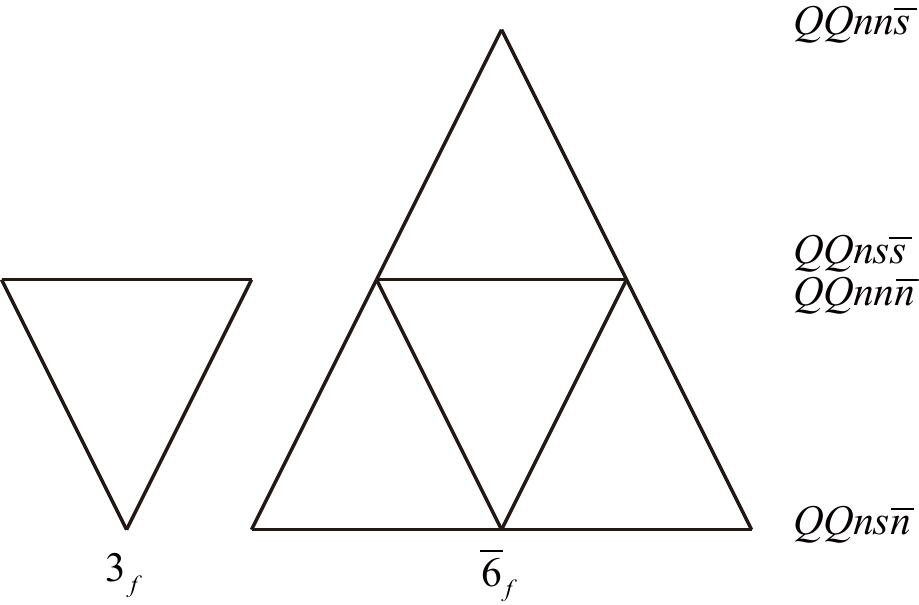}\\
(a) The two light quarks belong to $\bar{3}_f$. \\\\
\includegraphics[width=180pt]{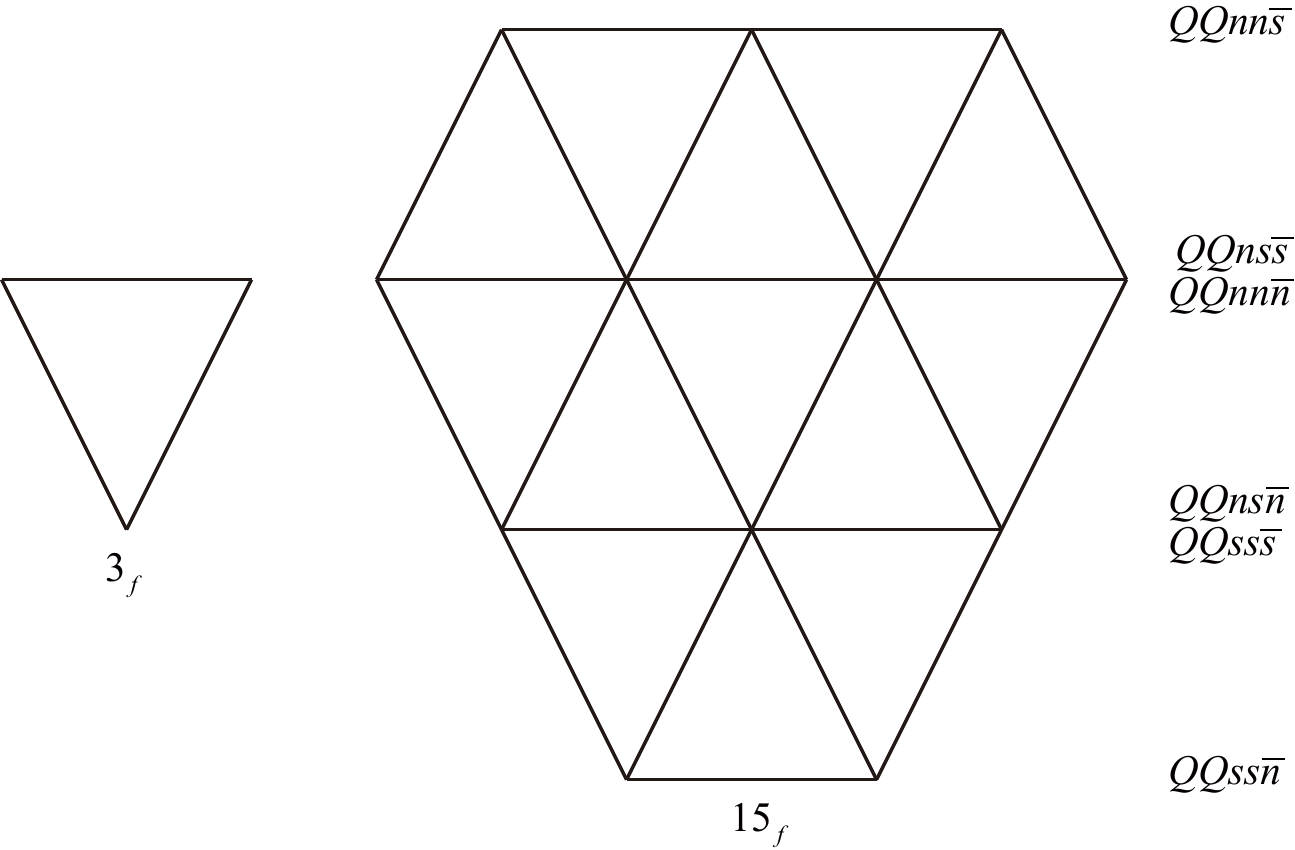}\\
(b) The two light quarks belong to $6_f$.
\end{tabular}
\caption{SU(3)$_f$ weight diagrams for the $QQqq\bar{q}$ pentaquark
states.}\label{fig1}
\end{figure}

In flavor space, the heavy quarks are treated as SU(3)$_f$ singlet
states and the light diquark may be in the flavor antisymmetric
$\bar{3}_f$ or symmetric $6_f$ representation. For the case of the
antisymmetric (symmetric) light diquark, the representations of the
pentaquarks are $\bar{6}_f$ and $3_f$ ($3_f$ and $15_f$). We plot
the SU(3)$_f$ weight diagrams for the $QQqq\bar{q}$ systems in Fig.
\ref{fig1}. The explicit wave functions are similar to the
$qq\bar{q}\bar{Q}$ tetraquark states presented in Ref.
\cite{Liu:2004kd}. Because of the unequal quark masses, we consider
SU(3)$_f$ symmetry breaking and the flavor mixing among different
representations occurs. The resulting systems we consider are:
$Q_1Q_2nn\bar{n}$, $Q_1Q_2nn\bar{s}$, $Q_1Q_2ns\bar{n}$,
$Q_1Q_2ns\bar{s}$, $Q_1Q_2ss\bar{n}$, and $Q_1Q_2ss\bar{s}$, where $n$ represents $u$ or $d$.

In color space, the Young diagrams tell us that the pentaquark
systems have three color singlets. Then we have three color wave
functions. The direct product for the representations can be written
as
\begin{eqnarray}\label{Eq2}
&&(3_{c}\otimes3_{c})\otimes(3_{c}\otimes3_{c})\otimes\bar{3}_{c}\nonumber\\
&=&(\bar{3}_{c}\oplus6_{c})\otimes(\bar{3}_{c}\oplus6_{c})\otimes\bar{3}_{c}\nonumber\\
&=&(\bar{3}_{c}\otimes\bar{3}_{c}\otimes\bar{3}_{c})\oplus(\bar{3}_{c}\otimes6_{c}\otimes\bar{3}_{c})\oplus(6_{c}\otimes\bar{3}_{c}\otimes\bar{3}_{c}).
\end{eqnarray}
In the last line, the representations in the parentheses are for the
heavy diquark, light diquark, and antiquark, respectively. Then the
color-singlet wave functions can be constructed as
\begin{eqnarray}
\phi^{AA}&=&[(Q_{1}Q_{2})^{\bar{3}_c}(q_{3}q_{4})^{\bar{3}_c}\bar{q}],\nonumber\\
\phi^{AS}&=&[(Q_{1}Q_{2})^{\bar{3}_c}(q_{3}q_{4})^{6_c}\bar{q}],\nonumber \\
\phi^{SA}&=&[(Q_{1}Q_{2})^{6_c}(q_{3}q_{4})^{\bar{3}_c}\bar{q}],
\end{eqnarray}
where $A$ ($S$) means antisymmetric (symmetric) for the diquarks.
Explicitly, we have

\begin{eqnarray}
\phi^{AA}&=&\frac{1}{2\sqrt{6}}\Big[(rbbg-rbgb+brgb-brbg+gbrb-gbbr\nonumber\\
&&+bgbr-bgrb)\bar{b}+(rbrg-rbgr+brgr-brrg\nonumber\\
&&+grrb-grbr+rgbr-rgrb)\bar{r}+(gbrg-gbgr\nonumber\\
&&+bggr-bgrg+grgb-grbg+rgbg-rggb)\bar{g}\Big],
\\
\phi^{AS}&=&\frac{1}{4\sqrt{3}}\Big[(2rgbb-2grbb-rbgb-rbbg+brgb\nonumber\\
&&+brbg+gbrb+gbbr-bgrb-bgbr)\bar{b}+(2gbrr \nonumber\\
&&-2bgrr-rbrg-rbgr+brrg+brgr-grrb-grbr\nonumber\\
&&+rgrb+rgbr)\bar{r}+(2brgg-2rbgg+gbrg+gbgr\nonumber\\
&&-bgrg-bggrg-grgb-grbg+rggb+rgbg)\bar{g}\Big],
\\
\phi^{SA}&=&\frac{1}{4\sqrt{3}}\Big[(2bbgr-2bbrg+gbrb-gbbr+bgrb-bgbr\nonumber\\
&&-rbgb+rbbg-brgb+brbg)\bar{b}+(2rrbg-2rrgb \nonumber\\
&&+rgrb-rgbr+grrb-grbr+rbgr-rbrg+brgr\nonumber\\
&&-brrg)\bar{r}+(2ggrb-2ggbr-rggb+rgbg\nonumber\\
&&-grgb+grbg+gbgr-gbrg+bggr-bgrg)\bar{g}\Big].
\end{eqnarray}

The spin wave functions for the pentaquark states are
\begin{eqnarray}
\chi^{SS}&:& \left\{\begin{array}{ccc}
\chi_{1}&=&[(Q_{1}Q_{2})_{1}(q_{3}q_{4})_{1}\bar{q}]_{2}^{\frac52},\\
\chi_{2}&=&[(Q_{1}Q_{2})_{1}(q_{3}q_{4})_{1}\bar{q}]_{2}^{\frac32},\\
\chi_{3}&=&[(Q_{1}Q_{2})_{1}(q_{3}q_{4})_{1}\bar{q}]_{1}^{\frac{3}{2}},\\
\chi_{4}&=&[(Q_{1}Q_{2})_{1}(q_{3}q_{4})_{1}\bar{q}]_{1}^{\frac{1}{2}},\\
\chi_{5}&=&[(Q_{1}Q_{2})_{1}(q_{3}q_{4})_{1}\bar{q}]_{0}^{\frac{1}{2}},
\end{array}\right.
\\
\chi^{SA}&:& \left\{\begin{array}{ccc}
\chi_{6}&=&[(Q_{1}Q_{2})_{1}(q_{3}q_{4})_{0}\bar{q}]_{1}^{\frac{3}{2}},\\
\chi_{7}&=&[(Q_{1}Q_{2})_{1}(q_{3}q_{4})_{0}\bar{q}]_{1}^{\frac{1}{2}},
\end{array}\right.
\\
\chi^{AS}&:& \left\{\begin{array}{ccc}
\chi_{8}&=&[(Q_{1}Q_{2})_{0}(q_{3}q_{4})_{1}\bar{q}]_{1}^{\frac{3}{2}},\\
\chi_{9}&=&[(Q_{1}Q_{2})_{0}(q_{3}q_{4})_{1}\bar{q}]_{1}^{\frac{1}{2}},
\end{array}\right.
\\
\chi^{AA}&:&
\chi_{10}=[(Q_{1}Q_{2})_{0}(q_{3}q_{4})_{0}\bar{q}]_{0}^{\frac{1}{2}}.
\end{eqnarray}
Here in the symbol
$[(Q_{1}Q_{2})_{spin}(q_{3}q_{4})_{spin}\bar{q}]_{j}^{total spin}$,
$j$ is the total spin of the first four quarks. The superscript $SA$
of $\chi$ means that the first two quarks are symmetric and the
second two quarks are antisymmetric. Other superscripts are
understood similarly.

Considering the Pauli principle, we obtain twelve types of total
wave functions $[\phi^{AA}\otimes\chi^{SS}]\delta_{34}^{A}$,
$[\phi^{AA}\otimes\chi^{SA}]\delta_{34}^{S}$,
$[\phi^{AA}\otimes\chi^{AS}]\delta_{12}\delta_{34}^{A}$,
$[\phi^{AA}\otimes\chi^{AA}]\delta_{12}\delta_{34}^{S}$,
$[\phi^{AS}\otimes\chi^{SS}]\delta_{34}^{S}$,
$[\phi^{AS}\otimes\chi^{SA}]\delta_{34}^{A}$,
$[\phi^{AS}\otimes\chi^{AS}]\delta_{12}\delta_{34}^{S}$,
$[\phi^{AS}\otimes\chi^{AA}]\delta_{12}\delta_{34}^{A}$,
$[\phi^{SA}\otimes\chi^{SS}]\delta_{12}\delta_{34}^{A}$,
$[\phi^{SA}\otimes\chi^{SA}]\delta_{12}\delta_{34}^{S}$,
$[\phi^{SA}\otimes\chi^{AS}]\delta_{34}^{A}$, and
$[\phi^{SA}\otimes\chi^{AA}]\delta_{34}^{S}$. Here, $\delta_{12}=0$
when the first two quarks are identical, or else $\delta_{12}=1$.
When the two light quarks are antisymmetric (symmetric) in the
flavor space, $\delta_{34}^{A}=0$ ($\delta_{34}^{S}=0$), or else
$\delta_{34}^{A}=1$ ($\delta_{34}^{S}=1$). Then the considered
pentaquark states are categorized into six classes:

1.The $(ccnn)^{I=1}\bar{q}, (bbnn)^{I=1}\bar{q}, (ccss)\bar{q}$, and
$(bbss)\bar{q}$ states with $\delta_{12}=\delta_{34}^{S}=0$;

2.The $(ccnn)^{I=0}\bar{q}$ and $(bbnn)^{I=0}\bar{q}$ states with
$\delta_{12}=\delta_{34}^{A}=0$;

3.The $(bcnn)^{I=1}\bar{q}$ and $(bcss)\bar{q}$ states with
$\delta_{12}=1$ and $\delta_{34}^{S}=0$;

4.The $(bcnn)^{I=0}\bar{q}$ states with $\delta_{12}=1$ and
$\delta_{34}^{A}=0$;

5.The $(ccns)\bar{q}$ and $(bbns)\bar{q}$ states with
$\delta_{12}=0$ and $\delta_{34}^{S}=\delta_{34}^{A}=1$;

6.The $(bcns)\bar{q}$ states with $\delta_{12}=\delta_{34}^{A}=\delta_{34}^{S}=1$.\\
In the following discussions, we also use the notation
$[(Q_{1}Q_{2})_{spin}^{color}(q_{3}q_{4})_{spin}^{color}\bar{q}]_{j}^{totalspin}$
to denote the total wave function.

\section{The Hamiltonian  expressions}\label{sec3}

With the constructed wave functions, we calculate color-magnetic
matrix elements on various bases. In this section, we present the
obtained Hamiltonians in the matrix form. To simplify the
expressions, {we use the variables defined in Table \ref{variable substitution}. }
\begin{table}[!h]
\caption{Defined variables to simplify the CMI expressions.}\label{variable substitution} \centering
\begin{tabular}{cc|cc}
\toprule[1pt]
\midrule[1pt]
Variable & Definition&Variable & Definition\\
\midrule[1pt]
$\alpha$&$C_{12}+C_{34}$&$\beta$&$C_{13}+C_{14}+C_{23}+C_{24}$\\
$\lambda$&$C_{15}+C_{25}$&$\gamma$&$C_{13}+C_{14}-C_{23}-C_{24}$\\
$\mu$&$C_{15}-C_{25}$&$\delta$&$C_{13}-C_{14}+C_{23}-C_{24}$\\
$\nu$&$C_{35}+C_{45}$&$\eta$&$C_{13}-C_{14}-C_{23}+C_{24}$\\
$\rho$&$C_{35}-C_{45}$\\
$\theta$&$C_{12}-3C_{34}$\\
$\tau$&$3C_{12}-C_{34}$\\

\midrule[1pt]
\bottomrule[1pt]
\end{tabular}
\end{table}

\subsection{$(ccnn)^{I=1}\bar{q}$, $(bbnn)^{I=1}\bar{q}$, $(ccss)\bar{q}$, and $(bbss)\bar{q}$ states in the first class}

Three types of basis vectors are involved in calculating the
relevant matrix elements:
$[\phi^{AA}\otimes\chi^{SS}]\delta_{34}^{A}$,
$[\phi^{AS}\otimes\chi^{SA}]\delta_{34}^{A}$, and
$[\phi^{SA}\otimes\chi^{AS}]\delta_{34}^{A}$.

For the $J^P=\frac{5}{2}^-$ states, there is only one basis vector
$[(QQ)_{1}^{\bar{3}}(q_{3}q_{4})_{1}^{\bar{3}}\bar{q}]_{2}^{\frac{5}{2}}$.
The obtained Hamiltonian is
\begin{eqnarray}\label{eq11}
\langle H_{CM}\rangle_{J=\frac{5}{2}}=\frac{2}{3}(4\alpha+\beta+2\lambda+2\nu).
\end{eqnarray}

For the $J^P=\frac{3}{2}^-$ states, we have four basis vectors,
$[(QQ)_{1}^{\bar{3}}(q_{3}q_{4})_{1}^{\bar{3}}\bar{q}]_{2}^{\frac{3}{2}}$,
$[(QQ)_{1}^{\bar{3}}(q_{3}q_{4})_{1}^{\bar{3}}\bar{q}]_{1}^{\frac{3}{2}}$,
$[(QQ)_{1}^{\bar{3}}(q_{3}q_{4})_{0}^{6}\bar{q}]_{1}^{\frac{3}{2}}$,
and
$[(QQ)_{0}^{6}(q_{3}q_{4})_{1}^{\bar{3}}\bar{q}]_{1}^{\frac{3}{2}}$.
The resulting Hamiltonian is
\begin{eqnarray}\scriptsize
\langle H_{CM}\rangle_{J=\frac{3}{2}}= \frac{2}{3}
\begin{pmatrix}
4\alpha+\beta-3(\lambda+\nu) & \sqrt{5}(\nu-\lambda) & 3\sqrt{5}\nu & 3\sqrt{5}\lambda\\
\sqrt{5}(\nu-\lambda) & 4\alpha-\beta+\lambda+\nu & 3(\beta-\nu) & 3(\lambda-\beta)\\
3\sqrt{5}\nu & 3(\beta-\nu) & \frac{1}{2}(9\alpha-\theta)-\lambda & -\frac{3}{2}\beta\\
3\sqrt{5}\lambda & 3(\lambda-\beta) & -\frac{3}{2}\beta &
\frac{1}{2}(9\alpha+\tau)-\nu
\end{pmatrix}.\nonumber\\
\end{eqnarray}

For the $J^P=\frac{1}{2}^-$ states, the basis vectors are
$[(QQ)_{1}^{\bar{3}}(q_{3}q_{4})_{1}^{\bar{3}}\bar{q}]_{1}^{\frac{1}{2}}$,
$[(QQ)_{1}^{\bar{3}}(q_{3}q_{4})_{0}^{6}\bar{q}]_{1}^{\frac{1}{2}}$,
$[(QQ)_{0}^{6}(q_{3}q_{4})_{1}^{\bar{3}}\bar{q}]_{1}^{\frac{1}{2}}$,
and
$[(QQ)_{1}^{\bar{3}}(q_{3}q_{4})_{1}^{\bar{3}}\bar{q}]_{0}^{\frac{1}{2}}$
and the Hamiltonian reads
\begin{eqnarray}\scriptsize
\langle H_{CM}\rangle_{J=\frac{1}{2}}=\frac{2}{3}
\begin{pmatrix}
4\alpha-\beta-2(\lambda+\nu) & 3(\beta+2\nu) & -3(\beta+2\lambda) & 2\sqrt{2}(\nu-\lambda)\\
3(\beta+2\nu) & \frac{1}{2}(9\alpha-\theta)+2\lambda & -\frac{3}{2}\beta & -3\sqrt{2}\nu\\
-3(\beta+2\lambda) & -\frac{3}{2}\beta & \frac{1}{2}(9\alpha+\tau)+2\nu & -3\sqrt{2}\lambda\\
2\sqrt{2}(\nu-\lambda) & -3\sqrt{2}\nu & -3\sqrt{2}\lambda &
4\alpha-2\beta
\end{pmatrix}.\nonumber\\
\end{eqnarray}

\subsection{$(ccnn)^{I=0}\bar{q}$ and $(bbnn)^{I=0}\bar{q}$ states in the second class}

In this case, we also have three types of basis vectors to consider:
$[\phi^{AA}\chi^{SA}]\delta_{34}^{S}$,
$[\phi^{AS}\chi^{SS}]\delta_{34}^{S}$, and
$[\phi^{SA}\chi^{AA}]\delta_{34}^{S}$.

For the $J^P=\frac{5}{2}^-$ states, the involved basis vector is
$[(QQ)_{1}^{\bar{3}}(nn)_{1}^{6}\bar{q}]_{2}^{\frac{5}{2}}$ and the
obtained Hamiltonian is
\begin{eqnarray}\label{eq14}
\langle
H_{CM}\rangle_{J=\frac{5}{2}}=\frac{1}{3}(3\tau-\alpha+5\beta-2\lambda+10\nu).
\end{eqnarray}

For the $J^P=\frac{3}{2}^-$ states, there are three basis vectors
$[(QQ)_{1}^{\bar{3}}(nn)_{1}^{6}\bar{q}]_{2}^{\frac{3}{2}}$,
$[(QQ)_{1}^{\bar{3}}(nn)_{1}^{6}\bar{q}]_{1}^{\frac{3}{2}}$, and
$[(QQ)_{1}^{\bar{3}}(nn)_{0}^{\bar{3}}\bar{q}]_1^{\frac32}$. We can
get the following Hamiltonian,
\begin{eqnarray}\scriptsize
\langle H_{CM}\rangle_{J=\frac{3}{2}}=\frac{1}{3}
\begin{pmatrix}
3\tau-\alpha+5\beta+3\lambda-15\nu & \sqrt{5}(\lambda+5\nu) & 6\sqrt{5}\nu\\
\sqrt{5}(\lambda+5\nu) & 3\tau-\alpha-5\beta-\lambda+5\nu & 6(\beta-\nu)\\
6\sqrt{5}\nu & 6(\beta-\nu) & 4(2\theta+\lambda)
\end{pmatrix}.\nonumber\\
\end{eqnarray}

For the $J^P=\frac{1}{2}^-$ states, we have four basis vectors
$[(QQ)_{1}^{\bar{3}}(nn)_{1}^{6}\bar{q}]_{1}^{\frac{1}{2}}$,
$[(QQ)_{1}^{\bar{3}}(nn)_{0}^{\bar{3}}\bar{q}]_{1}^{\frac{1}{2}}$,
$[(QQ)_{0}^{6}(nn)_{0}^{\bar{3}}\bar{q}]_{0}^{\frac{1}{2}}$, and
$[(QQ)_{1}^{\bar{3}}(nn)_{1}^{6}\bar{q}]_{0}^{\frac{1}{2}}$. Then
the Hamiltonian
\begin{eqnarray}\scriptsize
\langle H_{CM}\rangle_{J=\frac{1}{2}}=\frac{1}{3}
\begin{pmatrix}
3\tau-\alpha-5\beta+2\lambda-10\nu & 6(\beta+2\nu) & 0 & 2\sqrt{2}(\lambda+5\nu)\\
6(\beta+2\nu) & 8(\theta-\lambda) & 6\sqrt{6}\lambda & -6\sqrt{2}\nu\\
0 & 6\sqrt{6}\lambda & 3(3\theta+\alpha) & 3\sqrt{3}\beta\\
 2\sqrt{2}(\lambda+5\nu) & -6\sqrt{2}\nu & 3\sqrt{3}\beta & 3\tau-\alpha-10\beta
\end{pmatrix}\nonumber\\
\end{eqnarray}
can be obtained.

\subsection{$(cbnn)^{I=1}\bar{q}$ and $(cbss)\bar{q}$ states in the third class}

\begin{widetext}
Now, one does not need to consider the constraint for the heavy
diquark from the Pauli principle and we then have six types of basis
vectors, $[\phi^{AA}\chi^{SS}]\delta_{34}^{A}$,
$[\phi^{AA}\chi^{AS}]\delta_{12}\delta_{34}^{A}$,
$[\phi^{AS}\chi^{SA}]\delta_{34}^{A}$,
$[\phi^{AS}\chi^{AA}]\delta_{12}\delta_{34}^{A}$,
$[\phi^{SA}\chi^{SS}]\delta_{12}\delta_{34}^{A}$, and
$[\phi^{SA}\chi^{AS}]\delta_{34}^{A}$.

For the $J^P=\frac{5}{2}^-$ states, two basis vectors,
$[(cb)_1^{\bar 3}(q_3q_4)_1^{\bar 3}\bar{q}]_2^{\frac52}$ and
$[(cb)_1^6(q_3q_4)_1^{\bar{3}}\bar{q}]_2^{\frac52}$, are involved
and the obtained Hamiltonian is
\begin{eqnarray}
\langle H_{CM}\rangle_{J=\frac{5}{2}}=\frac{1}{3}
\begin{pmatrix}
2(4\alpha+\beta+2\lambda+2\nu) & 3\sqrt{2}(\gamma-2\mu)\\
3\sqrt{2}(\gamma-2\mu) & 5\beta+10\lambda-2\nu-\alpha-3\theta
\end{pmatrix}.
\end{eqnarray}

For the $J^P=\frac32^-$ states, there are seven basis vectors,
$[(cb)_1^{\bar{3}}(q_3q_4)_1^{\bar{3}}\bar{q}]_2^{\frac32}$,
$[(cb)_1^{\bar{3}}(q_3q_4)_1^{\bar{3}}\bar{q}]_1^{\frac32}$,
$[(cb)_0^{\bar{3}}(q_3q_4)_1^{\bar{3}}\bar{q}]_1^{\frac32}$,
$[(cb)_1^{\bar{3}}(q_3q_4)_0^{6}\bar{q}]_1^{\frac32}$,
$[(cb)_1^6(q_3q_4)_{1}^{\bar{3}}\bar{q}]_2^{\frac32}$,
$[(cb)_1^6(q_3q_4)_{1}^{\bar{3}}\bar{q}]_1^{\frac32}$, and
$[(cb)_0^6(q_3q_4)_{1}^{\bar{3}}\bar{q}]_1^{\frac32}$. One obtains
the Hamiltonian as follows,
\begin{eqnarray}
&&\langle H_{CM}\rangle_{J=\frac{3}{2}}=\frac{2}{3}
\begin{pmatrix}
\begin{smallmatrix}
\begin{pmatrix}\begin{smallmatrix}4\alpha+\beta\\-3(\lambda+\nu)\end{smallmatrix}\end{pmatrix} & \sqrt{5}(\nu-\lambda) & -\sqrt{10}\mu & 3\sqrt{5}\nu & \frac{3}{\sqrt{2}}(\gamma+3\mu) & \frac{3\sqrt{10}}{2}\mu & 3\sqrt{5}\lambda\\
\sqrt{5}(\nu-\lambda) & \begin{pmatrix}\begin{smallmatrix}4\alpha-\beta\\+\lambda+\nu \end{smallmatrix}\end{pmatrix} & -\sqrt{2}(\gamma+\mu) & 3(\beta-\nu) & \frac{3\sqrt{10}}{2}\mu & -\frac{3}{\sqrt{2}}(\gamma+\mu) & 3(\lambda-\beta)\\
-\sqrt{10}\mu & -\sqrt{2}(\gamma+\nu) & 2(\nu-2\tau) & \frac{3}{\sqrt{2}}\gamma & 3\sqrt{5}\lambda & 3(\lambda-\beta) & 0\\
3\sqrt{5}\nu & 3(\beta-\nu) & \frac{3}{\sqrt{2}}\gamma & \frac{1}{2}(9\alpha-\theta)-\lambda & 0 & -\frac{3}{\sqrt{2}}\gamma & -\frac{3}{2}\beta\\
\frac{3}{\sqrt{3}}(\gamma+3\mu) & \frac{3\sqrt{10}}{2}\mu & 3\sqrt{5}\lambda & 0 & \frac{1}{2}\begin{pmatrix}\begin{smallmatrix}5\beta-15\lambda+\\3\nu-\alpha-3\theta \end{smallmatrix}\end{pmatrix} & -\frac{\sqrt{5}}{2}(5\lambda+\nu) & -\frac{5\sqrt{10}}{2}\mu\\
\sqrt{5}(\nu-\lambda) & -\frac{3}{\sqrt{2}}(\gamma+\mu) & 3(\lambda-\beta) & -\frac{3}{\sqrt{2}}\gamma & -\frac{\sqrt{5}}{2}(5\lambda+\nu) & \frac{1}{2}\begin{pmatrix}\begin{smallmatrix}5\lambda-\alpha-3\theta\\-5\beta-\nu\end{smallmatrix}\end{pmatrix} & -\frac{5}{\sqrt{2}}(\gamma+\mu)\\
3\sqrt{5}\lambda & 3(\lambda-\beta) & 0 & -\frac{3}{2}\beta &
-\frac{5\sqrt{10}}{2}\mu & -\frac{5}{\sqrt{2}}(\gamma+\mu) &
\frac{1}{2}(9\alpha+\tau)-\nu
\end{smallmatrix}
\end{pmatrix}.
\end{eqnarray}

For the $J^P=\frac{1}{2}^-$ states, eight basis vectors are
involved,
$[(cb)_{1}^{\bar{3}}(q_{3}q_{4})_{1}^{\bar{3}}\bar{q}]_{1}^{\frac{1}{2}}$,
$[(cb)_{1}^{\bar{3}}(q_{3}q_{4})_{1}^{\bar{3}}\bar{q}]_{0}^{\frac{1}{2}}$,
$[(cb)_{0}^{\bar{3}}(q_{3}q_{4})_{1}^{\bar{3}}\bar{q}]_{1}^{\frac{1}{2}}$,
$[(cb)_{1}^{\bar{3}}(q_{3}q_{4})_{0}^{6}\bar{q}]_{1}^{\frac{1}{2}}$,
$[(cb)_{0}^{\bar{3}}(q_{3}q_{4})_{0}^{6}\bar{q}]_{0}^{\frac{1}{2}}$,
$[(cb)_{1}^{6}(q_{3}q_{4})_{1}^{\bar{3}}\bar{q}]_{1}^{\frac{1}{2}}$,
$[(cb)_{1}^{6}(q_{3}q_{4})_{1}^{\bar{3}}\bar{q}]_{0}^{\frac{1}{2}}$,
and
$[(cb)_{0}^{6}(q_{3}q_{4})_{1}^{\bar{3}}\bar{q}]_{1}^{\frac{1}{2}}$.
The resulting Hamiltonian is
\begin{eqnarray}
\langle H_{CM}\rangle_{J=\frac{1}{2}}=\frac{2}{3}
\begin{pmatrix}
\begin{smallmatrix}
\begin{pmatrix}\begin{smallmatrix}4\alpha-\beta-\\2\lambda-2\nu\end{smallmatrix}\end{pmatrix} & 2\sqrt{2}(\nu-\lambda) & \sqrt{2}(2\mu-\gamma) & 3(\beta+2\nu) & 0 & \frac{3}{\sqrt{2}}(2\mu-\gamma) & 6\mu & -3(\beta+2\lambda)\\
2\sqrt{2}(\nu-\lambda) & 2(2\alpha-\beta) & 2\mu & -3\sqrt{2}\nu & -\frac{3\sqrt{6}}{2}\gamma & 6\mu & -3\sqrt{2}\gamma & -3\sqrt{2}\lambda\\
\sqrt{2}(2\mu-\gamma) & -3\sqrt{2}\nu & -4(\tau+\nu) & \frac{3}{\sqrt{2}}\gamma & 3\sqrt{6}\nu & -3(\beta+2\lambda) & -3\sqrt{2}\lambda & 0\\
3(\beta+2\nu) & -\frac{3\sqrt{6}}{2}\gamma & \frac{3}{\sqrt{2}}\gamma & \frac{1}{2}(9\alpha-\theta)+2\lambda & \sqrt{3}\mu & -\frac{3}{\sqrt{2}}\gamma & 0 & -\frac{3}{2}\beta \\
0 & -\frac{3\sqrt{6}}{2}\gamma & 3\sqrt{6}\nu & \sqrt{3}\mu & \frac{3}{2}(\alpha-3\tau) & 0 & \frac{3\sqrt{3}}{2}\beta & 0\\
\frac{3}{\sqrt{2}}(2\mu-\gamma) & -3\sqrt{2}\gamma & -3(\beta+2\lambda) & -\frac{3}{\sqrt{2}}\gamma & 0 & \begin{pmatrix}\begin{smallmatrix}-\frac{1}{2}(\alpha+3\theta)-\\ \frac{5}{2}\beta-5\lambda+\nu\end{smallmatrix}\end{pmatrix} & -\sqrt{2}(5\lambda+\nu) & \frac{5}{\sqrt{2}}(2\mu-\gamma)\\
6\mu & -3\sqrt{2}\gamma & -3\sqrt{2}\lambda & 0 & \frac{3\sqrt{3}}{2}\beta & -\sqrt{2}(5\lambda+\nu) & -\frac{1}{2}(\alpha+3\theta)-5\beta & 5\mu\\
-3(\beta+2\lambda) & -3\sqrt{2}\lambda & 0 & -\frac{3}{2}\beta & 0 &
\frac{5}{\sqrt{2}}(2\mu-\gamma) & 5\mu &
\frac{1}{2}(9\alpha+\tau)+2\nu
\end{smallmatrix}
\end{pmatrix}.
\end{eqnarray}

\subsection{$(cbnn)^{I=0}\bar{q}$ states in the fourth class}

In this case, we also have six types of basis vectors,
$[\phi^{AA}\chi^{SA}]\delta_{34}^{S}$,
$[\phi^{AA}\chi^{AA}]\delta_{12}\delta_{34}^{S}$,
$[\phi^{AS}\chi^{SS}]\delta_{34}^{S}$,
$[\phi^{AS}\chi^{AS}]\delta_{12}\delta_{34}^{S}$,
$[\phi^{SA}\chi^{SA}]\delta_{12}\delta_{34}^{S}$, and
$[\phi^{SA}\chi^{AA}]\delta_{34}^{S}$.

For the $J^P=\frac{5}{2}^-$ states, there is only one basis vector
$[(cb)_{1}^{\bar{3}}(nn)_{1}^{6}\bar{q}]_{2}^{\frac{5}{2}}$. The
obtained Hamiltonian is
\begin{eqnarray}
\langle
H_{CM}\rangle_{J=\frac{5}{2}}=\frac{1}{3}(3\tau-\alpha+5\beta-2\lambda+10\nu).
\end{eqnarray}

For the $J^P=\frac32^-$ states, the involved basis vectors are
$[(cb)_{1}^{\bar{3}}(nn)_{0}^{\bar{3}}\bar{q}]_{1}^{\frac{3}{2}}$,
$[(cb)_{1}^{\bar{3}}(nn)_{1}^{6}\bar{q}]_{2}^{\frac{3}{2}}$,
$[(cb)_{1}^{\bar{3}}(nn)_{1}^{6}\bar{q}]_{1}^{\frac{3}{2}}$,
$[(cb)_{0}^{\bar{3}}(nn)_{1}^{6}\bar{q}]_{1}^{\frac{3}{2}}$, and
$[(cb)_{1}^{6}(nn)_{0}^{\bar{3}}\bar{q}]_{1}^{\frac{3}{2}}$. The
Hamiltonian can be written as
\begin{eqnarray}\scriptsize
\langle H_{CM}\rangle_{J=\frac{3}{2}}=\frac{2}{3}
\begin{pmatrix}
4\theta+2\lambda & 3\sqrt{5}\nu & 3(\beta-\nu) & \frac{3}{\sqrt{2}}\gamma & -3\sqrt{2}\mu\\
3\sqrt{5}\nu & \frac{1}{2}\begin{pmatrix}3\tau-\alpha+5\beta\\+3\lambda-15\nu\end{pmatrix} & \frac{\sqrt{5}}{2}(\lambda+5\nu) & \frac{\sqrt{10}}{2}\mu & 0\\
3(\beta-\nu) & \frac{\sqrt{5}}{2}(\lambda+5\nu) & \frac{1}{2}\begin{pmatrix}3\tau-\alpha-5\beta\\-\lambda+5\nu\end{pmatrix} & \frac{1}{\sqrt{2}}(\mu-5\gamma) & -\frac{3}{\sqrt{2}}\gamma\\
\frac{3}{\sqrt{2}}\gamma & \frac{\sqrt{10}}{2}\mu & \frac{1}{\sqrt{2}}(\mu-5\gamma) & 5\nu-\frac{1}{2}(9\alpha+5\tau) & -\frac{3}{2}\beta\\
-3\sqrt{2}\mu & 0 & -\frac{3}{\sqrt{2}}\gamma & -\frac{3}{2}\beta &
5\lambda+\frac{1}{2}(5\theta-9\alpha)
\end{pmatrix}.
\end{eqnarray}

For the $J^P=\frac{1}{2}^-$ states, we have seven basis vectors,
$[(cb)_{1}^{\bar{3}}(nn)_{0}^{\bar{3}}\bar{q}]_{1}^{\frac{1}{2}}$,
$[(cb)_{0}^{\bar{3}}(nn)_{0}^{\bar{3}}\bar{q}]_{0}^{\frac{1}{2}}$,
$[(cb)_{1}^{\bar{3}}(nn)_{1}^{6}\bar{q}]_{1}^{\frac{1}{2}}$,
$[(cb)_{1}^{\bar{3}}(nn)_{1}^{6}\bar{q}]_{0}^{\frac{1}{2}}$,
$[(cb)_{0}^{\bar{3}}(nn)_{1}^{6}\bar{q}]_{1}^{\frac{1}{2}}$,
$[(cb)_{1}^{6}(nn)_{0}^{\bar{3}}\bar{q}]_{1}^{\frac{1}{2}}$, and
$[(cb)_{0}^{6}(nn)_{0}^{\bar{3}}\bar{q}]_{0}^{\frac{1}{2}}$. The
obtained Hamiltonian reads
\begin{eqnarray}
\langle H_{CM}\rangle_{J=\frac{1}{2}}=\frac{2}{3}
\begin{pmatrix}
\begin{smallmatrix}
4(\theta-\lambda) & -2\sqrt{3}\mu & 3(\beta+2\nu) & -3\sqrt{2}\nu & \frac{3}{\sqrt{2}}\gamma & 6\sqrt{2}\mu & 3\sqrt{6}\lambda\\
-2\sqrt{3}\mu & -12\alpha & 0 & -\frac{3\sqrt{6}}{2}\gamma & 3\sqrt{6}\nu & 3\sqrt{6}\lambda & 0 \\
3(\beta+2\nu) & 0 & \frac{1}{2}\begin{pmatrix}\begin{smallmatrix}3\tau-\alpha-5\beta\\+2\lambda-10\nu\end{smallmatrix}\end{pmatrix} & \sqrt{2}(\lambda+5\nu) & -\frac{1}{\sqrt{2}}(5\gamma+2\mu) & -\frac{3}{\sqrt{2}}\gamma & 0\\
-3\sqrt{2}\nu & -\frac{3\sqrt{6}}{2}\gamma & \sqrt{2}(\lambda+5\nu) & \frac{1}{2}(3\tau-\alpha)-5\beta & -\mu & 0 & \frac{3\sqrt{3}}{2}\beta\\
\frac{3}{\sqrt{2}}\gamma & 3\sqrt{6}\nu &  -\frac{1}{\sqrt{2}}(5\gamma+2\mu) & -\mu & -\frac{1}{2}(9\alpha+5\tau)-10\nu & -\frac{3}{2}\beta & 0\\
6\sqrt{2}\mu & 3\sqrt{6}\lambda & -\frac{3}{\sqrt{2}}\gamma & 0 & -\frac{3}{2}\beta & \frac{1}{2}(5\theta-9\alpha)-10\lambda & -5\sqrt{3}\mu\\
3\sqrt{6}\lambda & 0 & 0 & \frac{3\sqrt{3}}{2}\beta & 0 &
-5\sqrt{3}\mu & \frac{3}{2}(3\theta+\alpha)
\end{smallmatrix}
\end{pmatrix}.
\end{eqnarray}

\subsection{$(ccns)\bar{q}$ and $(bbns)\bar{q}$ states in the fifth class}

In this case, again we have six types of basis vectors,
$[\phi^{AA}\chi^{SS}]\delta_{34}^{A}$,
$[\phi^{AA}\chi^{SA}]\delta_{34}^{S}$,
$[\phi^{AS}\chi^{SS}]\delta_{34}^{S}$,
$[\phi^{AS}\chi^{SA}]\delta_{34}^{A}$,
$[\phi^{SA}\chi^{AS}]\delta_{34}^{A}$, and
$[\phi^{SA}\chi^{AA}]\delta_{34}^{S}$.

For the $J^P=\frac{5}{2}^-$ states, the basis vectors are
$[(QQ)_{1}^{\bar{3}}(ns)_{1}^{\bar{3}}\bar{q}]_{2}^{\frac{5}{2}}$
and $[(QQ)_{1}^{\bar{3}}(ns)_{1}^{6}\bar{q}]_{2}^{\frac{5}{2}}$ and
the Hamiltonian is
\begin{eqnarray}
\langle H_{CM}\rangle_{J=\frac{5}{2}}=\frac{2}{3}
\begin{pmatrix}
4\alpha+\beta+2\lambda+2\nu & \frac{3}{\sqrt{2}}(\delta-2\rho)\\
\frac{3}{\sqrt{2}}(\delta-2\rho) &
\frac{1}{2}(3\tau-\alpha+5\beta-2\lambda+10\nu)
\end{pmatrix}.
\end{eqnarray}

For the $J^P=\frac{3}{2}^-$ states, the involved basis vectors are
$[(QQ)_{1}^{\bar{3}}(ns)_{1}^{\bar{3}}\bar{q}]_{2}^{\frac{3}{2}}$,
$[(QQ)_{1}^{\bar{3}}(ns)_{1}^{\bar{3}}\bar{q}]_{1}^{\frac{3}{2}}$,
$[(QQ)_{1}^{\bar{3}}(ns)_{0}^{\bar{3}}\bar{q}]_{1}^{\frac{3}{2}}$,
$[(QQ)_{1}^{\bar{3}}(ns)_{1}^{6}\bar{q}]_{2}^{\frac{3}{2}}$,
$[(QQ)_{1}^{\bar{3}}(ns)_{1}^{6}\bar{q}]_{1}^{\frac{3}{2}}$,
$[(QQ)_{1}^{\bar{3}}(ns)_{0}^{6}\bar{q}]_{1}^{\frac{3}{2}}$, and
$[(QQ)_{0}^{6}(ns)_{1}^{\bar{3}}\bar{q}]_{1}^{\frac{3}{2}}$. Then
one can get
\begin{eqnarray}
\langle H_{CM}\rangle_{J=\frac{3}{2}}=\frac{2}{3}
\begin{pmatrix}
\begin{smallmatrix}
\begin{pmatrix}\begin{smallmatrix}4\alpha+\beta-\\3(\lambda+\nu)\end{smallmatrix}\end{pmatrix} & \sqrt{5}(\nu-\lambda) & -\sqrt{10}\rho & \frac{3}{\sqrt{2}}(\delta+3\rho) & -\frac{3\sqrt{10}}{2}\rho & 3\sqrt{5}\nu & 3\sqrt{5}\lambda\\
\sqrt{5}(\nu-\lambda) & \begin{pmatrix}\begin{smallmatrix}4\alpha-\beta\\+\lambda+\nu\end{smallmatrix}\end{pmatrix} & \sqrt{2}(\delta+\rho) & -\frac{3\sqrt{10}}{2}\rho & -\frac{3}{\sqrt{2}}(\delta+\rho) & 3(\beta-\nu) & 3(\lambda-\beta)\\
-\sqrt{10}\rho & \sqrt{2}(\delta+\rho) & 4\theta+2\lambda & 3\sqrt{5}\nu & 3(\beta-\nu) & 0 & \frac{3}{\sqrt{2}}\delta\\
\frac{3}{\sqrt{2}}(\delta+3\rho) & -\frac{3\sqrt{10}}{2}\rho & 3\sqrt{5}\nu & \frac{1}{2}\begin{pmatrix}\begin{smallmatrix}3\tau-\alpha+5\beta\\+3\lambda-15\nu\end{smallmatrix}\end{pmatrix} & \frac{\sqrt{5}}{2}(\lambda+5\nu) & -\frac{5\sqrt{10}}{2}\rho & 0\\
-\frac{3\sqrt{10}}{2}\rho & -\frac{3}{\sqrt{2}}(\delta+\rho) & 3(\beta-\nu) & \frac{\sqrt{5}}{2}(\lambda+5\nu) & \frac{1}{2}\begin{pmatrix}\begin{smallmatrix}3\tau-\alpha-\lambda\\-5\beta+5\nu\end{smallmatrix}\end{pmatrix} & \frac{5}{\sqrt{2}}(\delta+\rho) & \frac{3}{\sqrt{2}}\delta\\
3\sqrt{5}\nu & 3(\beta-\nu) & 0 & -\frac{5\sqrt{10}}{2}\rho & \frac{5}{\sqrt{2}}(\delta+\rho) & \frac{1}{2}(9\alpha-\theta)-\lambda & -\frac{3}{2}\beta\\
3\sqrt{5}\lambda & 3(\lambda-\beta) & \frac{3}{\sqrt{2}}\delta & 0 &
\frac{3}{\sqrt{2}}\delta & -\frac{3}{2}\beta &
\frac{1}{2}(9\alpha+\tau)-\nu
\end{smallmatrix}
\end{pmatrix}.
\end{eqnarray}

For the $J^P=\frac{1}{2}^-$ states, the basis vectors are
$[(QQ)_{1}^{\bar{3}}(ns)_{1}^{\bar{3}}\bar{q}]_{1}^{\frac{1}{2}}$,
$[(QQ)_{1}^{\bar{3}}(ns)_{1}^{\bar{3}}\bar{q}]_{0}^{\frac{1}{2}}$,
$[(QQ)_{1}^{\bar{3}}(ns)_{0}^{\bar{3}}\bar{q}]_{1}^{\frac{1}{2}}$,
$[(QQ)_{1}^{\bar{3}}(ns)_{1}^{6}\bar{q}]_{1}^{\frac{1}{2}}$,
$[(QQ)_{1}^{\bar{3}}(ns)_{1}^{6}\bar{q}]_{0}^{\frac{1}{2}}$,
$[(QQ)_{1}^{\bar{3}}(ns)_{0}^{6}\bar{q}]_{1}^{\frac{1}{2}}$,
$[(QQ)_{0}^{6}(ns)_{1}^{\bar{3}}\bar{q}]_{1}^{\frac{1}{2}}$, and
$[(QQ)_{0}^{6}(ns)_{0}^{\bar{3}}\bar{q}]_{0}^{\frac{1}{2}}$. The
derived Hamiltonian reads
\begin{eqnarray}
\langle H_{CM}\rangle_{J=\frac{1}{2}}=\frac{2}{3}
\begin{pmatrix}
\begin{smallmatrix}
\begin{pmatrix}\begin{smallmatrix}4\alpha-\beta\\-2\lambda-2\nu\end{smallmatrix}\end{pmatrix} & 2\sqrt{2}(\nu-\lambda) & \sqrt{2}(\delta-2\rho) & \frac{3}{\sqrt{2}}(2\rho-\delta) &  -6\rho & 3(\beta+2\nu) & -3(\beta+2\lambda) & 0\\
2\sqrt{2}(\nu-\lambda) & 2(2\alpha-\beta) & 2\rho & -6\rho & -3\sqrt{2}\delta & -3\sqrt{2}\nu & -3\sqrt{2}\lambda & -\frac{3\sqrt{6}}{2}\delta\\
\sqrt{2}(\delta-2\rho) & 2\rho & 4(\theta-\lambda) & 3(\beta+2\nu) & -3\sqrt{2}\nu & 0 & \frac{3}{\sqrt{2}}\delta & 3\sqrt{6}\lambda\\
\frac{3}{\sqrt{2}}(2\rho-\delta) & -6\rho & 3(\beta+2\nu) & \frac{1}{2}\begin{pmatrix}\begin{smallmatrix}3\tau-\alpha-5\beta\\+2\lambda-10\nu\end{smallmatrix}\end{pmatrix} & \sqrt{2}(\lambda+5\nu) & \frac{5}{\sqrt{2}}(\delta-2\rho) & \frac{3}{\sqrt{2}}\delta & 0\\
 -6\rho & -3\sqrt{2}\delta & -3\sqrt{2}\nu & \sqrt{2}(\lambda+5\nu) & \frac{1}{2}(3\tau-\alpha)-5\beta & 5\rho & 0 & \frac{3\sqrt{3}}{2}\beta\\
 3(\beta+2\nu) & -3\sqrt{2}\nu & 0 & \frac{5}{\sqrt{2}}(\delta-2\rho) & 5\rho & \frac{1}{2}(9\alpha-\theta)+2\lambda & -\frac{3}{2}\beta & 0 \\
-3(\beta+2\lambda) & -3\sqrt{2}\lambda & \frac{3}{\sqrt{2}}\delta & \frac{3}{\sqrt{2}}\delta & 0 & -\frac{3}{2}\beta & \frac{1}{2}(9\alpha+\tau)+2\nu & \sqrt{3}\rho\\
0 & -\frac{3\sqrt{6}}{2}\delta & 3\sqrt{6}\lambda & 0 &
\frac{3\sqrt{3}}{2}\beta & 0 & \sqrt{3}\rho &
\frac{3}{2}(3\theta+\alpha)
\end{smallmatrix}
\end{pmatrix}.
\end{eqnarray}

\subsection{$(cbns)\bar{q}$ states in the sixth class}

\rotatebox[origin=c]{90}{
\begin{minipage}{\textheight}
\begin{eqnarray}\label{Hclass5}
\langle H_{CM}\rangle_{J=\frac{3}{2}}=\frac{2}{3}
\left(\begin{matrix}
\begin{smallmatrix}
\begin{pmatrix}\begin{smallmatrix}4\alpha+\beta\\-3\lambda-3\nu \end{smallmatrix}\end{pmatrix} & \sqrt{5}(\nu-\lambda) & -\sqrt{10}\rho & -\sqrt{10}\mu & \frac{3}{\sqrt{2}}(\delta+3\rho) & -\frac{3\sqrt{10}}{2}\rho &3\sqrt{5}\nu & 0 & \frac{3\sqrt{2}}{2}(\gamma+3\mu) & \frac{3\sqrt{10}}{2}\mu & 0 & 3\sqrt{5}\lambda\\
\sqrt{5}(\nu-\lambda) &
\begin{pmatrix}\begin{smallmatrix}4\alpha-\beta\\+\lambda+\nu\end{smallmatrix}\end{pmatrix}
& \sqrt{2}(\delta+\rho) & -\sqrt{2}(\gamma+\mu) &
-\frac{3\sqrt{10}}{2}\rho & -\frac{3\sqrt{2}}{2}(\delta+\rho)
&3(\beta-\nu) & -3\eta & \frac{3\sqrt{10}}{2}\mu & -\frac{3\sqrt{2}}{2}(\gamma+\mu) & 3\eta & 3(\lambda-\beta)\\
-\sqrt{10}\rho & \sqrt{2}(\delta+\rho) & 4\theta+2\lambda & \eta &
3\sqrt{5}\nu & 3(\beta-\nu)
&0 & \frac{3}{\sqrt{2}}\gamma & 0 & 3\eta & -3\sqrt{2}\mu & \frac{3}{\sqrt{2}}\delta\\
-\sqrt{10}\mu & -\sqrt{2}(\gamma+\mu) & \eta & 2\nu-4\tau & 0 &
-3\eta
&\frac{3}{\sqrt{2}}\gamma & -3\sqrt{2}\rho & 3\sqrt{5}\lambda & 3(\lambda-\beta) & \frac{3}{\sqrt{2}}\delta & 0\\
\frac{3}{\sqrt{2}}(\delta+3\rho) & -\frac{3\sqrt{10}}{2}\rho &
3\sqrt{5}\nu & 0 &
\frac{1}{2}\begin{pmatrix}\begin{smallmatrix}3\tau-\alpha+\\5\beta+3\lambda\\-15\nu\end{smallmatrix}\end{pmatrix}
& \frac{\sqrt{5}}{2}(\lambda+5\nu)
&-\frac{5\sqrt{10}}{2}\rho & \frac{\sqrt{10}}{2}\mu & -\frac{3}{2}\eta & 0 & 0 & 0\\
-\frac{3\sqrt{10}}{2}\rho & -\frac{3\sqrt{2}}{2}(\delta+\rho) &
3(\beta-\nu) & -3\eta & \frac{\sqrt{5}}{2}(\lambda+5\nu) &
\frac{1}{2}\begin{pmatrix}\begin{smallmatrix}3\tau-\alpha\\-5\beta-\lambda\\+5\nu\end{smallmatrix}\end{pmatrix}
&\frac{5}{\sqrt{2}}(\delta+\rho) & \frac{1}{\sqrt{2}}(\mu-5\gamma) & 0 & \frac{3}{2}\eta & -\frac{3}{\sqrt{2}}\gamma & \frac{3}{\sqrt{2}}\delta\\
3\sqrt{5}\nu & 3(\beta-\nu) & 0 & \frac{3}{\sqrt{2}}\gamma &
-\frac{5\sqrt{10}}{2}\rho & \frac{5}{\sqrt{2}}(\delta+\rho)
&\frac92\alpha-\frac12\theta-\lambda & \frac{5}{2}\eta & 0 & -\frac{3}{\sqrt{2}}\gamma & 0 & -\frac{3}{2}\beta\\
0 & -3\eta & \frac{3}{\sqrt{2}}\gamma & -3\sqrt{2}\rho &
\frac{\sqrt{10}}{2}\mu & \frac{1}{\sqrt{2}}(\mu-5\gamma)
&\frac{5}{2}\eta & 5\nu-\frac92\alpha-\frac52\tau & 0 & \frac{3}{\sqrt{2}}\delta & -\frac{3}{2}\beta & 0\\
\frac{3\sqrt{2}}{2}(\gamma+3\mu) & \frac{3\sqrt{10}}{2}\mu & 0 &
3\sqrt{5}\lambda & -\frac{3}{2}\eta & 0
&0 & 0 & \frac{1}{2}\begin{pmatrix}\begin{smallmatrix}5\beta-\alpha-\\3\theta+3\nu\\-15\lambda\end{smallmatrix}\end{pmatrix}& -\frac{\sqrt{5}}{2}(5\lambda+\nu) & \frac{\sqrt{10}}{2}\rho & -\frac{5\sqrt{10}}{2}\mu\\
\frac{3\sqrt{10}}{2}\mu & -\frac{3\sqrt{2}}{2}(\gamma+\mu) & 3\eta &
3(\lambda-\beta) & 0 & \frac{3}{2}\eta
&-\frac{3}{\sqrt{2}}\gamma & \frac{3}{\sqrt{2}}\delta & -\frac{\sqrt{5}}{2}(5\lambda+\nu) & \frac{1}{2}\begin{pmatrix}\begin{smallmatrix}5\lambda-3\theta\\-\alpha-\nu\\-5\beta\end{smallmatrix}\end{pmatrix} & \frac{1}{\sqrt{2}}(5\delta-\rho) & -\frac{5}{\sqrt{2}}(\gamma+\mu)\\
0 & 3\eta & -3\sqrt{2}\mu & \frac{3}{\sqrt{2}}\delta & 0 &
-\frac{3}{\sqrt{2}}\gamma
&0 & -\frac{3}{2}\beta & \frac{\sqrt{10}}{2}\rho & \frac{1}{\sqrt{2}}(5\delta-\rho) & 5\lambda+\frac52\theta-\frac92\alpha & \frac{5}{2}\eta\\
3\sqrt{5}\lambda & 3(\lambda-\beta) & \frac{3}{\sqrt{2}}\delta & 0 &
0 & \frac{3}{\sqrt{2}}\delta
&-\frac{3}{2}\beta & 0 & -\frac{5\sqrt{10}}{2}\mu & -\frac{5}{\sqrt{2}}(\gamma+\mu) & \frac{5}{2}\eta & \frac92\alpha+\frac12\tau-\nu\\
\end{smallmatrix}
\end{matrix}\right).\nonumber\\
\end{eqnarray}
\begin{eqnarray}\label{Hclass6}
&&\langle H_{CM}\rangle_{J=\frac{1}{2}}=\frac{2}{3}\times \nonumber\\
&&\left(\begin{matrix}
\begin{smallmatrix}
\begin{pmatrix}\begin{smallmatrix}4\alpha-\beta\\-2\lambda-2\nu\end{smallmatrix}\end{pmatrix} & 2\sqrt{2}(\nu-\lambda) & \sqrt{2}(\delta-2\rho) & \sqrt{2}(2\mu-\gamma) & 0 & \frac{3}{\sqrt{2}}(2\rho-\delta) & -6\rho
&3(\beta+2\nu) & -3\eta & 0 & \frac{3}{\sqrt{2}}(2\mu-\gamma) & 6\mu & 3\eta & -3(\beta+2\lambda) & 0\\
2\sqrt{2}(\nu-\lambda) & 2(2\alpha-\beta) & 2\rho & 2\mu &
-\sqrt{3}\eta & -6\rho & -3\sqrt{2}\delta
&-3\sqrt{2}\nu & 0 & -\frac{3\sqrt{6}}{2}\gamma & 6\mu & -3\sqrt{2}\gamma & 0 & -3\sqrt{2}\lambda & -\frac{3\sqrt{6}}{2}\delta\\
\sqrt{2}(\delta-2\rho) & 2\rho & 4(\theta-\lambda) & \eta &
-2\sqrt{3}\mu & 3(\beta+2\nu) & -3\sqrt{2}\nu
&0 & \frac{3}{\sqrt{2}}\gamma & 0 & 3\eta & 0 & 6\sqrt{2}\mu & \frac{3}{\sqrt{2}}\delta & 3\sqrt{6}\lambda\\
\sqrt{2}(2\mu-\gamma) & 2\mu & \eta & -4(\tau+\nu) & -2\sqrt{3}\rho
& -6\eta & 0
&\frac{3}{\sqrt{2}}\gamma & 6\sqrt{2}\rho & 3\sqrt{6}\nu & -3(\beta+2\lambda) & -3\sqrt{2}\lambda & \frac{3}{\sqrt{2}}\delta & 0 & 0\\
0 & -\sqrt{3}\eta & -2\sqrt{3}\mu & -2\sqrt{3}\rho & -12\alpha & 0 &
-\frac{3\sqrt{6}}{2}\gamma
&0 & 3\sqrt{6}\nu & 0 & 0 & -\frac{3\sqrt{6}}{2}\delta & 3\sqrt{6}\lambda & 0 & 0\\
\frac{3}{\sqrt{2}}(2\rho-\delta) & -6\rho & 3(\beta+2\nu) & -6\eta &
0 &
\frac{1}{2}\begin{pmatrix}\begin{smallmatrix}3\tau-\alpha-\\5\beta+2\lambda\\-10\nu\end{smallmatrix}\end{pmatrix}
& \sqrt{2}(\lambda+5\nu)
&\frac{5}{\sqrt{2}}(\delta-2\rho) & -\frac{1}{\sqrt{2}}(5\gamma-2\mu) & 0 & -\frac{3}{2}\eta & 0 & -\frac{3}{\sqrt{2}}\gamma & \frac{3}{\sqrt{2}}\delta & 0\\
-6\rho & -3\sqrt{2}\delta & -3\sqrt{2}\nu & 0 &
-\frac{3\sqrt{6}}{2}\gamma & \sqrt{2}(\lambda+5\nu) &
\frac32\tau-\frac12\alpha-5\beta
&5\rho & -\lambda & -\frac{5\sqrt{3}}{2}\eta & 0 & 3\eta & 0 & 0 & \frac{3\sqrt{3}}{2}\beta\\
3(\beta+2\nu) & -3\sqrt{2}\nu & 0 & \frac{3}{\sqrt{2}}\gamma & 0 &
\frac{5}{\sqrt{2}}(\delta-2\rho) & 5\rho
&\frac92\alpha-\frac12\theta+2\lambda & \frac{5}{2}\eta & \sqrt{3}\mu & -\frac{3}{\sqrt{2}}\gamma & 0 & 0 & -\frac{3}{2}\beta & 0\\
-3\eta & 0 & \frac{3}{\sqrt{2}}\gamma & 6\sqrt{2}\rho & 3\sqrt{6}\nu
& \sqrt{2}\mu-\frac{5}{\sqrt2}\gamma & -\lambda
&\frac{5}{2}\eta & -\frac92\alpha-\frac52\tau-10\nu & -5\sqrt{3}\rho & \frac{3}{\sqrt{2}}\delta & 0 & -\frac{3}{2}\beta & 0 & 0\\
0 & -\frac{3\sqrt{6}}{2}\gamma & 0 & 3\sqrt{6}\nu & 0 & 0 &
-\frac{5\sqrt{3}}{2}\eta
&\sqrt{3}\mu & -5\sqrt{3}\rho & \frac32\alpha-\frac92\tau & 0 & \frac{3\sqrt{3}}{2}\beta & 0 & 0 & 0\\
\frac{3}{\sqrt{2}}(2\mu-\gamma) & 6\mu & 3\eta & -3(\beta+2\lambda)
& 0 & -\frac{3}{2}\eta & 0
&-\frac{3}{\sqrt{2}}\gamma &\frac{3}{\sqrt{2}}\delta & 0 & \frac{1}{2}\begin{pmatrix}\begin{smallmatrix}-3\theta-\alpha\\+\nu-5\beta\\-10\lambda\end{smallmatrix}\end{pmatrix} & -\sqrt{2}(5\lambda+\nu) & \sqrt{2}(\frac{5}{2}\delta+\rho) & \frac{5}{\sqrt{2}}(2\mu-\gamma) & 0\\
6\mu & -3\sqrt{2}\gamma & 0 & -3\sqrt{2}\lambda &
-\frac{3\sqrt{6}}{2}\delta & 0 & 3\eta
&0 & 0 & \frac{3\sqrt{3}}{2}\beta & -\sqrt{2}(5\lambda+\nu) & -\frac32\theta-\frac12\alpha-5\beta & -\rho & 5\mu & -\frac{5\sqrt{3}}{2}\eta\\
3\eta & 0 & 6\sqrt{2}\mu & \frac{3}{\sqrt{2}}\delta &
3\sqrt{6}\lambda & -\frac{3}{\sqrt{2}}\gamma & 0
&0 & -\frac{3}{2}\beta & 0 & \sqrt{2}(\frac{5}{2}\delta+\rho) & -\rho & \frac52\theta-\frac92\alpha-10\lambda & \frac{5}{2}\eta & -5\sqrt{3}\mu\\
-3(\beta+2\lambda) & -3\sqrt{2}\lambda & \frac{3}{\sqrt{2}}\delta &
0 & 0 & \frac{3}{\sqrt{2}}\delta & 0
&-\frac{3}{2}\beta & 0 & 0 & \frac{5}{\sqrt{2}}(2\mu-\gamma) & 5\mu & \frac{5}{2}\eta & \frac92\alpha+\frac12\tau+2\nu & \sqrt{3}\rho\\
0 & -\frac{3\sqrt{6}}{2}\delta & 3\sqrt{6}\lambda & 0 & 0 & 0 &
\frac{3\sqrt{3}}{2}\beta
&0 & 0 & 0 & 0 & -\frac{5\sqrt{3}}{2}\eta & -5\sqrt{3}\mu & \sqrt{3}\rho & \frac32\alpha+\frac92\theta 
\end{smallmatrix}
\end{matrix}\right). \nonumber\\
\end{eqnarray}
\end{minipage}}

Since the Pauli principle has no effects in this case, the most
basis vectors are involved. There are twelve types of bases,
$[\phi^{AA}\chi^{SS}]\delta_{34}^{A}$,
$[\phi^{AA}\chi^{SA}]\delta_{34}^{S}$,
$[\phi^{AA}\chi^{AS}]\delta_{12}\delta_{34}^{A}$,
$[\phi^{AA}\chi^{AA}]\delta_{12}\delta_{34}^{S}$,
$[\phi^{AS}\chi^{SS}]\delta_{34}^{S}$,
$[\phi^{AS}\chi^{SA}]\delta_{34}^{A}$,
$[\phi^{AS}\chi^{AS}]\delta_{12}\delta_{34}^{S}$,
$[\phi^{AS}\chi^{AA}]\delta_{12}\delta_{34}^{A}$,
$[\phi^{SA}\chi^{SS}]\delta_{12}\delta_{34}^{A}$,
$[\phi^{SA}\chi^{SA}]\delta_{12}\delta_{34}^{S}$,
$[\phi^{SA}\chi^{AS}]\delta_{34}^{A}$, and
$[\phi^{SA}\chi^{AA}]\delta_{34}^{S}$.

For the $J^P=\frac{5}{2}^-$ states, the basis vectors are
$[(cb)_1^{\bar{3}}(ns)_{1}^{\bar{3}}\bar{q}]_{2}^{\frac{5}{2}}$,
$[(cb)_{1}^{\bar{3}}(ns)_{1}^{6}\bar{q}]_{2}^{\frac{5}{2}}$, and
$[(cb)_{1}^{6}(ns)_{1}^{\bar{3}}\bar{q}]_{2}^{\frac{5}{2}}$ and the
Hamiltonian is
\begin{eqnarray}
\langle H_{CM}\rangle_{J=\frac{5}{2}}=\frac{1}{3}
\begin{pmatrix}
2(4\alpha+\beta+2\lambda+2\nu) & 3\sqrt{2}(\delta-2\rho) & 3\sqrt{2}(\gamma-2\mu)\\
3\sqrt{2}(\delta-2\rho) & 3\tau-\alpha+5\beta-2\lambda+10\nu & -3\eta\\
3\sqrt{2}(\gamma-2\mu) & -3\eta &
5\beta+10\lambda-2\nu-(\alpha+3\theta)
\end{pmatrix}.
\end{eqnarray}
\end{widetext}

For the $J^P=\frac{3}{2}^-$ states, the basis vectors are
$[(cb)_{1}^{\bar{3}}(ns)_{1}^{\bar{3}}\bar{q}]_{2}^{\frac{3}{2}}$,
$[(cb)_{1}^{\bar{3}}(ns)_{1}^{\bar{3}}\bar{q}]_{1}^{\frac{3}{2}}$,
$[(cb)_{1}^{\bar{3}}(ns)_{0}^{\bar{3}}\bar{q}]_{1}^{\frac{3}{2}}$,
$[(cb)_{0}^{\bar{3}}(ns)_{1}^{\bar{3}}\bar{q}]_{1}^{\frac{3}{2}}$,
$[(cb)_{1}^{\bar{3}}(ns)_{1}^{6}\bar{q}]_{2}^{\frac{3}{2}}$,
$[(cb)_{1}^{\bar{3}}(ns)_{1}^{6}\bar{q}]_{1}^{\frac{3}{2}}$,
$[(cb)_{1}^{\bar{3}}(ns)_{0}^{6}\bar{q}]_{1}^{\frac{3}{2}}$,
$[(cb)_{0}^{\bar{3}}(ns)_{1}^{6}\bar{q}]_{1}^{\frac{3}{2}}$,
$[(cb)_{1}^{6}(ns)_{1}^{\bar{3}}\bar{q}]_{2}^{\frac{3}{2}}$,
$[(cb)_{1}^{6}(ns)_{1}^{\bar{3}}\bar{q}]_{1}^{\frac{3}{2}}$,
$[(cb)_{1}^{6}(ns)_{0}^{\bar{3}}\bar{q}]_{1}^{\frac{3}{2}}$, and
$[(cb)_{0}^{6}(ns)_{1}^{\bar{3}}\bar{q}]_{1}^{\frac{3}{2}}$. The
resulting Hamiltonian is given in Eq. \eqref{Hclass5}.

For the $J^P=\frac{1}{2}^-$ states, the fifteen basis vectors are
$[(cb)_{1}^{\bar{3}}(ns)_{1}^{\bar{3}}\bar{q}]_{1}^{\frac{1}{2}}$,
$[(cb)_{1}^{\bar{3}}(ns)_{1}^{\bar{3}}\bar{q}]_{0}^{\frac{1}{2}}$,
$[(cb)_{1}^{\bar{3}}(ns)_{0}^{\bar{3}}\bar{q}]_{1}^{\frac{1}{2}}$,
$[(cb)_{0}^{\bar{3}}(ns)_{1}^{\bar{3}}\bar{q}]_{1}^{\frac{1}{2}}$,
$[(cb)_{0}^{\bar{3}}(ns)_{0}^{\bar{3}}\bar{q}]_{0}^{\frac{1}{2}}$,
$[(cb)_{1}^{\bar{3}}(ns)_{1}^{6}\bar{q}]_{1}^{\frac{1}{2}}$,
$[(cb)_{1}^{\bar{3}}(ns)_{1}^{6}\bar{q}]_{0}^{\frac{1}{2}}$,
$[(cb)_{1}^{\bar{3}}(ns)_{0}^{6}\bar{q}]_{1}^{\frac{1}{2}}$,
$[(cb)_{0}^{\bar{3}}(ns)_{1}^{6}\bar{q}]_{1}^{\frac{1}{2}}$,
$[(cb)_{0}^{\bar{3}}(ns)_{0}^{6}\bar{q}]_{0}^{\frac{1}{2}}$,
$[(cb)_{1}^{6}(ns)_{1}^{\bar{3}}\bar{q}]_{1}^{\frac{1}{2}}$,
$[(cb)_{1}^{6}(ns)_{1}^{\bar{3}}\bar{q}]_{0}^{\frac{1}{2}}$,
$[(cb)_{1}^{6}(ns)_{0}^{\bar{3}}\bar{q}]_{1}^{\frac{1}{2}}$,
$[(cb)_{0}^{6}(ns)_{1}^{\bar{3}}\bar{q}]_{1}^{\frac{1}{2}}$, and
$[(cb)_{0}^{6}(ns)_{1}^{\bar{3}}\bar{q}]_{0}^{\frac{1}{2}}$. We
present the obtained Hamiltonian in Eq. \eqref{Hclass6}.

\section{The $QQqq\bar{q}$ pentaquark mass spectra}\label{sec4}


Now, we determine the values of the seventeen coupling parameters
($C_{nn}$, $C_{ns}$, $C_{ss}$, $C_{cn}$, $C_{bn}$, $C_{cs}$,
$C_{bs}$, $C_{bc}$, $C_{cc}$, $C_{bb}$, $C_{n\bar{n}}$,
$C_{s\bar{n}}=C_{n\bar{s}}$, $C_{s\bar{s}}$, $C_{c\bar{n}}$,
$C_{b\bar{n}}$, $C_{c\bar{s}}$, and $C_{b\bar{s}}$) and the four
effective quark masses ($m_n$, $m_s$, $m_c$, and $m_b$) in order to
estimate the pentaquark masses. The procedure to extract the
parameters has been illustrated in Ref. \cite{Wu:2016gas}. From the
calculated CMI matrix elements for ground state hadrons and their
mass splittings, we can get most values of the coupling parameters
which are shown in Table \ref{parameter}. To determine
$C_{s\bar{s}}$, one needs the mass of a ground pseudoscalar meson
having the same quark content with $\phi$. Since there is no such a
state, here we adopt approximately $C_{s\bar{s}}=C_{ss}$. Similarly,
we use the approximation $C_{QQ}=C_{Q\bar{Q}}$
($C_{bb}=C_{b\bar{b}}=2.9$ MeV, $C_{cc}=C_{c\bar{c}}=5.3$ MeV, and
$C_{bc}=C_{b\bar{c}}=3.3$ MeV) since only one doubly heavy baryon
$\Xi_{cc}$ is observed. In Table \ref{parameter}, the $B_{c}^{*}$
has not been observed yet and we take its mass from a model
calculation \cite{Godfrey:1985xj}. {The effective quark masses can be extracted from the ground state baryons after the
determination of the coupling parameters, and we present them in Table \ref{quark mass}.}
\begin{widetext}
\begin{center}
\begin{table}[htbp]
\caption{The extracted effective coupling parameters.}\label{parameter} \centering
\begin{tabular}{cccccc}
\toprule[1.5pt] \midrule[1pt]
Hadron&CMI&Hadron&CMI&Parameter(MeV)\\
\midrule[1pt]
$N$&$-8C_{nn}$&$\Delta$&$8C_{nn}$&$C_{nn}=18.4$\\
$\Sigma$&$\frac{8}{3}C_{nn}-\frac{32}{3}C_{n s}$&$\Sigma^*$&$\frac{8}{3}C_{nn}+\frac{16}{3}C_{n s}$&$C_{n s}=12.4$\\
$\Xi^0$&$\frac{8}{3}(C_{ss}-4C_{n s})$&$\Xi^{*0}$&$\frac{8}{3}(C_{ss}+C_{n s})$&\\
$\Omega$&8$C_{ss}$&&&$C_{ss}=6.5$\\
$\Lambda$&$-8C_{nn}$\\
$\pi^{0}$&$-16C_{n\bar{n}}$&$\rho$&$\frac{16}{3}C_{n\bar{n}}$&$C_{n\bar{n}}=30.0$\\
$K$ & $-16C_{n\bar{s}}$ & $K^{*}$ & $\frac{16}{3}C_{c\bar{s}}$ &$C_{n\bar{s}}=18.7$\\
$D$&$-16C_{c\bar{n}}$&$D^{*}$&$\frac{16}{3}C_{c\bar{n}}$&$C_{c\bar{n}}=6.7$\\
$D_s$&$-16C_{c\bar{s}}$&$D_{s}^{*}$&$\frac{16}{3}C_{c\bar{s}}$&$C_{c\bar{s}}$=6.7\\
$B$&$-16C_{b\bar{n}}$&$B^{*}$&$\frac{16}{3}C_{b\bar{n}}$&$C_{b\bar{n}}$=2.1\\
$B_s$&$-16C_{b\bar{s}}$&$B_s^{*}$&$\frac{16}{3}C_{b\bar{s}}$&$C_{b\bar{s}}$=2.3\\
$B_{c}$&$-16C_{b\bar{c}}$&$B_{c}^{*}$\cite{Godfrey:1985xj}&$\frac{16}{3}C_{b\bar{c}}$&$C_{b\bar{c}}=3.3$\\
$\eta_{c}$&$-16C_{c\bar{c}}$&$J/\psi$&$\frac{16}{3}C_{c\bar{c}}$&$C_{c\bar{c}}=5.3$\\
$\eta_{b}$&$-16C_{b\bar{b}}$&$\Upsilon$&$\frac{16}{3}C_{b\bar{b}}$&$C_{b\bar{b}}=2.9$\\
$\Sigma_{c}$&$\frac{8}{3}C_{nn}-\frac{32}{3}C_{cn}$&$\Sigma_{c}^{*}$&$\frac{8}{3}C_{nn}+\frac{16}{3}C_{cn}$&$C_{cn}=4.0$\\
$\Xi'_{c}$&$\frac{8}{3}C_{n s}-\frac{16}{3}C_{cn}-\frac{16}{3}C_{cs}$&$\Xi_{c}^*$&$\frac{8}{3}C_{n s}+\frac{8}{3}C_{cn}+\frac{8}{3}C_{cs}$&$C_{cs}=4.8$\\
$\Sigma_{b}$&$\frac{8}{3}C_{nn}-\frac{32}{3}C_{bn}$&$\Sigma_{b}^{*}$&$\frac{8}{3}C_{nn}+\frac{16}{3}C_{bn}$&$C_{bn}=1.3$\\
$\Xi'_{b}$&$\frac{8}{3}C_{n s}-\frac{16}{3}C_{bn}-\frac{16}{3}C_{bs}$&$\Xi_b^{*}$&
$\frac{8}{3}C_{n s}+\frac{8}{3}C_{bn}+\frac{8}{3}C_{bs}$&$C_{bs}=1.2$\\
\midrule[1pt] \bottomrule[1.5pt]
\end{tabular}
\end{table}
\end{center}
\begin{table}[!h]
\caption{The effective constituent quark masses extracted
from conventional baryons.}\label{quark mass} \centering
\begin{tabular}{lc}
\toprule[1pt]
Mass formula& Quark mass (MeV)\\
\midrule[1pt]
$M_{N}=3m_{n}-8C_{nn}$ & $m_{n}=361.8$\\
$M_{\Omega}=3m_{s}+8C_{ss}$ & $m_{s}=540.4$\\
$\begin{matrix}M_{\Sigma_{c}}=\frac{8}{3}C_{nn}-\frac{32}{3}C_{nc}+2m_{n}+m_{c}\\
M_{\Sigma_{c}^{*}}=\frac{8}{3}C_{nn}+\frac{16}{3}C_{nc}+2m_{n}+m_{c}\end{matrix}$ & $m_{c}=1724.8$\\
$\begin{matrix}M_{\Sigma_{b}}=\frac{8}{3}C_{nn}-\frac{32}{3}C_{nb}+2m_{n}+m_{b}\\
M_{\Sigma_{b}^{*}}=\frac{8}{3}C_{nn}+\frac{16}{3}C_{nb}+2m_{n}+m_{b}\end{matrix}$ & $m_{b}=5052.9$\\
\bottomrule[1pt]
\end{tabular}
\end{table}

With these parameters, we can estimate the pentaquark masses in two
ways. In the first method, one substitutes the relevant parameters
into $M=\sum_i m_i +\langle H_{CM}\rangle$. In the second method, we
employ the formula $M=M_{ref}-\langle H_{CM}\rangle_{ref}+\langle
H_{CM}\rangle$, where $M_{ref}=M_{baryon}+M_{meson}$ is a reference
mass scale and $\langle H_{CM}\rangle_{ref}=\langle
H_{CM}\rangle_{baryon}+\langle H_{CM}\rangle_{meson}$. The reference
baryon and meson system should have the same constituent quarks as
the considered system \cite{Hyodo:2012pm}. {Although the mass formula in the second method is from that in the first method, one should note the difference in adopting them. When applying the first formula to conventional hadrons, the resulting masses are usually higher than the experimental measurements, which is illustrated in table \ref{comp}. This indicates that the simple model does not incorporate attraction sufficiently.
As a result, we may treat the pentaquark masses estimated with the first method as theoretical upper limits. In the second method, we use the realistic values rather than the calculated values for the hadron masses of the reference system. The attraction that the model does not incorporate is somehow phenomenologically compensated in this procedure. The estimated masses in the second method should be more reasonable than those in the first method.} In the following parts, we will present numerical results obtained in both methods. {To understand the decay properties in the following discussions, we will adopt some masses of the not-yet-observed doubly heavy baryons, which were obtained from several theoretical calculations. They are presented in table \ref{doubly heavy baryons mass}.}

\begin{table}[htbp]
\caption{Mass differences ($\Delta M=M_{Th.}-M_{Ex.}$) between the calculated values (Th.) and experimental values (Ex.) for conventional hadrons in units of MeV.}\label{comp}
\begin{tabular}{cccccccc}\hline
Hadron&$\Delta M$&Hadron&$\Delta M$&Hadron&$\Delta M$&Hadron&$\Delta M$\\\hline
$\pi$&109.5&$\rho$&107.2&  $N$&0&$\Delta$&0\\
$K$  &110.6&$K^*$&105.3&   $\Sigma$&-12.4&$\Sigma^*$&-5.4\\

$\omega$&99.8&$\phi$&96.0 &       $\Xi$&9.4&$\Xi^*$&-7.3\\
$D$&112.2&$D^{*}$&113.7&   $\Lambda$&1.1&$\Omega$&0\\
$D_s$&189.7&$D_{s}^{*}$&188.7&  $\Sigma_{c}$&0&$\Sigma_{c}^*$&0\\
$B$&101.6&$B^{*}$&101.2 &       $\Lambda_{c}$&14.7& $\Xi_{c}$&58.4\\
$B_s$&189.6&$B_s^*$&190.2&      $\Xi'_{c}$&35&$\Xi_{c}^*$&37.6\\
$\eta_{c}$&380.9&$J/\psi$&381.0&  $\Omega_c$&76.5&$\Omega_c^*$&82.6\\
$\eta_{b}$&660.0&$\Upsilon$&661.0&  $\Sigma_b$&0&$\Sigma_{b}^{*}$&0\\
$B_c$&450.0 &&&        $\Lambda_b$&9.7& $\Xi_b$&62.7\\
&&&&$\Xi'_{b}$&39.8&$\Xi_b^*$&45.0\\
&&&&$\Omega_b$&92.1 & $\Xi_{cc}$&161.5\\\hline
\end{tabular}
\end{table}

\begin{table}[!h]
\caption{The adopted masses of the not-yet-observed doubly heavy baryons from several methods: RQM (relativized quark model), ECM (extended chromomagnetic model), FH (Feynman-Hellmann mass formulas), and NRM (nonrelativistic potential model).}\label{doubly heavy baryons mass} \centering
\begin{tabular}{ccccc}
\toprule[1pt]
\midrule[1pt]
Baryon & $\quad$ & Mass & $\quad$ & Theoretical model\\
\midrule[1pt]
$\Xi_{bb}$ & $\quad$ & 10138 & $\quad$ & RQM\,\,\cite{Lu:2017meb}\\
$\Xi_{bb}^{*}$ & $\quad$ & 10169 & $\quad$ & RQM\,\,\cite{Lu:2017meb}\\
$\Omega_{cc}$ & $\quad$ & 3715 & $\quad$ & RQM\,\,\cite{Lu:2017meb}\\
$\Omega_{cc}^{*}$ & $\quad$ & 3772 & $\quad$ & RQM\,\,\cite{Lu:2017meb}\\
$\Omega_{bb}$ & $\quad$ & 10230 & $\quad$ & RQM\,\,\cite{Lu:2017meb}\\
$\Omega_{bb}^{*}$ & $\quad$ & 10258 & $\quad$ & RQM\,\,\cite{Lu:2017meb}\\
$\Xi_{cb}$ & $\quad$ & 6922 & $\quad$ & ECM\,\,\cite{Weng:2018mmf}\\
$\Xi_{cb}^{'}$ & $\quad$ & 6948 & $\quad$ & ECM\,\,\cite{Weng:2018mmf}\\
$\Xi_{cb}^{*}$ & $\quad$ & 6973 & $\quad$ & ECM\,\,\cite{Weng:2018mmf}\\
$\Omega_{cb}$ & $\quad$ & 7011 & $\quad$ & ECM\,\,\cite{Weng:2018mmf}\\
$\Omega_{cb}^{'}$ & $\quad$ & 7047 & $\quad$ & ECM\,\,\cite{Weng:2018mmf}\\
$\Omega_{cb}^{*}$ & $\quad$ & 7066 & $\quad$ & ECM\,\,\cite{Weng:2018mmf}\\
$\Xi_{bb}$ & $\quad$ & 10340 & $\quad$ & FH\,\,\cite{Roncaglia:1995az}\\
$\Xi_{bb}^{*}$ & $\quad$ & 10370 & $\quad$ & FH\,\,\cite{Roncaglia:1995az}\\
$\Xi_{bb}$ & $\quad$ & 10340 & $\quad$ & NRM\,\,\cite{Roberts:2007ni}\\
$\Xi_{bb}^{*}$ & $\quad$ & 10367 & $\quad$ & NRM\,\,\cite{Roberts:2007ni}\\
\midrule[1pt]
\bottomrule[1pt]
\end{tabular}
\end{table}

\subsection{The $ccnn\bar{q}$, $ccss\bar{q}$, $bbnn\bar{q}$, and $bbss\bar{q}$ pentaquark states}

\begin{table}[htbp]
\centering \caption{The estimated masses for the $ccnn\bar{q}$
systems in units of MeV. The values in the third column are obtained
with the effective quark masses and are theoretical upper limits.
The masses after this column are determined with relevant
thresholds.}\label{mass-ccnnqbar}
\begin{tabular}{ccccc|ccccc}
\toprule[1.5pt] \midrule[1pt] \multicolumn{5}{l}{$ccnn\bar{n}$
$(I_{nn}=1,I=\frac{1}{2},\frac{3}{2})$}&
\multicolumn{5}{l}{$ccnn\bar{s}$ $(I=1)$}\\
$J^{P}$ &Eigenvalue&Mass&$(\Sigma_{c}D)$ & $\Xi_{cc}\pi$& $J^{P}$
&Eigenvalue&Mass&$(\Sigma_{c}D_{s})$ & $(\Xi_{cc}K)$\\ \midrule[1pt]
$\frac{5}{2}^{-}$ & 171.7 & 4706.7 & 4591.3 & 4436.7&
$\frac{5}{2}^{-}$ & 141.6 & 4855.2 & 4664.6 & 4584.4\\

$\frac{3}{2}^{-}$&
$\begin{pmatrix}288.6\\127.7\\35.6\\-314.3\end{pmatrix}$&
$\begin{pmatrix}4823.6\\4662.7\\4570.6\\4220.7\end{pmatrix}$&
$\begin{pmatrix}4708.2\\4547.3\\4455.2\\4105.3\end{pmatrix}$&
$\begin{pmatrix}4553.5\\4392.6\\4300.5\\3950.6\end{pmatrix}$&

$\frac{3}{2}^{-}$&
$\begin{pmatrix}205.9\\98.8\\54.9\\-176.8\end{pmatrix}$&
$\begin{pmatrix}4919.5\\4812.4\\4768.5\\4536.8\end{pmatrix}$&
$\begin{pmatrix}4728.9\\4621.8\\4577.9\\4346.3\end{pmatrix}$&
$\begin{pmatrix}4648.7\\4541.6\\4497.7\\4266.0\end{pmatrix}$\\

$\frac{1}{2}^{-}$&
$\begin{pmatrix}350.5\\197.9\\40.1\\-336.1\end{pmatrix}$&
$\begin{pmatrix}4885.5\\4732.9\\4575.1\\4198.9\end{pmatrix}$&
$\begin{pmatrix}4770.1\\4617.5\\4459.7\\4083.5\end{pmatrix}$&
$\begin{pmatrix}4615.4\\4462.8\\4305.0\\3928.8\end{pmatrix}$&

$\frac{1}{2}^{-}$&
$\begin{pmatrix}271.5\\162.7\\12.5\\-194.2\end{pmatrix}$&
$\begin{pmatrix}4985.1\\4876.3\\4726.1\\4519.4\end{pmatrix}$&
$\begin{pmatrix}4794.5\\4685.7\\4535.5\\4328.8\end{pmatrix}$&
$\begin{pmatrix}4714.3\\4605.5\\4455.3\\4248.6\end{pmatrix}$\\

\multicolumn{5}{l}{$ccnn\bar{n}$ $(I_{nn}=0,I=\frac{1}{2})$}&
\multicolumn{5}{l}{$ccnn\bar{s}$ $(I=0)$}\\

$\frac{5}{2}^{-}$ & 207.3 & 4742.3 & 4626.9 & 4472.3&
$\frac{5}{2}^{-}$ & 132.0 & 4845.6 & 4655.0 & 4574.8\\

$\frac{3}{2}^{-}$&
$\begin{pmatrix}191.6\\46.5\\-565.2\end{pmatrix}$&
$\begin{pmatrix}4726.6\\4581.5\\3969.8\end{pmatrix}$&
$\begin{pmatrix}4611.2\\4466.1\\3854.4\end{pmatrix}$&
$\begin{pmatrix}4456.6\\4311.5\\3699.7\end{pmatrix}$&

$\frac{3}{2}^{-}$&
$\begin{pmatrix}118.8\\-12.1\\-358.4\end{pmatrix}$&
$\begin{pmatrix}4832.4\\4701.5\\4355.2\end{pmatrix}$&
$\begin{pmatrix}4641.8\\4510.9\\4164.7\end{pmatrix}$&
$\begin{pmatrix}4561.6\\4430.7\\4084.4\end{pmatrix}$\\

$\frac{1}{2}^{-}$&
$\begin{pmatrix}169.4\\43.4\\-134.5\\-665.0\end{pmatrix}$&
$\begin{pmatrix}4704.4\\4578.4\\4400.6\\3870.0\end{pmatrix}$&
$\begin{pmatrix}4589.0\\4463.0\\4285.2\\3754.6\end{pmatrix}$&
$\begin{pmatrix}4434.3\\4308.3\\4130.5\\3600.0\end{pmatrix}$&

$\frac{1}{2}^{-}$&
$\begin{pmatrix}96.7\\-11.0\\-134.2\\-462.8\end{pmatrix}$ &
$\begin{pmatrix}4810.3\\4702.6\\4579.4\\4250.8\end{pmatrix}$ &
$\begin{pmatrix}4619.8\\4512.0\\4388.8\\4060.3\end{pmatrix}$ &
$\begin{pmatrix}4539.5\\4431.8\\4308.6\\3980.0\end{pmatrix}$\\
\midrule[1pt] \bottomrule[1.5pt]
\end{tabular}
\end{table}

\begin{table}[htbp]
\centering \caption{The estimated masses for the $bbnn\bar{q}$
systems in units of MeV. The values in the third column are obtained
with the effective quark masses and are theoretical upper limits.
The masses after this column are determined with relevant
thresholds.}\label{mass-bbnnqbar}
\begin{tabular}{cccc|cccc}
\toprule[1.5pt] \midrule[1pt]

\multicolumn{4}{l}{$bbnn\bar{n}$
$(I_{nn}=1,I=\frac{1}{2},\frac{3}{2})$} &
\multicolumn{2}{l}{$bbnn\bar{s}$ $(I=1)$}\\
$J^{P}$ &Eigenvalue&Mass&$(\Sigma_{b}\bar{B})$ & $J^{P}$
&Eigenvalue&Mass&$(\Sigma_{b}\bar{B}_{s})$\\ \midrule[1pt]

$\frac{5}{2}^{-}$ & 145.9 & 11337.1 & 11234.9 &

$\frac{5}{2}^{-}$ & 116.3 & 11486.1 & 11296.0\\

$\frac{3}{2}^{-}$ &
$\begin{pmatrix}291.7\\131.1\\21.2\\-316.8\end{pmatrix}$ &
$\begin{pmatrix}11482.9\\11322.3\\11212.4\\10874.4\end{pmatrix}$ &
$\begin{pmatrix}11380.7\\11220.1\\11110.2\\10772.2\end{pmatrix}$ &

$\frac{3}{2}^{-}$ &
$\begin{pmatrix}207.2\\101.2\\36.5\\-173.3\end{pmatrix}$ &
$\begin{pmatrix}11577.0\\11471.0\\11406.3\\11196.5\end{pmatrix}$ &
$\begin{pmatrix}11386.9\\11280.9\\11216.3\\11006.4\end{pmatrix}$ \\

$\frac{1}{2}^{-}$ &
$\begin{pmatrix}309.7\\157.3\\104.2\\-326.0\end{pmatrix}$&
$\begin{pmatrix}11500.9\\11348.5\\11295.4\\10865.2\end{pmatrix}$&
$\begin{pmatrix}11398.7\\11246.3\\11193.2\\10763.0\end{pmatrix}$&

$\frac{1}{2}^{-}$ &
$\begin{pmatrix}226.4\\128.0\\71.9\\-181.2\end{pmatrix}$&
$\begin{pmatrix}11596.2\\11497.8\\11441.7\\11188.6\end{pmatrix}$&
$\begin{pmatrix}11406.1\\11307.8\\11251.6\\10998.6\end{pmatrix}$\\

\multicolumn{4}{l}{$bbnn\bar{n}$ $(I_{nn}=0,I=\frac{1}{2})$} & \multicolumn{4}{l}{$bbnn\bar{s}$ $(I=0)$}\\

$\frac{5}{2}^{-}$ & 189.1 & 11380.3 & 11278.1 &

$\frac{5}{2}^{-}$ & 113.5 & 11483.3 & 11293.2 \\

$\frac{3}{2}^{-}$ &
$\begin{pmatrix}180.7\\47.5\\-592.9\end{pmatrix}$ &
$\begin{pmatrix}11371.9\\11238.7\\10598.3\end{pmatrix}$ &
$\begin{pmatrix}11269.7\\11136.5\\10496.1\end{pmatrix}$&

$\frac{3}{2}^{-}$ &
$\begin{pmatrix}105.9\\-8.4\\-386.1\end{pmatrix}$ &
$\begin{pmatrix}11475.7\\11361.4\\10983.7\end{pmatrix}$ &
$\begin{pmatrix}11285.7\\11171.4\\10793.6\end{pmatrix}$\\

$\frac{1}{2}^{-}$ &
$\begin{pmatrix}174.2\\43.2\\-136.4\\-624.1\end{pmatrix}$ &
$\begin{pmatrix}11365.4\\11234.4\\11054.8\\10567.1\end{pmatrix}$ &
$\begin{pmatrix}11263.2\\11132.3\\10952.6\\10464.9\end{pmatrix}$&

$\frac{1}{2}^{-}$ &
$\begin{pmatrix}99.6\\-12.6\\-136.6\\-418.9\end{pmatrix}$ &
$\begin{pmatrix}11469.4\\11357.2\\11233.2\\10950.9\end{pmatrix}$ &
$\begin{pmatrix}11279.3\\11167.1\\11043.1\\10760.8\end{pmatrix}$\\
\midrule[1pt] \bottomrule[1.5pt]
\end{tabular}
\end{table}

\begin{table}[htbp]
\centering \caption{The estimated masses for the $ccss\bar{q}$
systems in units of MeV. The values in the third column are obtained
with the effective quark masses and are theoretical upper limits.
The masses after this column are determined with relevant
thresholds.}\label{mass-ccssqbar}
\begin{tabular}{cccc|cccc}
\toprule[1.5pt] \midrule[1pt] \multicolumn{4}{l}{$ccss\bar{n}$
$(I=\frac{1}{2})$}&
\multicolumn{4}{l}{$ccss\bar{s}$ $(I=0)$}\\
$J^{P}$ &Eigenvalue&Mass&$(\Omega_{c}D)$ & $J^{P}$
&Eigenvalue&Mass&$(\Omega_{c}D_{s})$\\ \midrule[1pt]

$\frac{5}{2}^{-}$ & 112.0 & 5004.2 & 4813.1 &

$\frac{5}{2}^{-}$ & 79.5 & 5150.3 & 4875.5 \\

$\frac{3}{2}^{-}$ &
$\begin{pmatrix}163.7\\62.4\\26.3\\-212.5\end{pmatrix}$ &
$\begin{pmatrix}5055.9\\4954.6\\4918.5\\4679.7\end{pmatrix}$&
$\begin{pmatrix}4864.8\\4763.5\\4727.4\\4488.6\end{pmatrix}$&

$\frac{3}{2}^{-}$ &
$\begin{pmatrix}85.5\\-82.7\\61.3\\24.7\end{pmatrix}$&
$\begin{pmatrix}5156.3\\5132.1\\5095.5\\4988.1\end{pmatrix}$&
$\begin{pmatrix}4881.5\\4857.3\\4820.8\\4713.3\end{pmatrix}$\\

$\frac{1}{2}^{-}$ &
$\begin{pmatrix}240.1\\126.6\\-24.7\\-238.8\end{pmatrix}$&
$\begin{pmatrix}5132.3\\5018.9\\4867.5\\4653.4\end{pmatrix}$ &
$\begin{pmatrix}4941.2\\4827.8\\4676.4\\4462.3\end{pmatrix}$ &

$\frac{1}{2}^{-}$ &
$\begin{pmatrix}166.8\\77.0\\-33.3\\-107.3\end{pmatrix}$ &
$\begin{pmatrix}5237.7\\5147.8\\5037.5\\4963.5\end{pmatrix}$&
$\begin{pmatrix}4962.9\\4873.0\\4762.7\\4688.7\end{pmatrix}$\\
\midrule[1pt] \bottomrule[1.5pt]
\end{tabular}
\end{table}

\begin{table}[!h]
\centering \caption{The estimated masses for the $bbss\bar{q}$
systems in units of MeV. The values in the third column are obtained
with the effective quark masses and are theoretical upper limits.
The masses after this column are determined with relevant
thresholds.}\label{mass-bbssqbar}
\begin{tabular}{cccc|cccc}
\toprule[1.5pt] \midrule[1pt] \multicolumn{4}{l}{$bbss\bar{n}$
$(I=\frac{1}{2})$} &
\multicolumn{4}{l}{$bbss\bar{s}$ $(I=0)$}\\
$J^{P}$ &Eigenvalue&Mass&$(\Omega_{b}\bar{B})$ & $J^{P}$
&Eigenvalue&Mass&$(\Omega_{b}\bar{B}_{s})$ \\ \midrule[1pt]
$\frac{5}{2}^{-}$ & 83.7& 11632.1 & 11438.5 &
$\frac{5}{2}^{-}$ & 51.7 & 11778.7 & 11515.2\\

$\frac{3}{2}^{-}$ &
$\begin{pmatrix}165.6\\69.8\\4.6\\-210.4\end{pmatrix}$&
$\begin{pmatrix}11714.0\\11618.2\\11553.0\\11338.0\end{pmatrix}$ &
$\begin{pmatrix}11520.3\\11424.6\\11359.4\\11144.4\end{pmatrix}$ &

$\frac{3}{2}^{-}$ &
$\begin{pmatrix}75.1\\38.8\\22.1\\-58.4\end{pmatrix}$ &
$\begin{pmatrix}11802.1\\11765.8\\11749.1\\11668.6\end{pmatrix}$&
$\begin{pmatrix}11538.5\\11502.3\\11485.6\\11405.1\end{pmatrix}$\\

$\frac{1}{2}^{-}$ &
$\begin{pmatrix}183.3\\94.6\\43.1\\-217.8\end{pmatrix}$&
$\begin{pmatrix}11731.7\\11643.0\\11591.5\\11330.6\end{pmatrix}$&
$\begin{pmatrix}11538.1\\11449.4\\11397.9\\11136.9\end{pmatrix}$&

$\frac{1}{2}^{-}$ &
$\begin{pmatrix}95.6\\60.7\\9.4\\-62.5\end{pmatrix}$&
$\begin{pmatrix}11822.6\\11787.7\\11736.4\\11664.5\end{pmatrix}$&
$\begin{pmatrix}11559.1\\11524.2\\11472.9\\11400.9\end{pmatrix}$\\
\midrule[1pt] \bottomrule[1.5pt]
\end{tabular}
\end{table}

For the $ccnn\bar{q}$ ($q=n,s$) systems, we can use two types of
threshold to estimate their masses: (charmed baryon)-(charmed meson)
and (doubly charmed baryon)-(light meson). We will use
$M_{\Xi_{cc}}=3621.4$ MeV from the LHCb Collaboration
\cite{Aaij:2017ueg} in the latter case. For the $bbnn\bar{q}$
systems, we only use the (bottom baryon)-(bottom meson) threshold
since no doubly bottom baryon has been observed. For the
$ccss\bar{q}$ and $bbss\bar{q}$ systems, only (heavy baryon)-(heavy
meson) type thresholds are adopted because of the same reason. We
present the estimated masses for the $ccnn\bar{q}$, $bbnn\bar{q}$,
$ccss\bar{q}$, and $bbss\bar{q}$ pentaquark states in Tables
\ref{mass-ccnnqbar}, \ref{mass-bbnnqbar}, \ref{mass-ccssqbar}, and
\ref{mass-bbssqbar}, respectively. From these tables, it is obvious
that different estimation approaches give different masses. The
reason is that the model does not involve dynamics and contributions
from other terms in the potential are not elaborately considered.
For the $ccnn\bar{n}$ and $bbnn\bar{n}$ systems with $I_{nn}=1$, we
get the same spectra for the case of the total isospin
$I=\frac{1}{2}$ and $\frac{3}{2}$, which comes from the fact that
the color-magnetic interaction for a quark and an antiquark is
irrelevant with the isospin.

Table \ref{mass-ccnnqbar} shows us that the pentaquark masses obtained with $\Xi_{cc}\pi$ and $\Xi_{cc}K$ are lower than those with $\Sigma_{c}D$ and $\Sigma_{c}D_{s}$, respectively. This feature is consistent with the observation that more effects contribute to the effective attractions in {the former systems, which can be seen from the inequalities $\Delta M_{\Xi_{cc}}+\Delta M_{\pi}>\Delta M_{\Sigma_c}+\Delta M_{D}$ and $\Delta M_{\Xi_{cc}}+\Delta M_{K}>\Delta M_{\Sigma_c}+\Delta M_{D_s}$ according to table \ref{comp}. If the adopted model could reproduce all the hadron masses accurately, all the mentioned approaches would give consistent pentaquark masses. 
At present, we are not sure which type of threshold results in more appropriate pentaquark masses. For a multiquark hadron, the effective attraction is probably not strong and maybe a higher mass is more reasonable.} We plot the relative positions for the $ccnn\bar{n}$, $ccnn\bar{s}$, $bbnn\bar{n}$, $bbnn\bar{s}$, $ccss\bar{n}$, $ccss\bar{s}$, $bbss\bar{n}$, and $bbss\bar{s}$ systems in diagrams (a)-(h) of Fig. \ref{fig-ccnnqbar-bbnnqbar-ccssqbar-bbssqbar}, respectively. Here we select the masses obtained with the thresholds of $\Sigma_{c}D$, $\Sigma_{c}D_{s}$, $\Sigma_{b}\bar{B}$, $\Sigma_{b}\bar{B}_{s}$, $\Omega_{c}D$, $\Omega_{c}D_{s}$, $\Omega_{b}\bar{B}$, and $\Omega_{b}\bar{B}_{s}$, respectively. The thresholds relevant with rearrangement decay patterns are also displayed in the figure. The following discussions are based on the assumption that the obtained positions in this figure are all reasonable. {For the figures in the other systems, we will also adopt pentaquark masses estimated with higher thresholds. One should note that the figures show only rough positions of the pentaquarks. Their properties may be changed accordingly once the positions for states in a system are determined by an observed pentaquark. However, the mass splittings should not be affected.}

\begin{figure}[htbp]
\begin{tabular}{ccc}
\includegraphics[width=220pt]{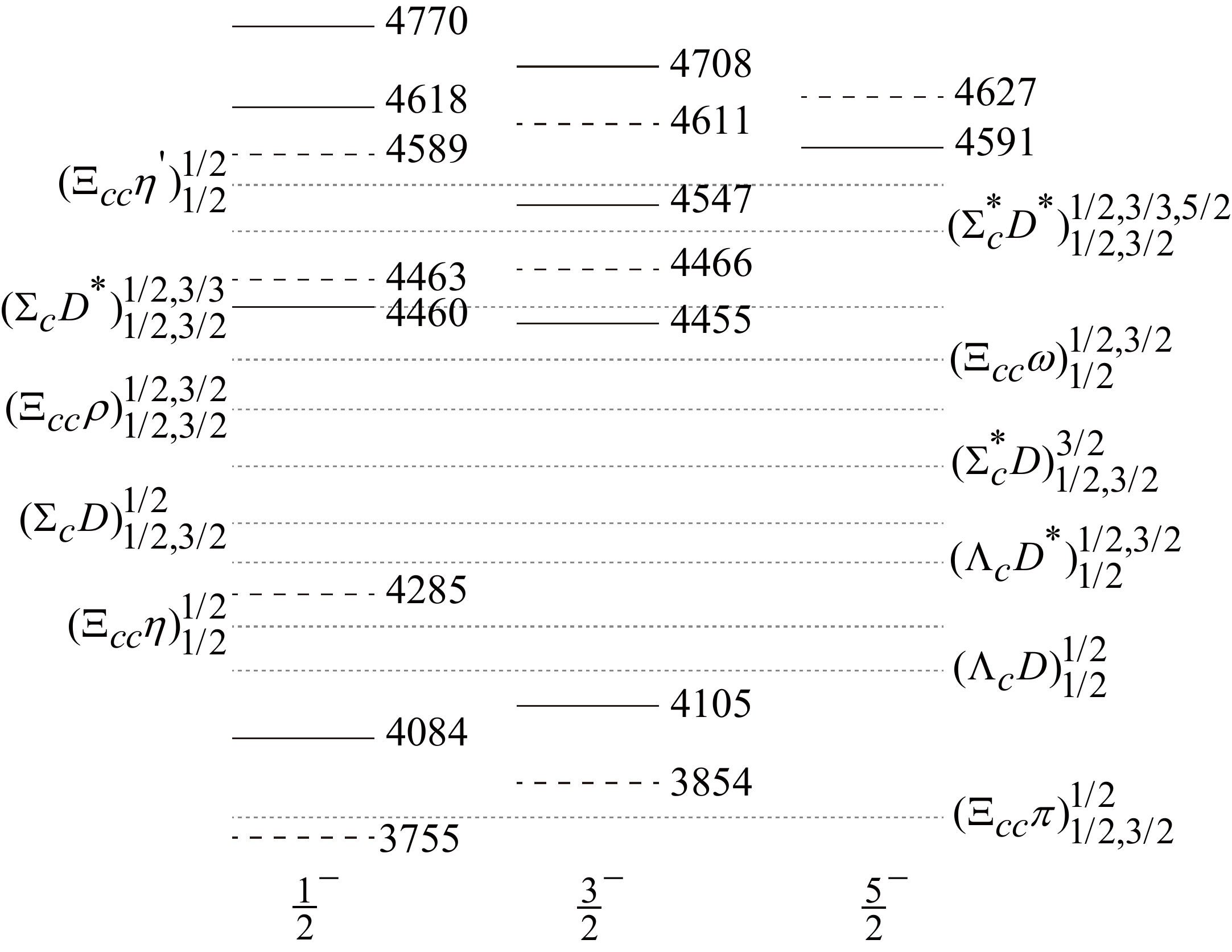}&$\qquad$&
\includegraphics[width=220pt]{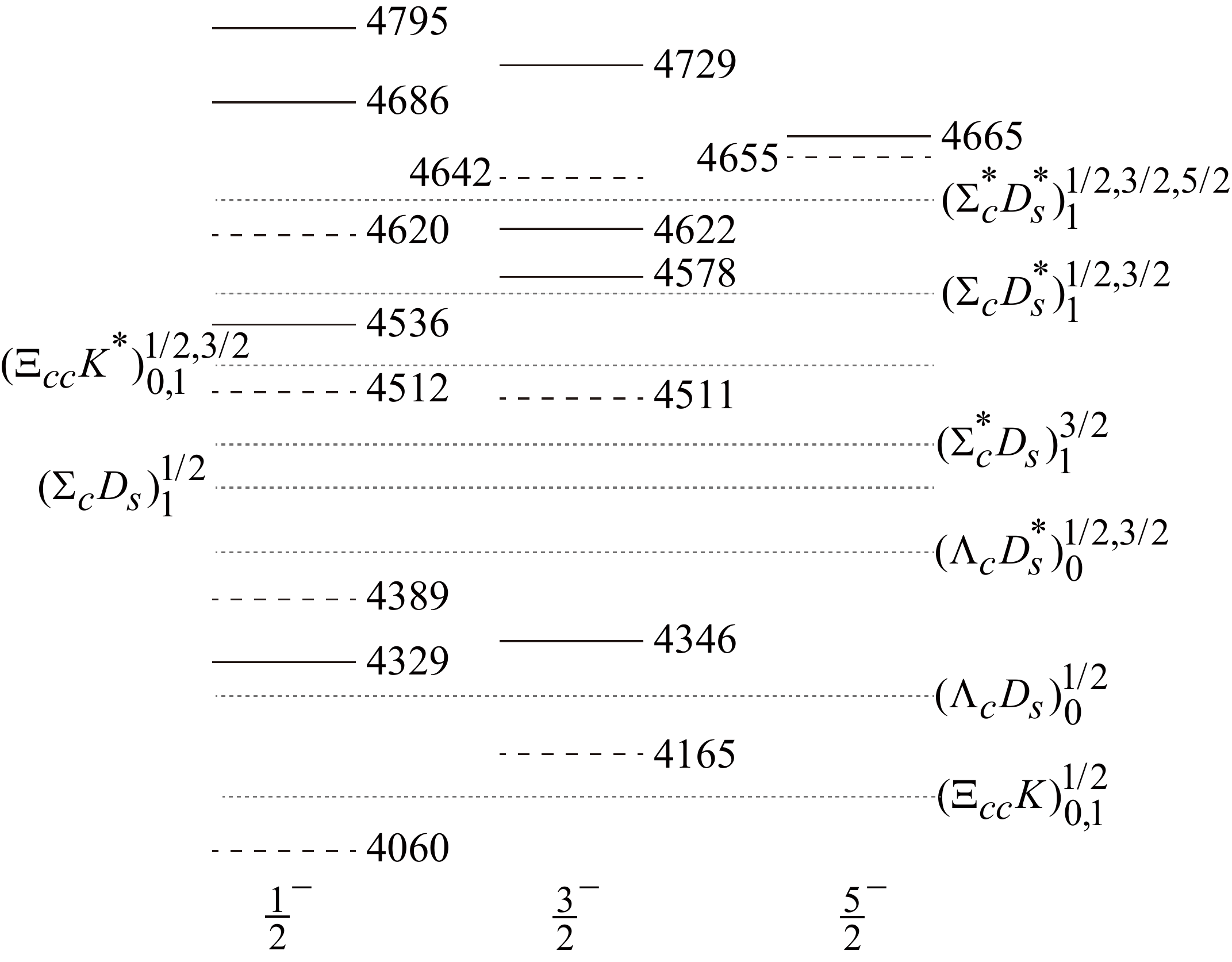}\\
(a) \begin{tabular}{c}($I_{nn}=1, I=\frac32 \& I=\frac12$) (solid) and\\ ($I_{nn}=0, I=\frac12$) (dashed) $ccnn\bar{n}$ states\end{tabular} &&(b) $I=1$ (solid) and $I=0$ (dashed) $ccnn\bar{s}$ states\\
&&\\
\includegraphics[width=220pt]{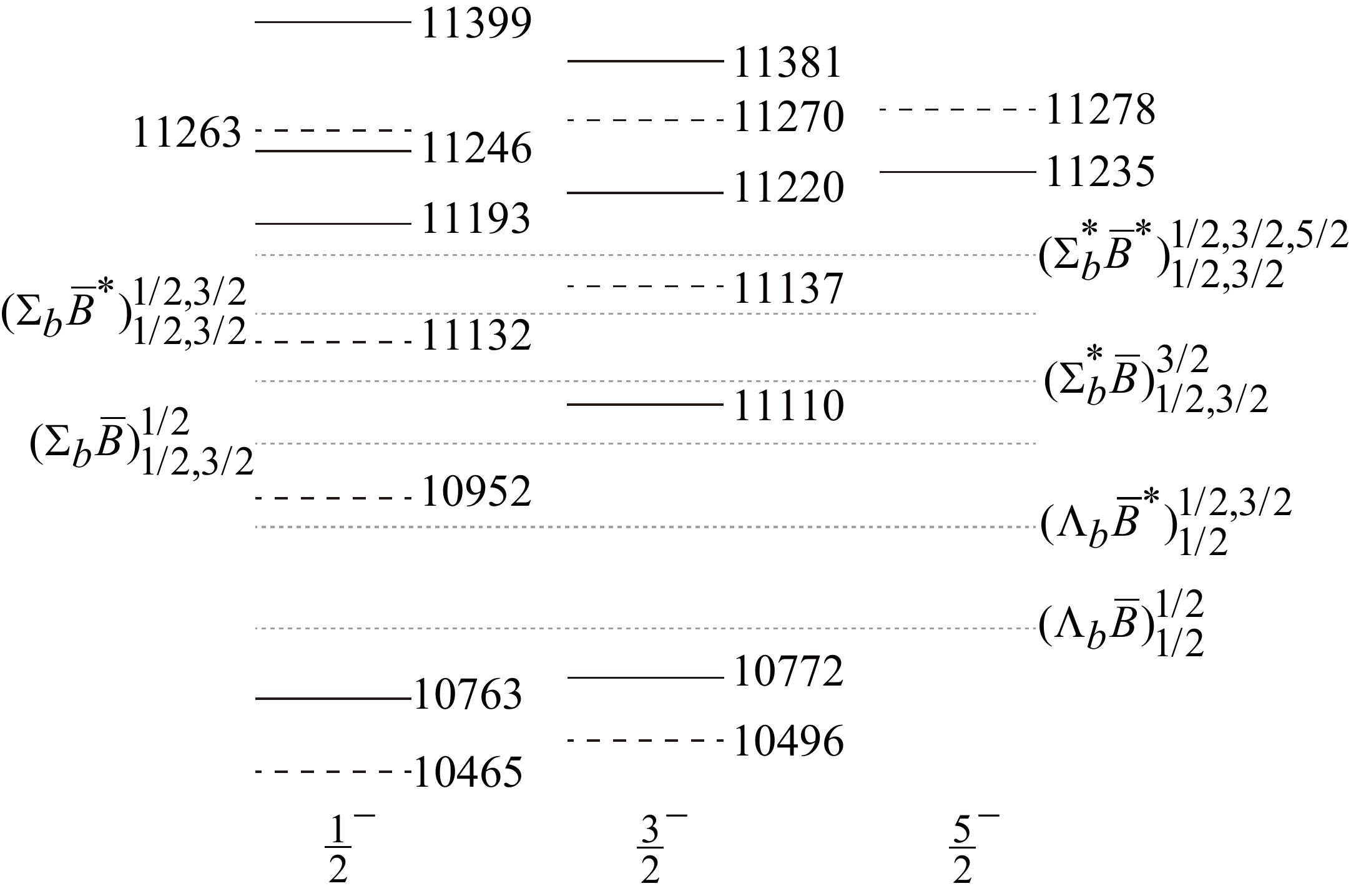}&$\qquad$&
\includegraphics[width=220pt]{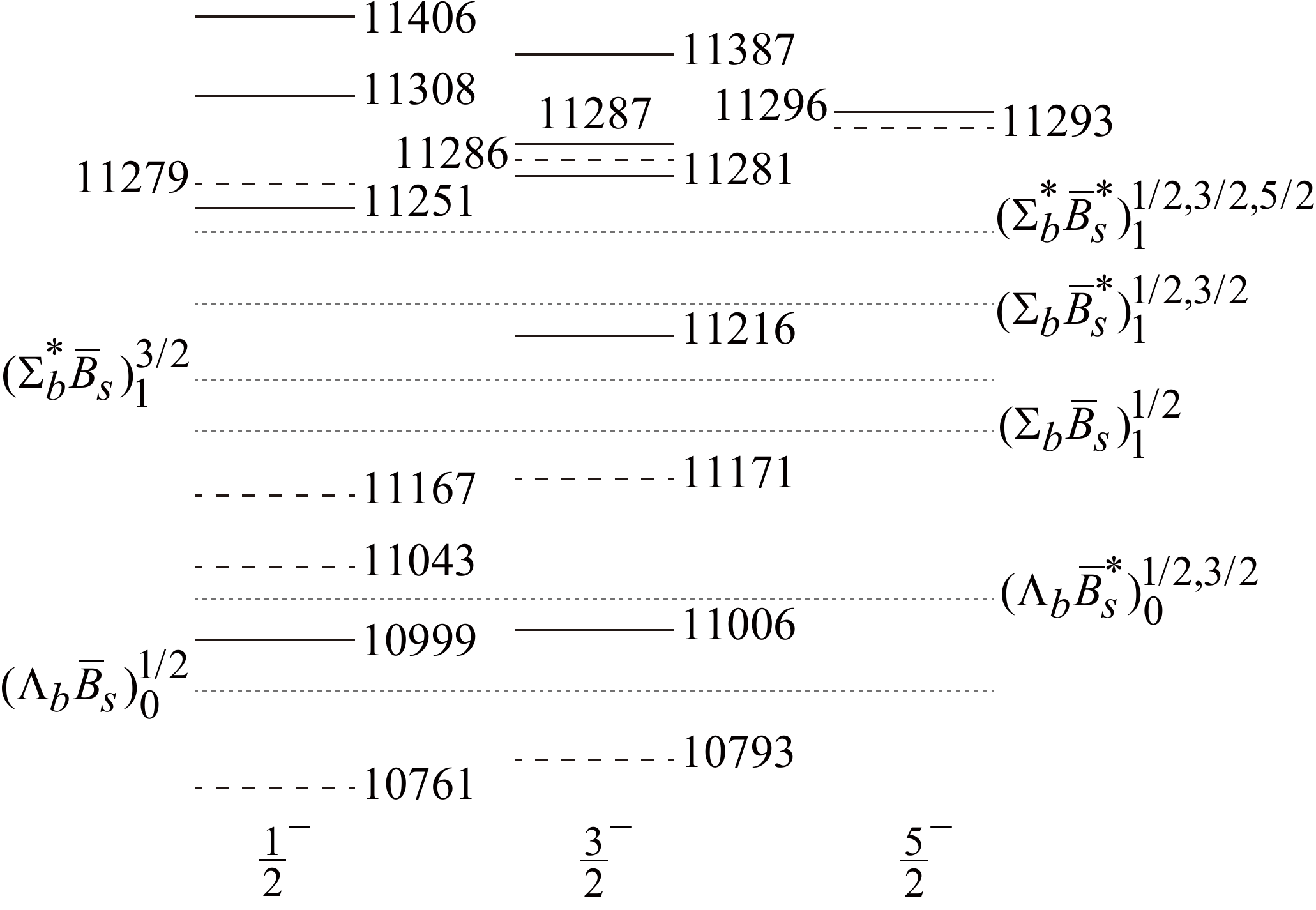}\\
(c) \begin{tabular}{c}($I_{nn}=1, I=\frac32 \& I=\frac12$) (solid) and\\ ($I_{nn}=0, I=\frac12$) (dashed) $bbnn\bar{n}$ states \end{tabular}&&(d) $I=1$ (solid) and $I=0$ (dashed) $bbnn\bar{s}$ states\\
&&\\
\includegraphics[width=220pt]{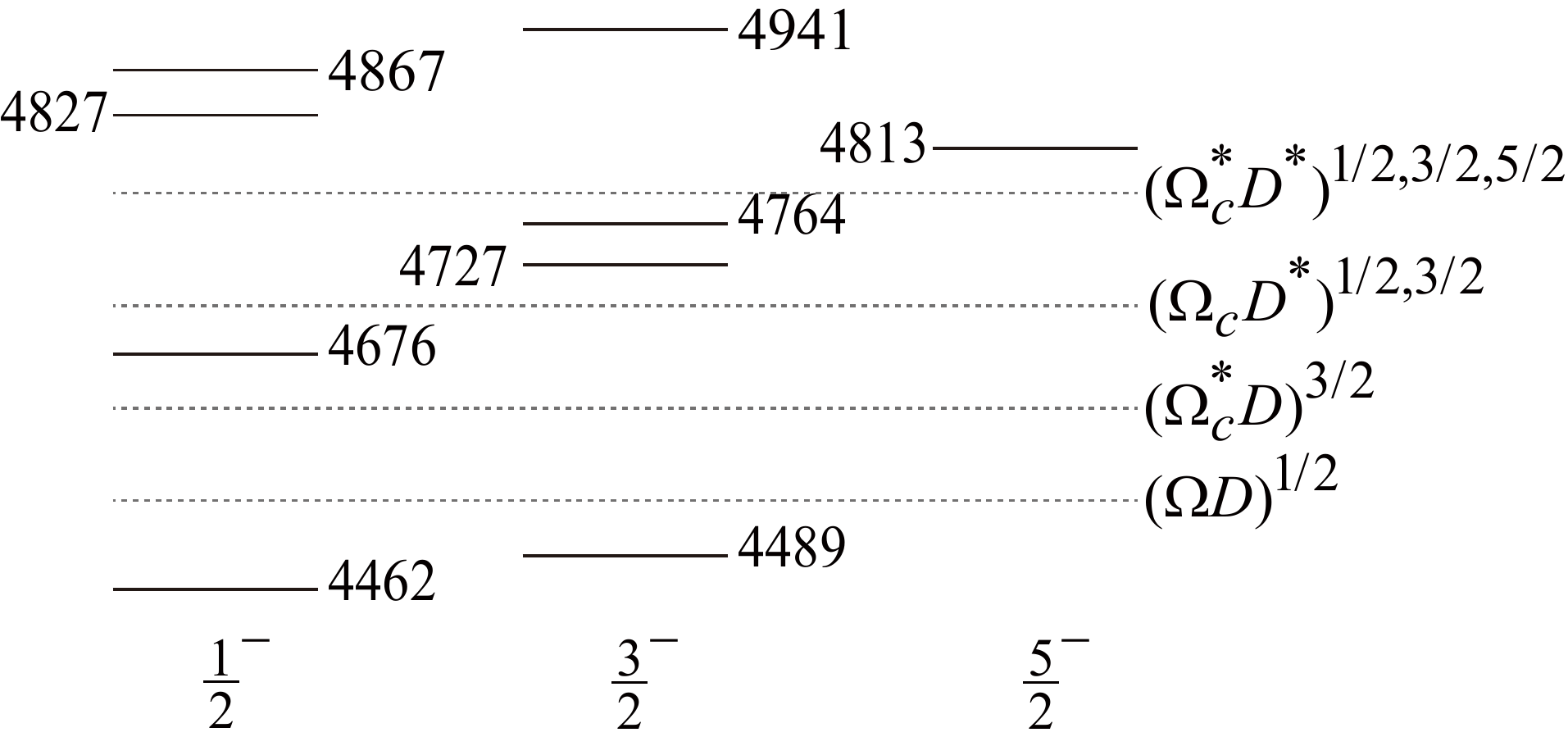}&$\qquad$&
\includegraphics[width=220pt]{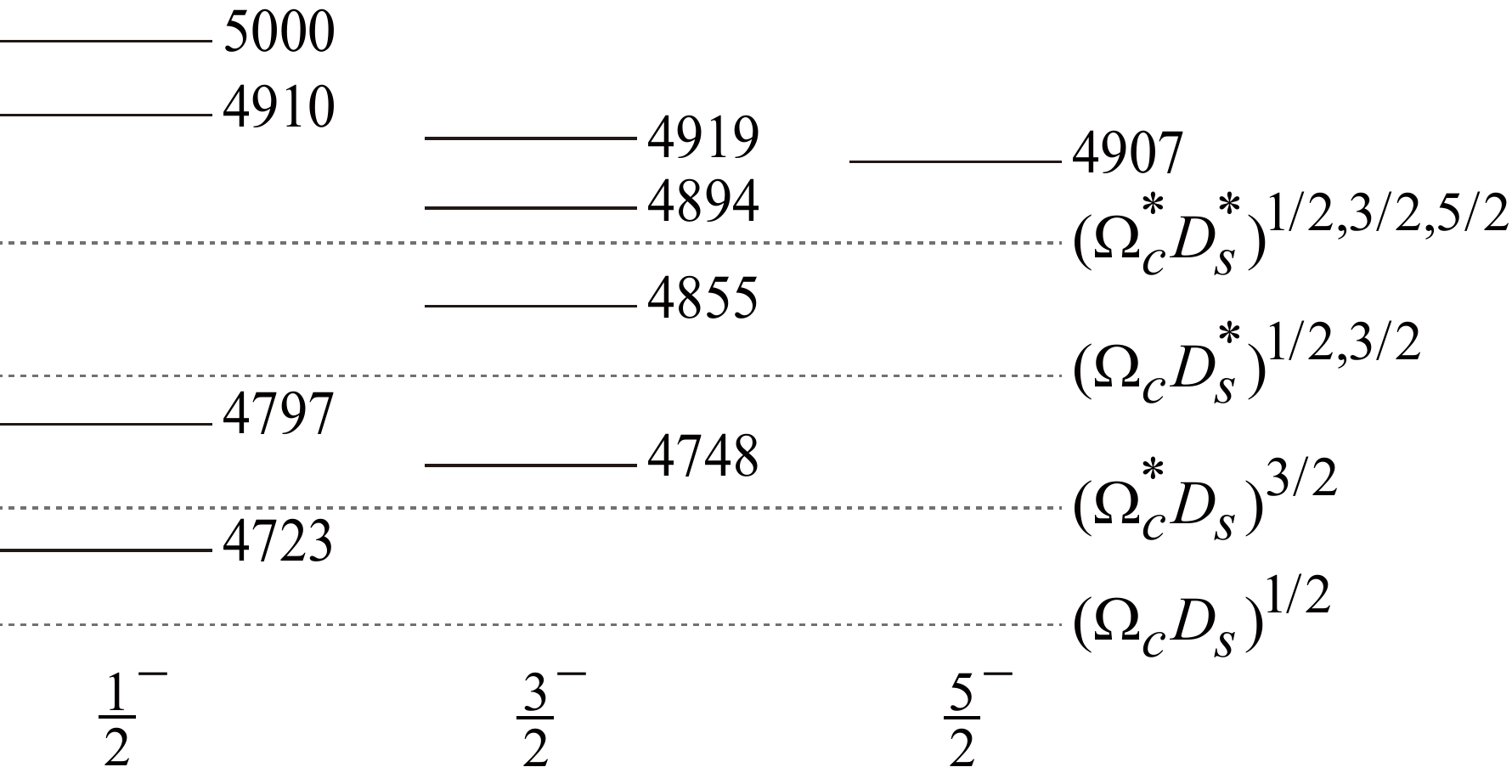}\\
(e) $I=\frac12$ (solid) $ccss\bar{n}$ states &&(f) $I=0$ (solid) $ccss\bar{s}$ states\\
&&\\
\includegraphics[width=220pt]{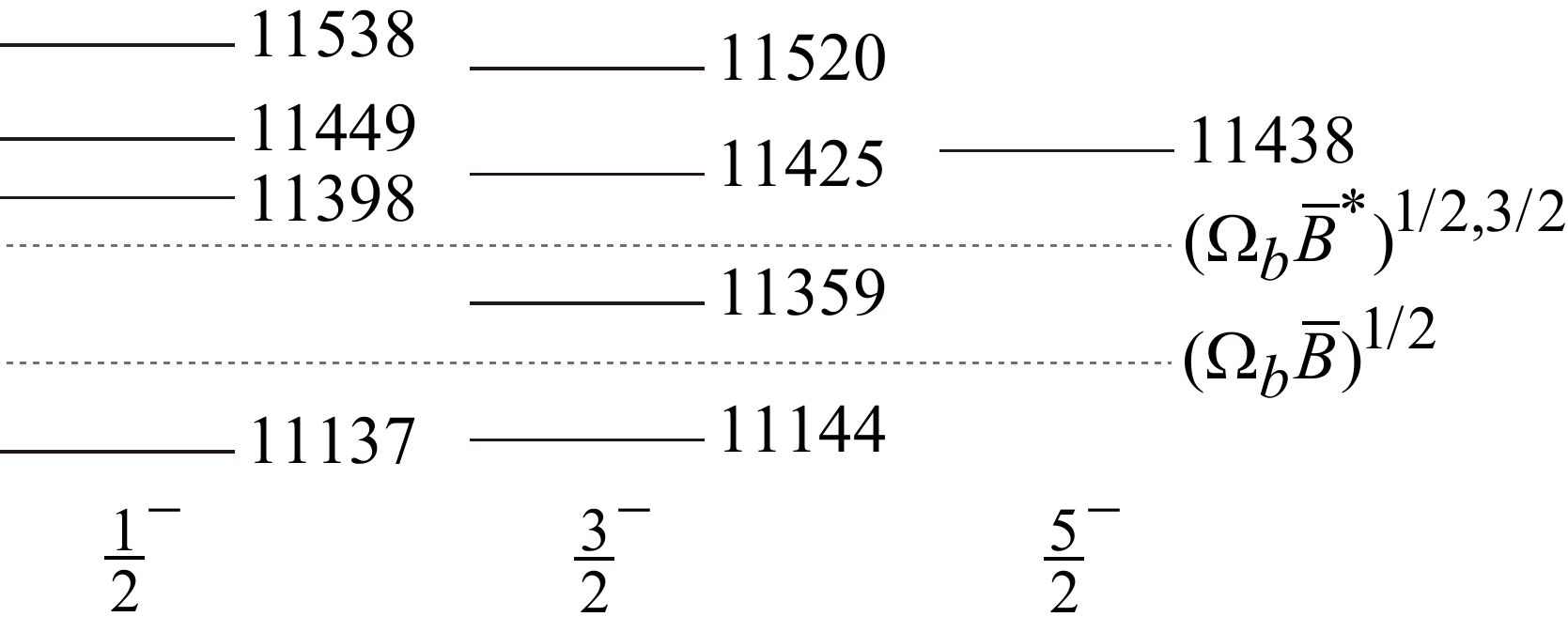}&$\qquad$&
\includegraphics[width=220pt]{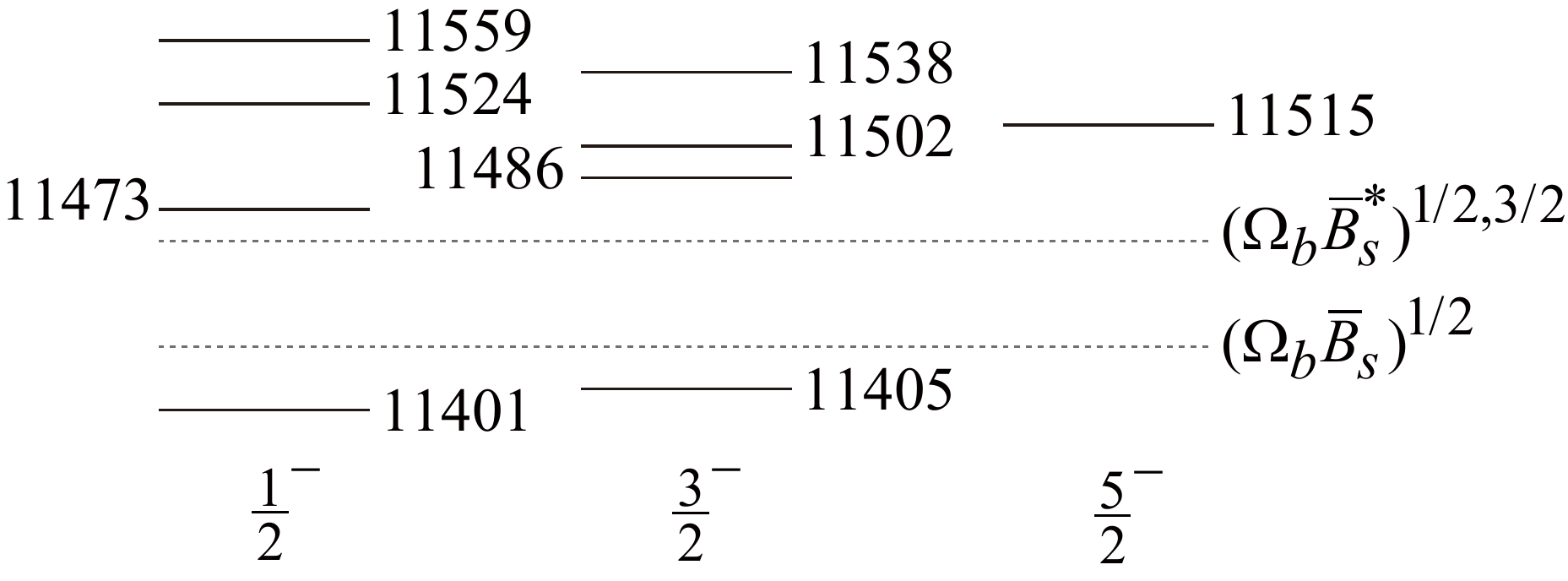}\\
(g) $I=\frac12$ (solid) $bbss\bar{n}$ states &&(h) $I=0$ (solid) $bbss\bar{s}$ states\\
\end{tabular}
\caption{Relative positions (units: MeV) for the $ccnn\bar{q}$,
$bbnn\bar{q}$, $ccss\bar{q}$, and $bbss\bar{q}$ pentaquark states
labeled with solid and dashed lines. The dotted lines indicate
various baryon-meson thresholds. The $I=\frac{3}{2}$ and
$\frac{1}{2}$ $ccnn\bar{n}$ states with $I_{nn}=1$ have the same
mass spectrum and are shown in the diagram (a) with solid lines. The
doubly bottom analog is shown in the diagram (c). When the isospin
(spin) of an initial pentaquark state is equal to a number in the
subscript (superscript) of a baryon-meson state, its decay into that
baryon-meson channel through $S$- or $D$-wave is allowed by the isospin
(angular momentum) conservation. {We have adopted
the masses estimated with the reference thresholds of (a) $\Sigma_{c}D$, (b) $\Sigma_{c}D_{s}$, (c) $\Sigma_{b}\bar{B}$, (d) $\Sigma_{b}\bar{B}_{s}$, (e) $\Omega_{c}D$, (f) $\Omega_{c}D_{s}$, (g) $\Omega_{b}\bar{B}$, and (h) $\Omega_{b}\bar{B}_{s}$.}
}\label{fig-ccnnqbar-bbnnqbar-ccssqbar-bbssqbar}
\end{figure}
\end{widetext}

For the $ccnn\bar{n}$ system, the $I_{nn}=0$ states are generally
lower than the $I_{nn}=1$ states and the lowest state is around the
$\Xi_{cc}\pi$ threshold. This pentaquark is in the mass range of
excited $\Xi_{cc}$ states \cite{Lu:2017meb}. It is highly probable
that an observed excited $\Xi_{cc}$ gets contributions from coupled
channel effects. An inverted mass order that the $I_{nn}=0$ state is
heavier is observed for the $J=\frac52$ states. This feature exists
because of the stronger $n\bar{n}$ interaction in the $I_{nn}=0$
state, which can be understood from the comparison between Eqs.
\eqref{eq14} and \eqref{eq11}. The $bbnn\bar{n}$ system should have
similar properties. From the mass distributions in diagrams (a) and
(c), we may guess roughly the mass of $\Xi_{bb}$,
$m_{\Xi_{bb}}\approx10465-135=10330$ MeV, a value consistent with
Refs. \cite{Roncaglia:1995az,Roberts:2007ni}. Replacing the
antiquark with an $\bar{s}$, we get the spectra of $QQnn\bar{s}$ in
the diagrams (b) and (d). The difference from the $QQnn\bar{n}$ case
lies only in the interaction strengths between the antiquark and
other quarks. The remaining systems are obtained by exchanging $s$
and $n$. All the lowest states have the quantum numbers
$J^P=\frac12^-$. In these systems, the $QQnn\bar{s}$, $QQss\bar{n}$,
and $I=\frac32$ $QQnn\bar{n}$ states are explicitly exotic.

Now we move on to the possible rearrangement decays of the
pentaquarks, which may occur through $S$-wave or $D$-wave, depending
on the conservation laws. The mass, total angular momentum, isospin,
and parity all together determine  whether the relevant decay
channels are open or not. For convenience, we label in Fig.
\ref{fig-ccnnqbar-bbnnqbar-ccssqbar-bbssqbar} the spin and isospin
of the baryon-meson states in the superscripts and subscripts of
their symbols, respectively. From the quantum numbers of the decay
product, it is possible to find pentaquark candidates. First, we
take a look at the $ccnn\bar{n}$ system. In the case of
$I(J^P)=\frac12(\frac52^-)$, the possible $S$-wave decay channel is
just $\Sigma_{c}^{*}D^{*}$. In the case of
$I(J^P)=\frac12(\frac32^-)$, the possible $S$-wave channels are
$\Sigma_{c}^{*}D^{*}$, $\Sigma_{c}D^{*}$, $\Sigma_{c}^{*}D$,
$\Lambda_cD^*$, $\Xi_{cc}\rho$, and $\Xi_{cc}\omega$. In the case of
$I(J^P)=\frac12(\frac12^-)$, the possible $S$-wave channels are
$\Sigma_{c}^{*}D^{*}$, $\Sigma_{c}D^{*}$, $\Sigma_cD$,
$\Lambda_cD^*$, $\Lambda_cD$, $\Xi_{cc}\rho$, $\Xi_{cc}\omega$,
$\Xi_{cc}\eta$, and $\Xi_{cc}\pi$. More channels will open if one
includes the $D$-wave decay modes. However, only the observation of
these decay patterns cannot prove the existence of a pentaquark
state consisting of $ccnn\bar{n}$ because the initial state may also
be an excited $\Xi_{cc}$. In this case, the mixing between 3$q$ state
and 5$q$ state is probably important. In the case of $I=\frac32$, an
observed state would be a good pentaquark candidate. The
$J^P=\frac52^-$ state with either isospin is probably not a very
broad pentaquark. For the $bbnn\bar{n}$ system, the situation is
similar to the $ccnn\bar{n}$ system. For the $ccss\bar{s}$ and
$bbss\bar{s}$ systems, the identification of a pentaquark state is
not so easy. On the contrary, the pentaquark states $ccnn\bar{s}$,
$ccss\bar{n}$, $bbnn\bar{s}$, and $bbss\bar{n}$ are easier to
identify since the quantum numbers are not allowed for the
conventional baryons. For example, if we observed a state in the
decay pattern $\Xi_{cc}K$, $\Xi_{cc}K^{*}$, $\Lambda_{c}D_{s}$,
$\Lambda_{c}D_{s}^{*}$, $\Sigma_{c}D_{s}$, $\Sigma_{c}D_{s}^{*}$,
$\Sigma_{c}^{*}D_{s}$, or $\Sigma_{c}^{*}D_{s}^{*}$, it would be a
good candidate of a $ccnn\bar{s}$ pentaquark state. From the
diagrams in Fig. \ref{fig-ccnnqbar-bbnnqbar-ccssqbar-bbssqbar}, the
lowest $ccnn\bar{s}$ pentaquark may be stable and the lowest one
with $J=\frac32$ is also relatively stable. {Because of the difference in coupling constants, the lowest two $I=0$ $bbnn\bar{s}$ pentaquarks, $J^P=1/2^-$ and $3/2^-$, probably both have strong decay patterns. This can be seen with the values $m_{\Xi_{bb}}=10138$ MeV and $m_{\Xi_{bb}*}=10169$ MeV obtained in Ref. \cite{Lu:2017meb}. If such masses are not far from the realistic values, the decay into $\Xi_{bb}K$ or $\Xi_{bb}^*K$ may occur. Once the $\Xi_{bb}$ ($\Xi_{bb}^*$) state is observed, the search for pentaquark candidates in the $\Xi_{bb}K$ ($\Xi_{bb}^*K$) channel may be performed. For the exotic $ccss\bar{n}$ and
$bbss\bar{n}$ pentaquarks, only the isospin $I=1/2$ is allowed. The lowest $J=1/2$ state and the lowest $J=3/2$ state in both systems are lower than the $(Qss)$-$(Q\bar{n})$ type thresholds and such decay patterns are forbidden. However, one finds that the decay for the $J=1/2$ ($3/2$) $ccss\bar{n}$ pentaquark into $\Omega_{cc}\bar{K}$ ($\Omega_{cc}^*\bar{K}$) is possible if the mass $m_{\Omega_{cc}}=3715$ MeV ($m_{\Omega_{cc}^*}=3772$ MeV) obtained in Ref. \cite{Lu:2017meb} is close to the realistic mass. Similarly, the decay for the $J=1/2$ ($3/2$) $bbss\bar{n}$ pentaquark into $\Omega_{bb}\bar{K}$ ($\Omega_{bb}^*\bar{K}$) is possible if one checks the threshold with $m_{\Omega_{bb}}=10230$ MeV ($m_{\Omega_{bb}^*}=10258$ MeV). With the $(QQs)$-$(s\bar{n})$ type channels, the identification of $ccss\bar{n}$ and $bbss\bar{n}$ pentaquarks may be performed in the future measurements.}

\begin{widetext}
\subsection{The $bcnn\bar{q}$ and $bcss\bar{q}$ pentaquark states}


\begin{center}
\begin{table}[htbp]
\caption{The estimated masses for the $bcnn\bar{q}$
systems in units of MeV. The values in the third column are obtained
with the effective quark masses and are theoretical upper limits.
The masses after this column are determined with relevant
thresholds.}\label{mass-bcnnqbar}
\begin{tabular}{ccccc|ccccc}
\toprule[1.5pt] \midrule[1pt] \multicolumn{5}{l}{$bcnn\bar{n}$
$(I_{nn}=1,I=\frac{1}{2},\frac{3}{2})$} &
\multicolumn{5}{l}{$bcnn\bar{s}$ $(I=1)$}\\
$J^{P}$ &Eigenvalue&Mass&$(\Sigma_{c}\bar{B})$ & $(\Sigma_{b}D)$ &
$J^{P}$ &Eigenvalue&Mass&$(\Sigma_{c}\bar{B}_{s})$ &
$(\Sigma_{b}D_{s})$ \\ \midrule[1pt]

$\frac{5}{2}^{-}$ & $\begin{pmatrix}156.9\\51.4\end{pmatrix}$ &
$\begin{pmatrix}7993.0\\7887.5\end{pmatrix}$&
$\begin{pmatrix}7917.4\\7811.9\end{pmatrix}$&
$\begin{pmatrix}7905.1\\7799.5\end{pmatrix}$&

$\frac{5}{2}^{-}$ & $\begin{pmatrix}127.2\\67.0\end{pmatrix}$&
$\begin{pmatrix}8141.9\\8081.7\end{pmatrix}$&
$\begin{pmatrix}7978.4\\7918.2\end{pmatrix}$&
$\begin{pmatrix}7978.8\\7918.6\end{pmatrix}$\\

$\frac{3}{2}^{-}$&
$\begin{pmatrix}288.4\\161.1\\131.4\\73.0\\37.6\\-46.4\\-319.6\end{pmatrix}$&
$\begin{pmatrix}8124.5\\7997.2\\7967.5\\7909.1\\7873.7\\7789.7\\7516.5\end{pmatrix}$&
$\begin{pmatrix}8048.8\\7921.6\\7891.8\\7833.5\\7798.0\\7714.1\\7440.9\end{pmatrix}$&
$\begin{pmatrix}8036.5\\7909.3\\7879.5\\7821.2\\7785.7\\7701.7\\7428.6\end{pmatrix}$&

$\frac{3}{2}^{-}$&
$\begin{pmatrix}204.9\\131.2\\102.3\\52.9\\46.3\\-31.2\\-181.9\end{pmatrix}$&
$\begin{pmatrix}8219.6\\8145.9\\8117.0\\8067.6\\8061.0\\7983.5\\7832.8\end{pmatrix}$&
$\begin{pmatrix}8056.1\\7982.4\\7953.5\\7904.1\\7897.5\\7820.0\\7669.3\end{pmatrix}$&
$\begin{pmatrix}8056.4\\7982.8\\7953.9\\7904.5\\7897.8\\7820.3\\7669.7\end{pmatrix}$\\

$\frac{1}{2}^{-}$&
$\begin{pmatrix}330.0\\266.9\\178.8\\117.1\\69.6\\-68.4\\-333.8\\-364.2\end{pmatrix}$&
$\begin{pmatrix}8166.1\\8103.0\\8014.9\\7953.2\\7905.7\\7767.8\\7502.3\\7471.9\end{pmatrix}$&
$\begin{pmatrix}8090.5\\8027.4\\7939.3\\7877.6\\7830.1\\7692.1\\7426.7\\7396.3\end{pmatrix}$&
$\begin{pmatrix}8078.2\\8015.0\\7926.9\\7865.2\\7817.7\\7679.8\\7414.3\\7384.0\end{pmatrix}$&

$\frac{1}{2}^{-}$&
$\begin{pmatrix}248.3\\186.5\\146.1\\85.7\\39.8\\-46.5\\-191.3\\-228.1\end{pmatrix}$&
$\begin{pmatrix}8263.0\\8201.3\\8160.8\\8100.4\\8054.5\\7968.2\\7823.4\\7786.6\end{pmatrix}$&
$\begin{pmatrix}8099.5\\8037.7\\7997.3\\7936.9\\7891.0\\7804.7\\7659.9\\7623.1\end{pmatrix}$&
$\begin{pmatrix}8099.9\\8038.1\\7997.7\\7937.3\\7891.4\\7805.1\\7660.3\\7623.5\end{pmatrix}$\\

\multicolumn{5}{l}{$bcnn\bar{n}$ $(I_{nn}=1,I=\frac{1}{2})$} & \multicolumn{5}{l}{$bcnn\bar{s}$ $(I=0)$}\\

$\frac{5}{2}^{-}$ & 196.1 & 8032.2 & 7956.6 & 7944.2 &

$\frac{5}{2}^{-}$ & 120.6 & 8135.3 & 7971.8 & 7972.2\\

$\frac{3}{2}^{-}$ &
$\begin{pmatrix}183.5\\150.4\\45.1\\-123.0\\-581.4\end{pmatrix}$&
$\begin{pmatrix}8019.6\\7986.5\\7881.2\\7713.1\\7254.7\end{pmatrix}$&
$\begin{pmatrix}7944.0\\7910.9\\7805.6\\7637.4\\7179.1\end{pmatrix}$&
$\begin{pmatrix}7931.6\\7898.5\\7793.3\\7625.1\\7166.7\end{pmatrix}$&

$\frac{3}{2}^{-}$&
$\begin{pmatrix}109.4\\75.8\\-12.4\\-122.4\\-374.7\end{pmatrix}$&
$\begin{pmatrix}8124.1\\8090.5\\8002.3\\7892.3\\7640.0\end{pmatrix}$&
$\begin{pmatrix}7960.6\\7926.9\\7838.8\\7728.8\\7476.4\end{pmatrix}$&
$\begin{pmatrix}7961.0\\7927.3\\7839.2\\7729.2\\7476.8\end{pmatrix}$\\

$\frac{1}{2}^{-}$&
$\begin{pmatrix}170.1\\41.9\\15.2\\-128.6\\-221.1\\-623.0\\-663.8\end{pmatrix}$&
$\begin{pmatrix}8006.2\\7878.0\\7851.3\\7707.5\\7615.0\\7213.1\\7172.3\end{pmatrix}$&
$\begin{pmatrix}7930.6\\7802.4\\7775.7\\7631.9\\7539.4\\7137.5\\7096.7\end{pmatrix}$&
$\begin{pmatrix}7918.2\\7790.1\\7763.3\\7619.6\\7527.0\\7125.2\\7084.3\end{pmatrix}$&

$\frac{1}{2}^{-}$&
$\begin{pmatrix}96.3\\-13.0\\-39.9\\-129.6\\-218.6\\-418.0\\-462.1\end{pmatrix}$&
$\begin{pmatrix}8111.0\\8001.7\\7974.8\\7885.1\\7796.1\\7596.7\\7552.6\end{pmatrix}$&
$\begin{pmatrix}7947.5\\7838.1\\7811.3\\7721.6\\7632.6\\7433.2\\7389.1\end{pmatrix}$&
$\begin{pmatrix}7947.9\\7838.5\\7811.7\\7721.9\\7633.0\\7433.6\\7389.5\end{pmatrix}$\\
\midrule[1pt] \bottomrule[1.5pt]
\end{tabular}
\end{table}
\end{center}


\begin{table}[htbp]
\centering \caption{The estimated masses for the $bcss\bar{q}$
systems in units of MeV. The values in the third column are obtained
with the effective quark masses and are theoretical upper limits.
The masses after this column are determined with relevant
thresholds.}\label{mass-bcssqbar}
\begin{tabular}{ccccc|ccccc}
\toprule[1.5pt] \midrule[1pt] \multicolumn{5}{l}{$bcss\bar{n}$
$(I=\frac{1}{2})$} &
\multicolumn{5}{l}{$bcss\bar{s}$ $(I=0)$}\\
$J^{P}$ &Eigenvalue&Mass&$(\Omega_{c}\bar{B})$ & $(\Omega_{b}D)$ &
$J^{P}$ &Eigenvalue&Mass&$(\Omega_{c}\bar{B}_{s})$ &
$(\Omega_{b}D_{s})$ \\ \midrule[1pt]

$\frac{5}{2}^{-}$& $\begin{pmatrix}95.9\\37.2\end{pmatrix}$&
$\begin{pmatrix}8289.2\\8230.5\end{pmatrix}$&
$\begin{pmatrix}8137.9\\8079.2\end{pmatrix}$&
$\begin{pmatrix}8109.8\\8051.1\end{pmatrix}$&

$\frac{5}{2}^{-}$&
$\begin{pmatrix}64.0\\53.7\end{pmatrix}$&
$\begin{pmatrix}8435.9\\8425.6\end{pmatrix}$&
$\begin{pmatrix}8196.7\\8186.4\end{pmatrix}$&
$\begin{pmatrix}8181.3\\8171.1\end{pmatrix}$\\

$\frac{3}{2}^{-}$&
$\begin{pmatrix}163.8\\100.4\\70.5\\24.7\\9.2\\-61.9\\-219.4\end{pmatrix}$&
$\begin{pmatrix}8357.1\\8293.7\\8263.8\\8218.0\\8202.5\\8131.4\\7974.0\end{pmatrix}$&
$\begin{pmatrix}8205.8\\8142.4\\8112.5\\8066.7\\8051.2\\7980.1\\7822.6\end{pmatrix}$&
$\begin{pmatrix}8177.7\\8114.3\\8084.4\\8038.6\\8023.1\\7952.0\\7794.5\end{pmatrix}$&

$\frac{3}{2}^{-}$&
$\begin{pmatrix}85.2\\62.2\\44.9\\36.4\\-11.6\\-24.9\\-105.8\end{pmatrix}$&
$\begin{pmatrix}8457.1\\8434.1\\8416.8\\8408.3\\8360.3\\8347.0\\8266.1\end{pmatrix}$&
$\begin{pmatrix}8217.9\\8194.9\\8177.6\\8169.1\\8121.1\\8107.8\\8026.8\end{pmatrix}$&
$\begin{pmatrix}8202.6\\8179.6\\8162.2\\8153.7\\8105.8\\8092.5\\8011.5\end{pmatrix}$\\

$\frac{1}{2}^{-}$&
$\begin{pmatrix}211.3\\148.7\\111.1\\46.3\\5.6\\-78.9\\-232.2\\-266.1\end{pmatrix}$&
$\begin{pmatrix}8404.6\\8342.0\\8304.4\\8239.6\\8198.9\\8114.4\\7961.1\\7927.2\end{pmatrix}$&
$\begin{pmatrix}8253.3\\8190.7\\8153.1\\8088.3\\8047.6\\7963.1\\7809.8\\7775.9\end{pmatrix}$&
$\begin{pmatrix}8225.2\\8162.6\\8125.0\\8060.2\\8019.5\\7935.0\\7781.7\\7747.8\end{pmatrix}$&

$\frac{1}{2}^{-}$&
$\begin{pmatrix}129.7\\89.8\\49.4\\10.3\\-17.5\\-31.5\\-87.8\\-148.4\end{pmatrix}$&
$\begin{pmatrix}8501.6\\8461.7\\8421.3\\8382.2\\8354.4\\8340.4\\8284.1\\8223.5\end{pmatrix}$&
$\begin{pmatrix}8262.4\\8222.5\\8182.1\\8143.0\\8115.2\\8101.2\\8044.9\\7984.3\end{pmatrix}$&
$\begin{pmatrix}8247.1\\8207.1\\8166.7\\8127.6\\8099.8\\8085.9\\8029.6\\7968.9\end{pmatrix}$\\
\midrule[1pt] \bottomrule[1.5pt]
\end{tabular}
\end{table}

\begin{figure}[htbp]
\begin{tabular}{ccc}
\includegraphics[width=220pt]{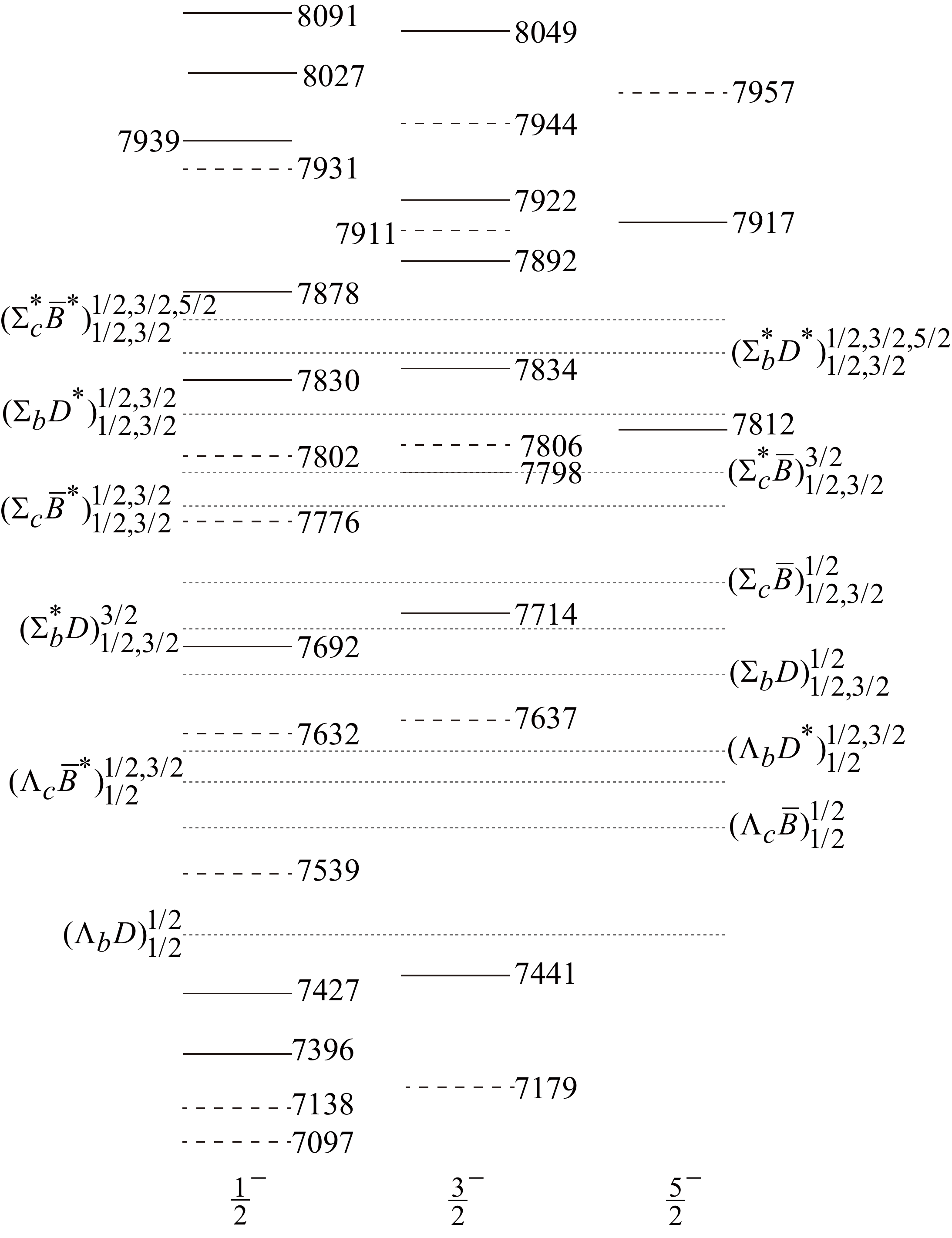}&$\qquad$&
\includegraphics[width=220pt]{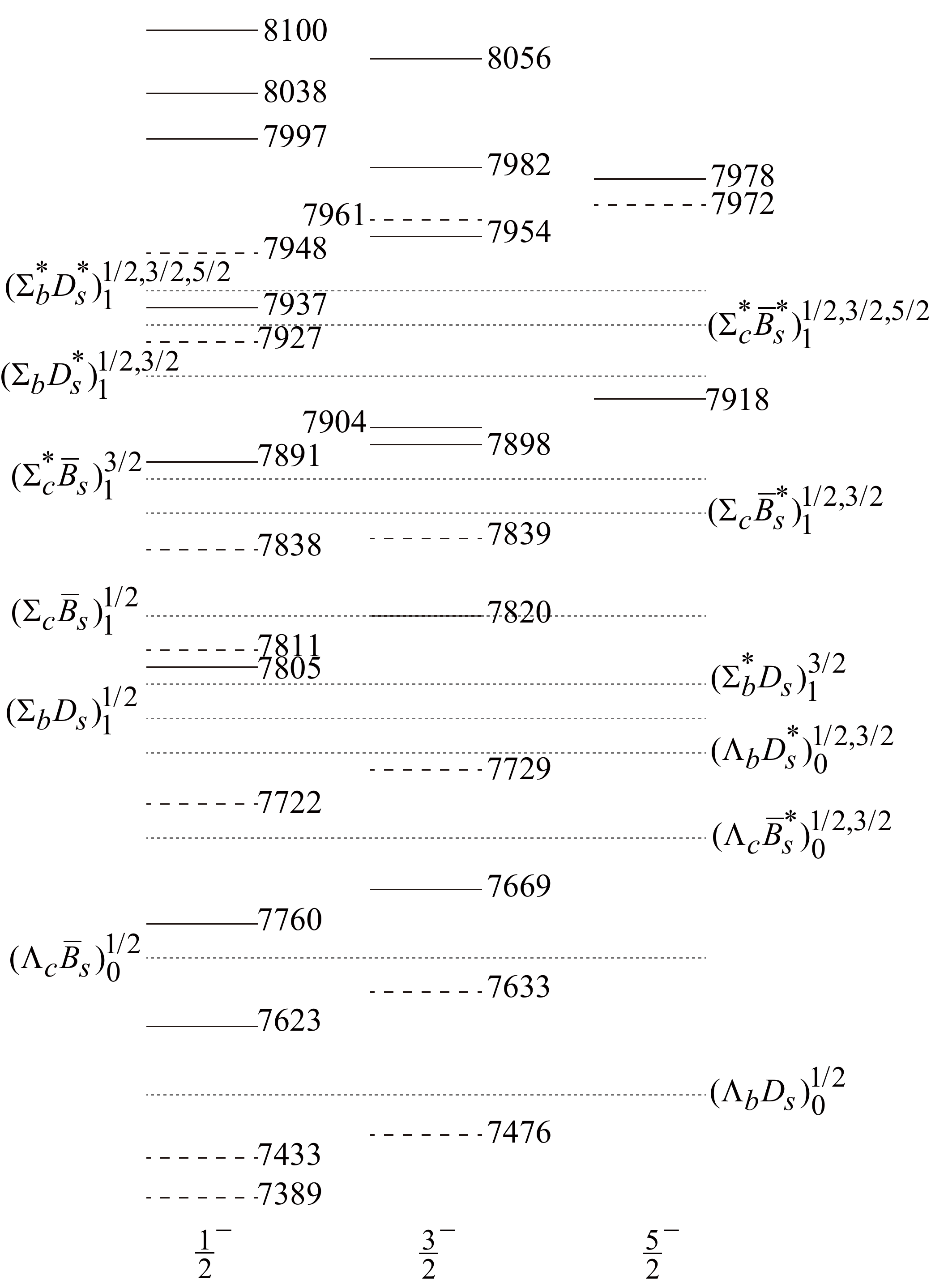}\\
(a) \begin{tabular}{c}($I_{nn}=1, I=\frac32 \& I=\frac{1}{2}$) (solid) and\\ ($I_{nn}=0, I=\frac12$) (dashed) $bcnn\bar{n}$ states\end{tabular} &&(b) $I=1$ (solid) and $I=0$ (dashed) $bcnn\bar{s}$ states\\
&&\\
\includegraphics[width=220pt]{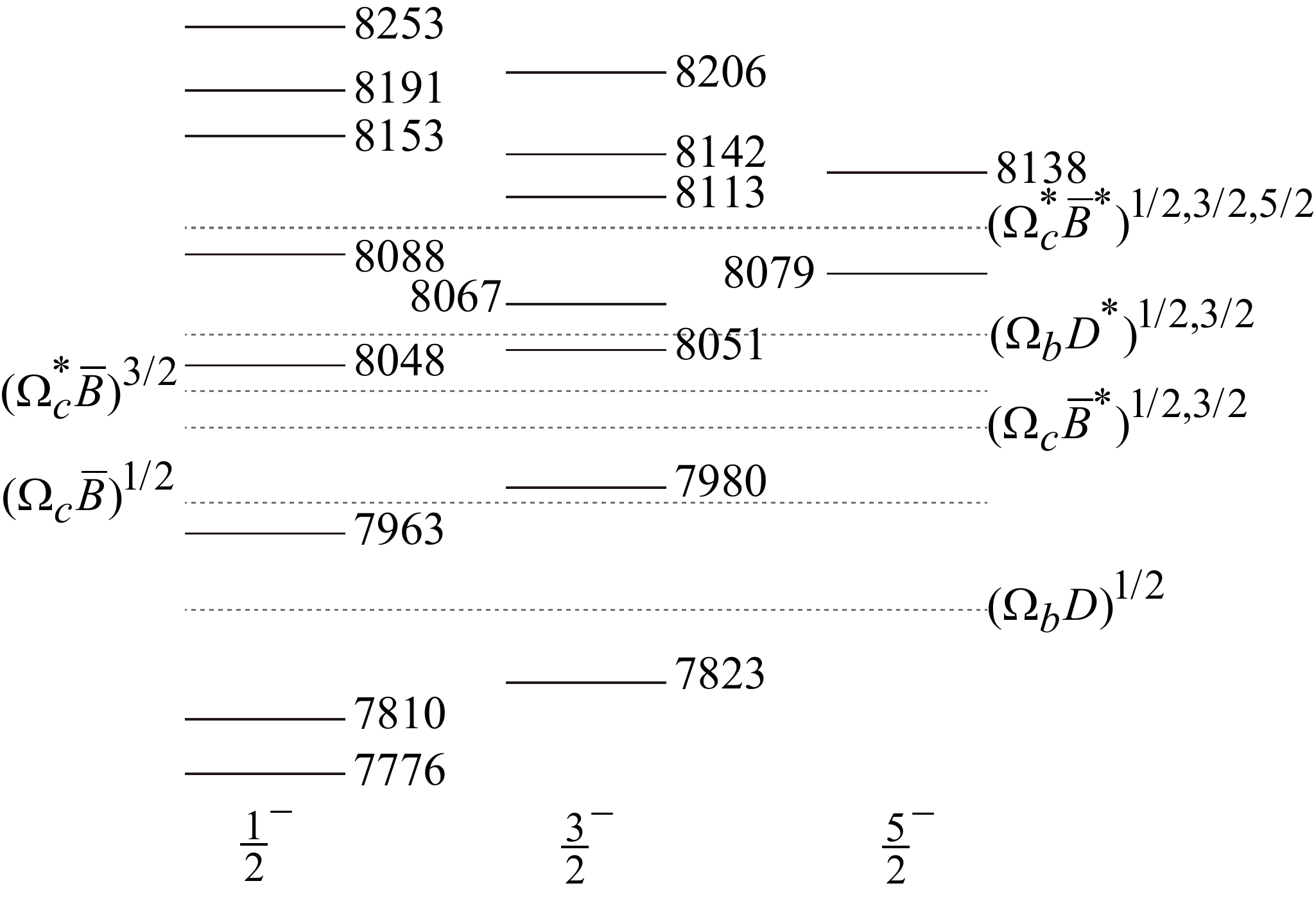}&$\qquad$&
\includegraphics[width=220pt]{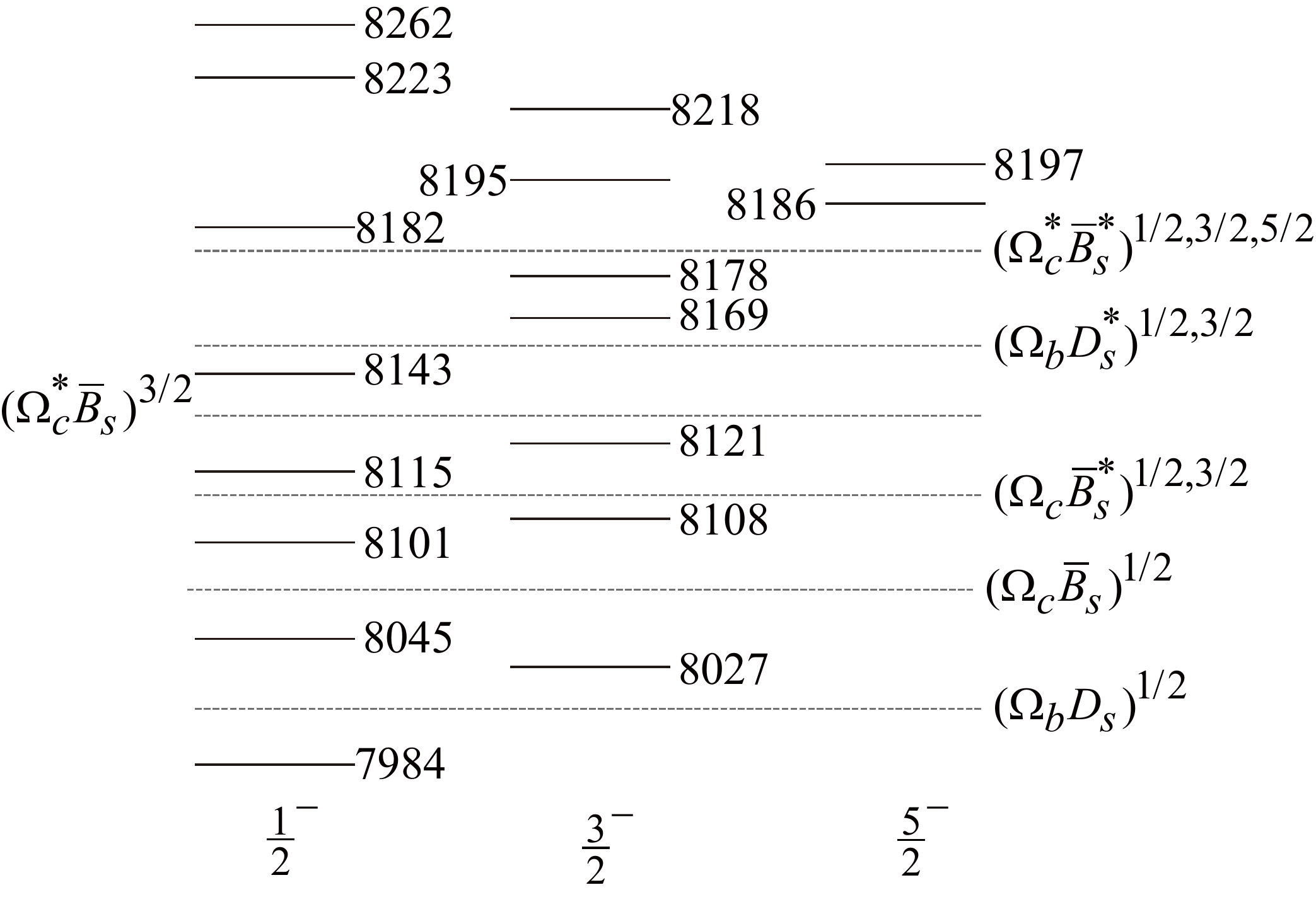}\\
(c) $I=\frac12$ (solid) $bcss\bar{n}$ states &&(d) $I=0$ (solid) $bcss\bar{s}$ states\\
\end{tabular}
\caption{Relative positions (units: MeV) for the $bcnn\bar{q}$ and
$bcss\bar{q}$ pentaquark states labeled with solid and dashed lines.
The dotted lines indicate various baryon-meson thresholds. The
$I=\frac{3}{2}$ and $\frac{1}{2}$ $bcnn\bar{n}$ states with
$I_{nn}=1$ have the same mass spectrum and are shown in the diagram
(a) with solid lines. When the isospin (spin) of an initial
pentaquark state is equal to a number in the subscript (superscript)
of a baryon-meson state, its decay into that baryon-meson channel
through $S$- or $D$-wave is allowed by the isospin (angular momentum)
conservation.  {We have adopted
the masses estimated with the reference thresholds of (a) $\Sigma_{c}\bar{B}$, (b) $\Sigma_{c}\bar{B}_{s}$, (c) $\Omega_{c}\bar{B}$, and (d) $\Omega_{c}\bar{B}_{s}$.}}\label{fig-bcnnqbar-bcssqbar}
\end{figure}


To estimate the masses of the $bcnn\bar{q}$ and $bcss\bar{q}$ states
($q=n,s$), we can also use two types of thresholds: (charmed
baryon)-(bottom meson) and (bottom baryon)-(charmed meson). The
results and relevant reference systems are presented in Tables
\ref{mass-bcnnqbar} and \ref{mass-bcssqbar}. The masses obtained
with the two types of thresholds are slightly different. We use
results estimated with the (charmed baryon)-(bottom meson) type
threshold for further discussions. In Fig.
\ref{fig-bcnnqbar-bcssqbar}, the relative positions for these
pentaquark states and relevant baryon-meson thresholds are plotted.
For the $bcss\bar{n}$ and $bcss\bar{s}$ states, only one value of
isospin is possible and we do not label the subscripts of the
baryon-meson states into which the pentaquarks may decay.

From the diagrams (a) and (d) of Fig. \ref{fig-bcnnqbar-bcssqbar},
the $bcnn\bar{n}$ system has more than 12 possible rearrangement
decay channels and the $bcss\bar{s}$ system has more than 6.
However, one cannot simply distinguish a pentaquark from a
conventional baryon or from a 3$q$ and 5$q$ mixed state just from these
decay channels if the isospin is not $3/2$. The discussions are
similar to the previous systems. On the other hand, in the
$bcnn\bar{s}$ and $bcss\bar{n}$ cases, good pentaquark candidates
may be searched for in their relevant decay patterns shown in the
diagrams (b) and (c) of Fig. \ref{fig-bcnnqbar-bcssqbar}. If we use
$m_{\Xi_{bc}}=6922$ MeV, $m_{\Xi'_{bc}}=6948$ MeV, and $m_{\Xi_{bc}^*}=6973$ MeV \cite{Weng:2018mmf}, one finds that the
lowest two $bcnn\bar{s}$ pentaquarks should be stable and the lowest $J^P=3/2^-$ state is probably narrow. Since the
three $I=3/2$ $bcnn\bar{n}$ states are more than 350 MeV lower than
the $\Lambda_bD$ threshold and just above the $\Xi_{bc}\pi$
threshold, they probably have narrow widths and we may use the
$\Xi_{bc}\pi$ channels to identify such pentaquarks. Similarly, the
$\Omega_{bc}\bar{K}$ channels may be used to identify the
$bcss\bar{n}$ pentaquarks if $m_{\Omega_{bc}}$ is around 7011 MeV
\cite{Weng:2018mmf}.

\subsection{The $ccns\bar{q}$ and $bbns\bar{q}$ pentaquark states}

\begin{table}[!h]
\centering \caption{The estimated masses for the $ccns\bar{q}$
systems in units of MeV. The values in the third column are obtained
with the effective quark masses and are theoretical upper limits.
The masses after this column are determined with relevant
thresholds.}\label{mass-ccnsqbar}
\begin{tabular}{ccccc|ccccc}
\toprule[1.5pt] \midrule[1pt] \multicolumn{5}{l}{$ccns\bar{n}$
$(I=0,1)$} &
\multicolumn{5}{l}{$ccns\bar{s}$ $(I=\frac{1}{2})$}\\
$J^{P}$ &Eigenvalue&Mass&$(\Xi_{c}^{'}D)$ & $(\Xi_{cc}K)$ & $J^{P}$
&Eigenvalue&Mass&$(\Xi_{c}^{'}D_{s})$ & $(\Xi_{cc}\phi)$ \\
\midrule[1pt]

$\frac{5}{2}^{-}$& $\begin{pmatrix}200.3\\121.7\end{pmatrix}$&
$\begin{pmatrix}4913.9\\4835.3\end{pmatrix}$&
$\begin{pmatrix}4764.1\\4685.5\end{pmatrix}$&
$\begin{pmatrix}4643.1\\4564.6\end{pmatrix}$&

$\frac{5}{2}^{-}$& $\begin{pmatrix}143.2\\69.2\end{pmatrix}$&
$\begin{pmatrix}5035.4\\4961.4\end{pmatrix}$&
$\begin{pmatrix}4810.5\\4736.4\end{pmatrix}$&
$\begin{pmatrix}4777.9\\4703.9\end{pmatrix}$\\

$\frac{3}{2}^{-}$&
$\begin{pmatrix}230.9\\173.6\\88.2\\46.5\\29.5\\-231.9\\-473.9\end{pmatrix}$&
$\begin{pmatrix}4944.5\\4887.2\\4801.8\\4760.1\\4743.1\\4481.7\\4239.7\end{pmatrix}$&
$\begin{pmatrix}4794.7\\4737.4\\4652.0\\4610.3\\4593.3\\4331.9\\4089.9\end{pmatrix}$&
$\begin{pmatrix}4673.7\\4616.4\\4531.0\\4489.3\\4472.3\\4210.9\\3968.9\end{pmatrix}$&

$\frac{3}{2}^{-}$&
$\begin{pmatrix}151.7\\106.9\\62.9\\45.2\\-13.3\\-89.6\\-275.6\end{pmatrix}$&
$\begin{pmatrix}5043.9\\4999.1\\4955.1\\4937.4\\4878.9\\4802.6\\4616.6\end{pmatrix}$&
$\begin{pmatrix}4818.9\\4774.2\\4730.2\\4712.5\\4653.9\\4577.6\\4391.6\end{pmatrix}$&
$\begin{pmatrix}4786.4\\4741.7\\4697.7\\4680.0\\4621.4\\4545.1\\4359.1\end{pmatrix}$\\

$\frac{1}{2}^{-}$&
$\begin{pmatrix}296.1\\169.8\\140.7\\48.9\\5.3\\-89.6\\-263.2\\-575.8\end{pmatrix}$&
$\begin{pmatrix}5009.7\\4883.4\\4854.3\\4762.5\\4718.9\\4624.0\\4450.4\\4137.8\end{pmatrix}$&
$\begin{pmatrix}4859.9\\4733.6\\4704.5\\4612.7\\4569.1\\4474.2\\4300.6\\3988.0\end{pmatrix}$&
$\begin{pmatrix}4739.0\\4612.6\\4583.5\\4491.7\\4448.1\\4353.3\\4179.7\\3867.0\end{pmatrix}$&

$\frac{1}{2}^{-}$&
$\begin{pmatrix}219.1\\126.4\\76.2\\-5.5\\-14.8\\-80.3\\-135.0\\-375.7\end{pmatrix}$&
$\begin{pmatrix}5111.3\\5018.6\\4968.4\\4886.7\\4877.4\\4812.0\\4757.3\\4516.5\end{pmatrix}$&
$\begin{pmatrix}4886.3\\4793.7\\4743.4\\4661.8\\4652.5\\4587.0\\4532.3\\4291.5\end{pmatrix}$&
$\begin{pmatrix}4853.8\\4761.2\\4710.9\\4629.3\\4620.0\\4554.5\\4499.8\\4259.0
\end{pmatrix}$\\
\midrule[1pt] \bottomrule[1.5pt]
\end{tabular}
\end{table}

\begin{table}[!h]
\centering \caption{The estimated masses for the $bbns\bar{q}$
systems in units of MeV. The values in the third column are obtained
with the effective quark masses and are theoretical upper limits.
The masses after this column are determined with relevant
thresholds.}\label{mass-bbnsqbar}
\begin{tabular}{cccc|cccc}
\toprule[1.5pt] \midrule[1pt] \multicolumn{4}{l}{$bbns\bar{n}$
$(I=0,1)$} &
\multicolumn{4}{l}{$bbns\bar{s}$ $(I=\frac{1}{2})$}\\
$J^{P}$ &Eigenvalue&Mass&$(\Xi_{b}^{'}\bar{B})$ & $J^{P}$
&Eigenvalue&Mass&$(\Xi_{b}^{'}\bar{B}_{s})$\\ \midrule[1pt]

$\frac{5}{2}^{-}$& $\begin{pmatrix}175.5\\98.2\end{pmatrix}$&
$\begin{pmatrix}11545.3\\11468.0\end{pmatrix}$&
$\begin{pmatrix}11400.7\\11323.4\end{pmatrix}$&

$\frac{5}{2}^{-}$& $\begin{pmatrix}116.4\\47.9\end{pmatrix}$&
$\begin{pmatrix}11664.8\\11596.3\end{pmatrix}$&
$\begin{pmatrix}11432.3\\11363.8\end{pmatrix}$\\

$\frac{3}{2}^{-}$&
$\begin{pmatrix}231.0\\166.8\\85.7\\55.2\\12.6\\-233.8\\-502.8\end{pmatrix}$&
$\begin{pmatrix}11600.8\\11536.6\\11455.5\\11425.0\\11382.4\\11136.0\\10867.0\end{pmatrix}$&
$\begin{pmatrix}11456.2\\11392.0\\11310.9\\11280.4\\11237.8\\10991.4\\10722.4\end{pmatrix}$&

$\frac{3}{2}^{-}$&
$\begin{pmatrix}144.8\\105.8\\41.2\\28.2\\4.9\\-83.3\\-301.6\end{pmatrix}$&
$\begin{pmatrix}11693.2\\11654.2\\11589.6\\11576.6\\11553.3\\11465.1\\11246.8\end{pmatrix}$&
$\begin{pmatrix}11460.7\\11421.7\\11357.1\\11344.1\\11320.8\\11232.7\\11014.3\end{pmatrix}$\\

$\frac{1}{2}^{-}$&
$\begin{pmatrix}248.5\\162.7\\118.3\\65.0\\51.5\\-89.5\\-245.4\\-529.9\end{pmatrix}$&
$\begin{pmatrix}11618.3\\11532.5\\11488.1\\11434.8\\11421.3\\11280.3\\11124.4\\10839.9\end{pmatrix}$&
$\begin{pmatrix}11473.7\\11387.9\\11343.5\\11290.2\\11276.7\\11135.7\\10979.8\\10695.3\end{pmatrix}$&

$\frac{1}{2}^{-}$&
$\begin{pmatrix}163.5\\109.2\\74.1\\21.6\\3.4\\-80.8\\-104.1\\-328.0\end{pmatrix}$&
$\begin{pmatrix}11711.9\\11657.6\\11622.5\\11570.0\\11551.8\\11467.6\\11444.3\\11220.4\end{pmatrix}$&
$\begin{pmatrix}11479.4\\11425.1\\11390.0\\11337.5\\11319.3\\11235.1\\11211.8\\10987.9\end{pmatrix}$\\
\midrule[1pt] \bottomrule[1.5pt]
\end{tabular}
\end{table}

In the mass estimation for the $ccns\bar{q}$ ($q=n,s$) system, we
use two types of thresholds: (charmed baryon)-(charmed meson) and
(doubly charmed baryon)-(light meson). For the $bbns\bar{q}$ system,
we only adopt the (bottom baryon)-(bottom meson) type threshold. The
pentaquark masses estimated with the help of the doubly charmed
baryon are smaller than those with the (charmed baryon)-(charmed
meson) type threshold. We present the numerical results for the $ccns\bar{q}$ and
$bbns\bar{q}$ systems in Tables \ref{mass-ccnsqbar} and
\ref{mass-bbnsqbar}, respectively. The relative positions for these
pentaquark states and the relevant rearrangement decay states are
shown in Fig. \ref{fig-ccnsqbar-bbnsqbar}. From the figure, we can
see that both the heaviest state and the lightest state are the
$J^P=\frac12^-$ pentaquarks in each system. Because all these
systems contain a quark-antiquark pair, it is not easy to
distinguish a pentaquark state from a 3$q$ baryon state if the isospin
of the decay product is less than 1. Also, the widths of the lowest
pentaquark states are probably not narrow if we take
$m_{\Omega_{cc}}=3715$ MeV and $m_{\Omega_{bb}}=10230$ MeV
\cite{Lu:2017meb}. In Ref. \cite{Guo:2017vcf}, a bound state with
$I=0$ below the $\Xi_{cc}\bar{K}$ threshold is predicted. If
experiments observed one state with the quark content $ccns\bar{n}$,
irrespective of its nature, its partner states could also be
searched for in the $\Omega_{cc}\pi$, $\Omega_{cc}K$,
$\Omega_{bb}\pi$, and $\Omega_{bb}K$ channels and whether they exist
or not can test the simple model we use.

\begin{figure}[!ht]
\begin{tabular}{ccc}
\includegraphics[width=220pt]{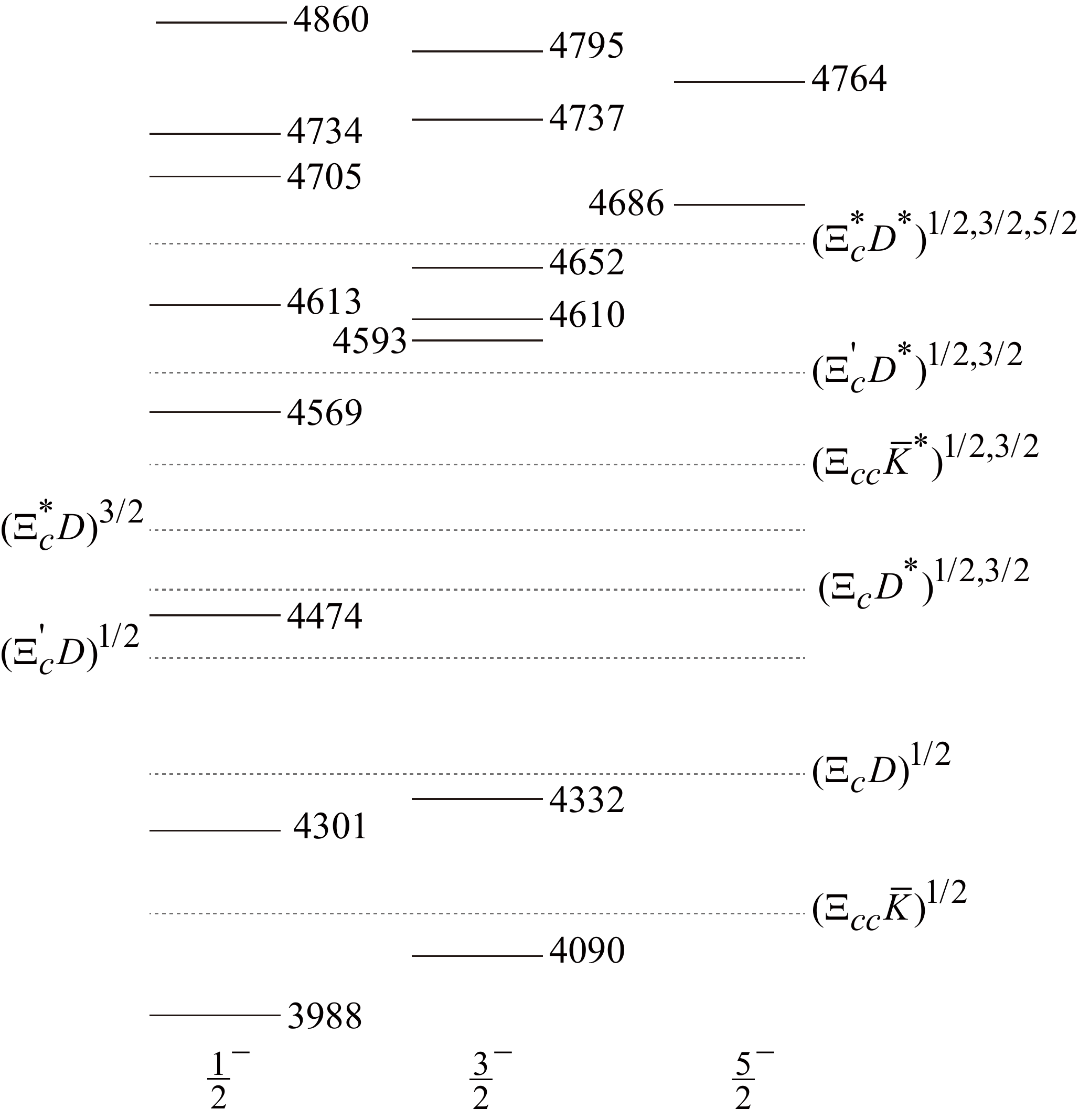}&$\qquad$&
\includegraphics[width=220pt]{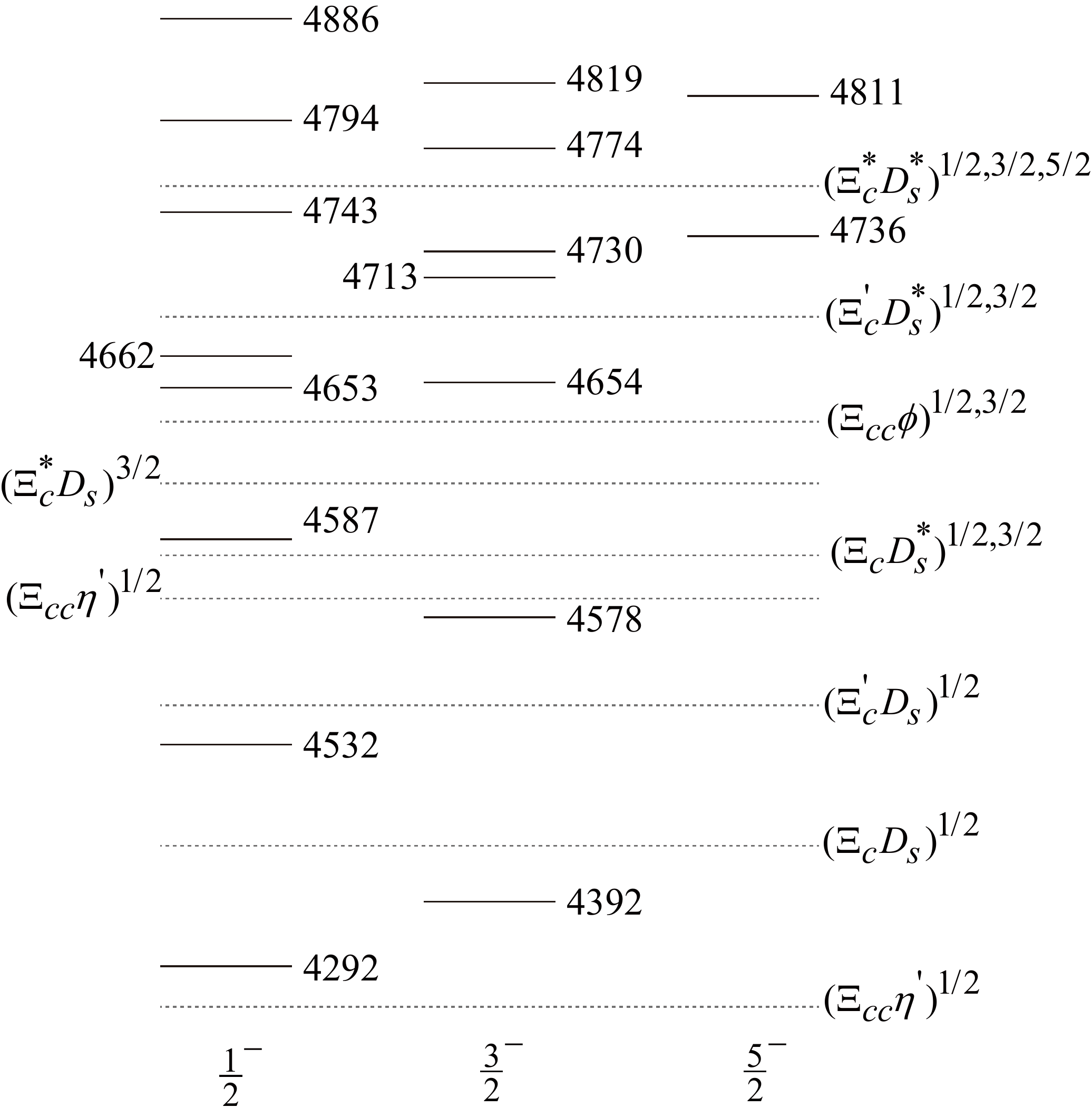}\\
(a) $I=0$ and $I=1$ (solid) $ccns\bar{n}$ states &&(b) $I=\frac{1}{2}$ (solid)$ccns\bar{s}$ states\\
&&\\
\includegraphics[width=220pt]{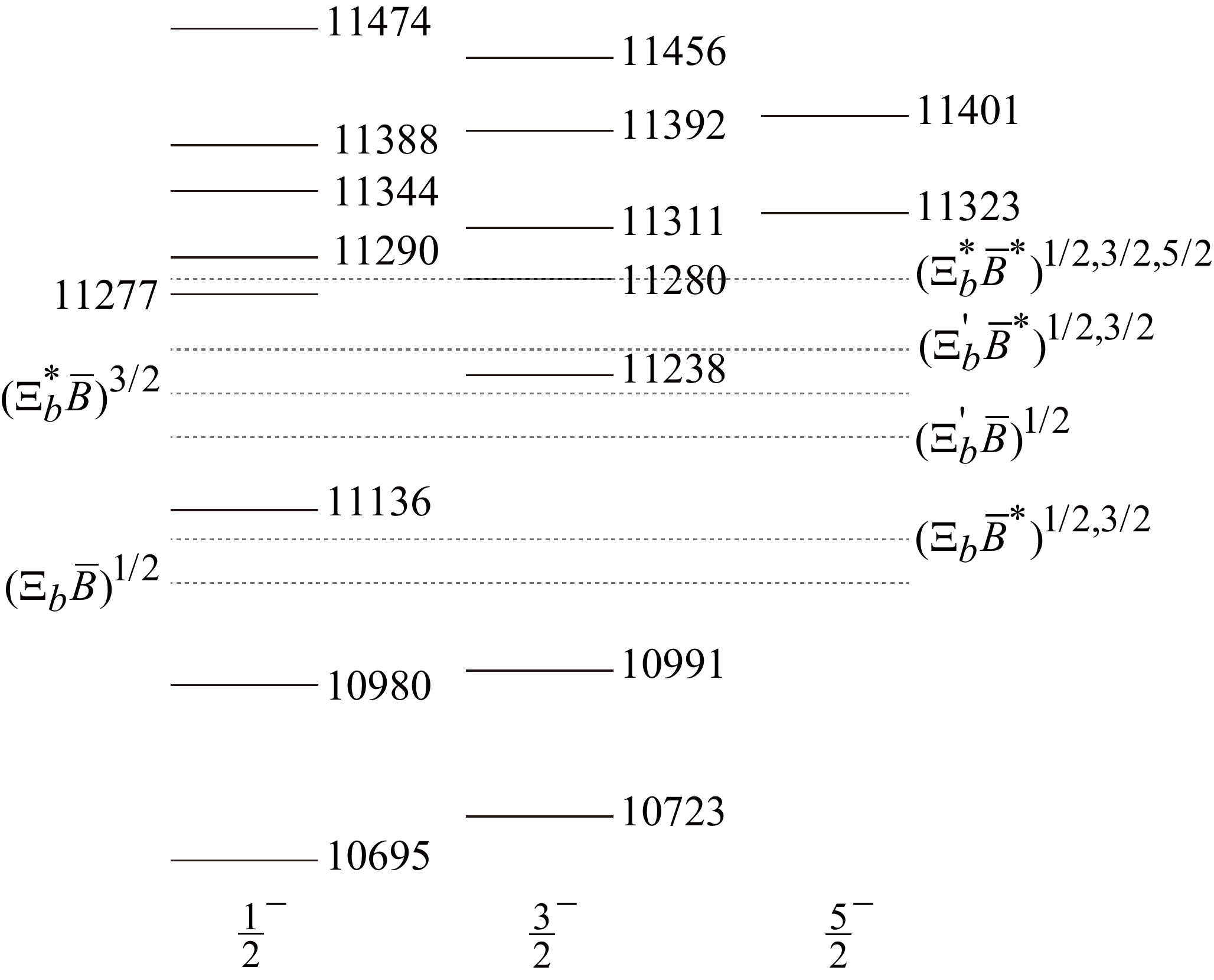}&$\qquad$&
\includegraphics[width=220pt]{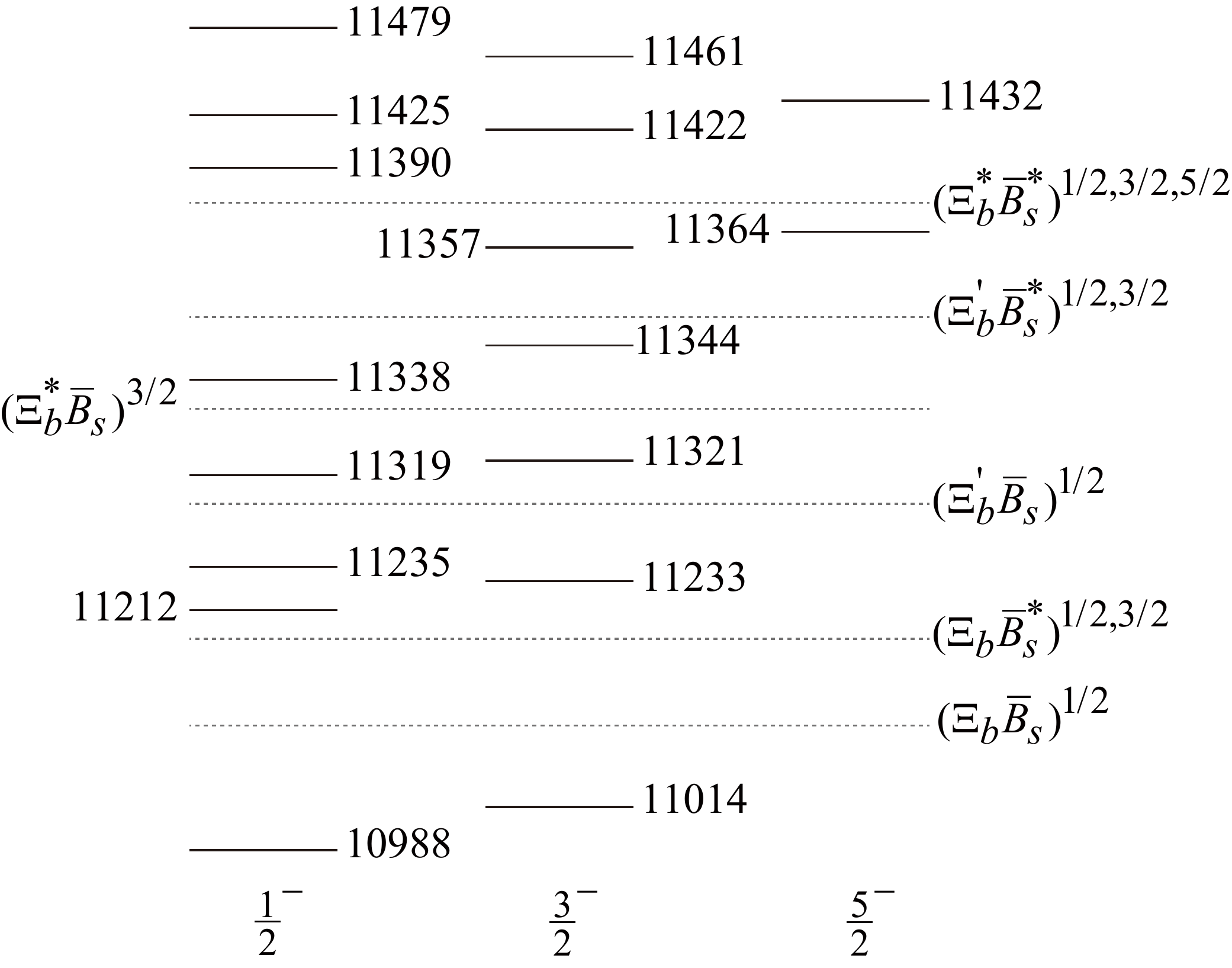}\\
(c) $I=0$ and $I=1$ (solid) $bbns\bar{n}$ states &&(d) $I=\frac{1}{2}$ (solid)$bbns\bar{s}$ states\\
\end{tabular}
\caption{Relative positions (units: MeV) for the $ccns\bar{q}$ and
$bbns\bar{q}$ pentaquark states labeled with solid lines. The dotted
lines indicate various baryon-meson thresholds. The $I=0$ and $I=1$
$ccns\bar{n}$ states have the same mass spectrum and are shown in
the diagram (a). The doubly bottom analog is shown in the diagram
(c). When the spin of an initial pentaquark state is equal to a
number in the superscript of a baryon-meson state, its decay into
that baryon-meson channel through $S$- or $D$-wave is allowed by the
angular momentum conservation. {We have adopted
the masses estimated with the reference thresholds of (a) $\Xi_{c}^{'}D$, (b) $\Xi_{c}^{'}D_{s}$, (c) $\Xi_{b}^{'}\bar{B}$, and (d) $\Xi_{b}^{'}\bar{B}_{s}$.}}\label{fig-ccnsqbar-bbnsqbar}
\end{figure}

\subsection{The $bcns\bar{q}$ pentaquark states}

\begin{table}[!h]
\centering \caption{The estimated masses for the $bcns\bar{q}$
systems in units of MeV. The values in the third column are obtained
with the effective quark masses and are theoretical upper limits.
The masses after this column are determined with relevant
thresholds.}\label{mass-bcnsqbar}
\begin{tabular}{ccccc|ccccc}
\toprule[1.5pt] \midrule[1pt]
\multicolumn{5}{l}{$bcns\bar{n}$ $(I=0,1)$} &
\multicolumn{5}{l}{$bcns\bar{s}$ $(I=\frac{1}{2})$}\\
$J^{P}$ &Eigenvalue&Mass&$(\Xi_{c}^{'}\bar{B})$ &$(\Xi_{b}^{'}D)$&
$J^{P}$
&Eigenvalue&Mass&$(\Xi_{c}^{'}\bar{B}_{s})$&$(\Xi_{b}^{'}D_{s})$\\
\midrule[1pt]

$\frac{5}{2}^{-}$& $\begin{pmatrix}185.8\\108.0\\44.2\end{pmatrix}$&
$\begin{pmatrix}8227.5\\8149.7\\8085.9\end{pmatrix}$&
$\begin{pmatrix}8088.3\\8010.5\\7946.6\end{pmatrix}$&
$\begin{pmatrix}8073.1\\7995.4\\7931.5\end{pmatrix}$&

$\frac{5}{2}^{-}$& $\begin{pmatrix}127.8\\61.2\\55.8\end{pmatrix}$&
$\begin{pmatrix}8348.1\\8281.5\\8276.1\end{pmatrix}$&
$\begin{pmatrix}8121.0\\8054.4\\8049.0\end{pmatrix}$&
$\begin{pmatrix}8118.6\\8051.9\\8046.6\end{pmatrix}$\\

$\frac{3}{2}^{-}$&
$\begin{pmatrix}229.0\\169.4\\144.8\\119.1\\84.3\\50.9\\34.7\\28.6\\-54.2\\-77.1\\-238.7\\-490.7\end{pmatrix}$&
$\begin{pmatrix}8270.7\\8211.1\\8186.5\\8160.8\\8126.0\\8092.6\\8076.4\\8070.3\\7987.5\\7964.6\\7803.0\\7551.0\end{pmatrix}$&
$\begin{pmatrix}8131.5\\8071.8\\8047.3\\8021.5\\7986.8\\7953.4\\7937.2\\7931.1\\7848.3\\7825.4\\7663.8\\7411.8\end{pmatrix}$&
$\begin{pmatrix}8116.3\\8056.7\\8032.1\\8006.4\\7971.6\\7938.2\\7922.0\\7915.9\\7833.1\\7810.2\\7648.6\\7396.6\end{pmatrix}$&

$\frac{3}{2}^{-}$&
$\begin{pmatrix}147.0\\107.6\\102.9\\66.2\\48.0\\33.6\\13.1\\-10.0\\-34.5\\-78.2\\-104.6\\-291.3\end{pmatrix}$&
$\begin{pmatrix}8367.3\\8327.9\\8323.2\\8286.5\\8268.3\\8253.9\\8233.4\\8210.3\\8185.8\\8142.1\\8115.8\\7929.0\end{pmatrix}$&
$\begin{pmatrix}8140.2\\8100.8\\8096.1\\8059.4\\8041.1\\8026.8\\8006.3\\7983.2\\7958.7\\7915.0\\7888.6\\7701.9\end{pmatrix}$&
$\begin{pmatrix}8137.8\\8098.4\\8093.7\\8057.0\\8038.7\\8024.4\\8003.9\\7980.8\\7956.3\\7912.6\\7886.2\\7699.4\end{pmatrix}$\\

$\frac{1}{2}^{-}$&
$\begin{pmatrix}272.1\\209.2\\160.8\\132.5\\83.9\\50.4\\33.4\\22.1\\-73.8\\-83.1\\-171.6\\-258.1\\-287.5\\-530.0\\-574.3\end{pmatrix}$&
$\begin{pmatrix}8313.8\\8251.0\\8202.5\\8174.2\\8125.6\\8092.1\\8075.1\\8063.8\\7967.9\\7958.6\\7870.1\\7783.6\\7754.2\\7511.7\\7467.4\end{pmatrix}$&
$\begin{pmatrix}8174.5\\8111.7\\8063.3\\8034.9\\7986.4\\7952.8\\7935.8\\7924.6\\7828.7\\7819.4\\7730.9\\7644.4\\7615.0\\7372.5\\7328.2\end{pmatrix}$&
$\begin{pmatrix}8159.4\\8096.6\\8048.1\\8019.8\\7971.2\\7937.7\\7920.7\\7909.4\\7813.5\\7804.3\\7715.7\\7629.3\\7599.8\\7357.4\\7313.0\end{pmatrix}$&

$\frac{1}{2}^{-}$&
$\begin{pmatrix}189.9\\137.1\\107.7\\71.7\\52.1\\3.0\\-3.7\\-13.3\\-51.6\\-79.7\\-110.5\\-144.4\\-186.4\\-330.8\\-375.5\end{pmatrix}$&
$\begin{pmatrix}8410.2\\8357.4\\8328.0\\8292.0\\8272.4\\8223.3\\8216.6\\8207.0\\8168.7\\8140.6\\8109.8\\8075.9\\8033.9\\7889.5\\7844.8\end{pmatrix}$&
$\begin{pmatrix}8183.1\\8130.3\\8100.9\\8064.9\\8045.3\\7996.2\\7989.5\\7979.9\\7941.6\\7913.5\\7882.7\\7848.8\\7806.8\\7662.4\\7617.7\end{pmatrix}$&
$\begin{pmatrix}8180.7\\8127.9\\8098.5\\8062.5\\8042.8\\7993.8\\7987.1\\7977.4\\7939.1\\7911.1\\7880.3\\7846.4\\7804.3\\7660.0\\7615.2\end{pmatrix}$\\
\midrule[1pt] \bottomrule[1.5pt]
\end{tabular}
\end{table}

For the $bcns\bar{q}$ states, the wave functions do not get
constraints from the Pauli principle and the number of wave function
bases for a given quantum number is bigger than that for other
systems. After diagonalizing the Hamiltonian, one gets numbers of
possible pentaquark states. Here we use two types of thresholds to
estimate their masses: (charmed baryon)-(bottom meson) and (bottom
baryon)-(charmed meson). The results are presented in Table
\ref{mass-bcnsqbar}. One finds that these two types of thresholds
lead to comparable values. With the masses from the (charmed
baryon)-(bottom meson) type thresholds, we plot the relative
positions for these pentaquarks and their relevant decay patterns in
Fig. \ref{fig-bcnsqbar}. The quantum numbers of the heaviest state
and the lightest state are both  $J^P=\frac12^-$. The mass of the
lightest state for the $bcns\bar{n}$ system is around 7313 MeV which
is above the thresholds of $\Omega_{bc}\pi$, $\Omega_{bc}'\pi$, and
$\Omega_{bc}^*\pi$ and is much lower than other two-body
baryon-meson thresholds if we adopt the masses obtained in Ref.
\cite{Weng:2018mmf}. This feature is helpful for us to identify
compact $I=1$ pentaquarks once the $bcs$ type baryons can be used to
spectrum reconstruction. On the other hand, the identification of a
$bcns\bar{s}$ pentaquark is not easy since it may share the same
decay products with an excited $\Xi_{bc}$ baryon.

\begin{figure}[!h]
\begin{tabular}{ccc}
\includegraphics[width=220pt]{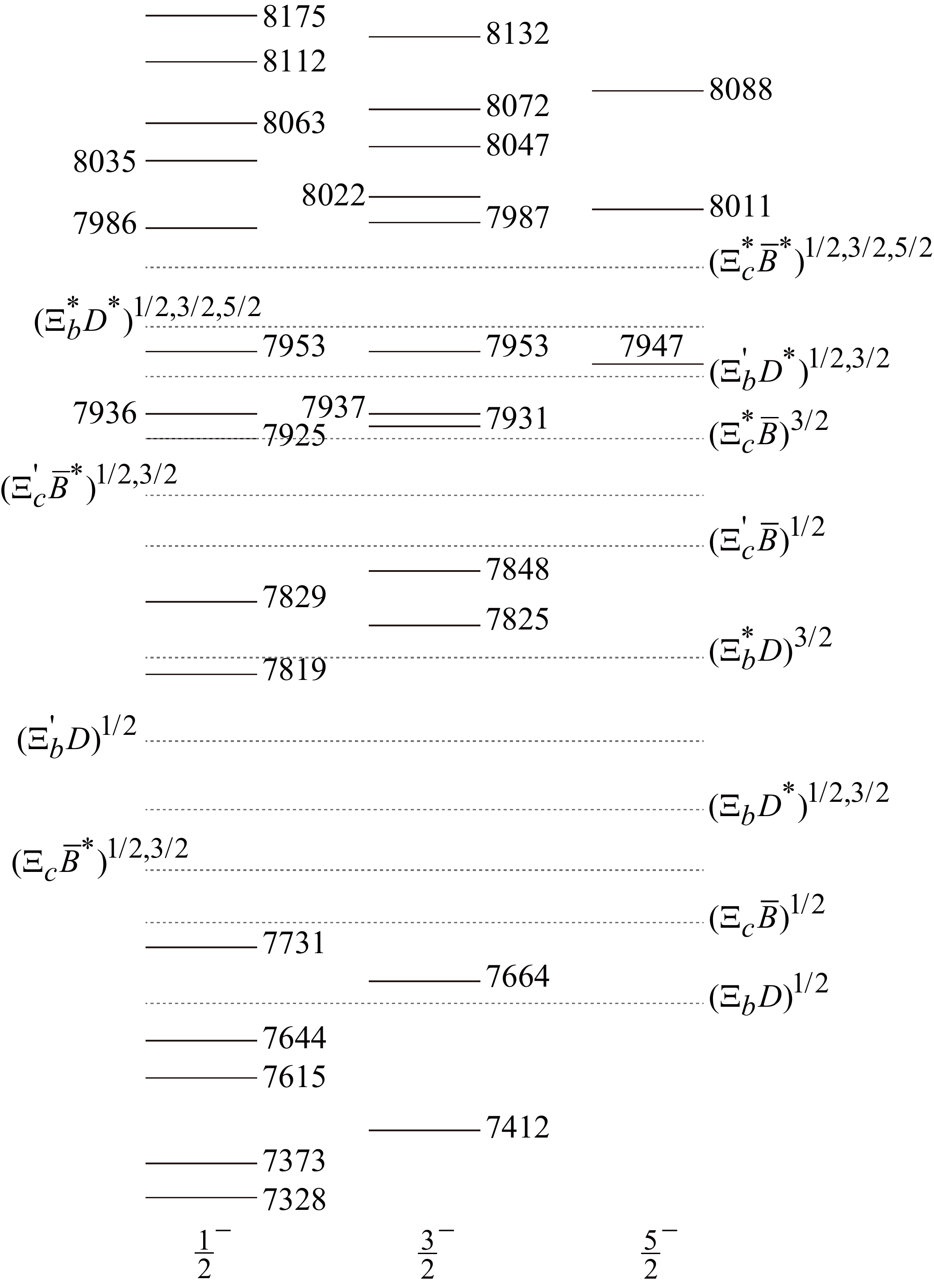}&$\qquad$&
\includegraphics[width=220pt]{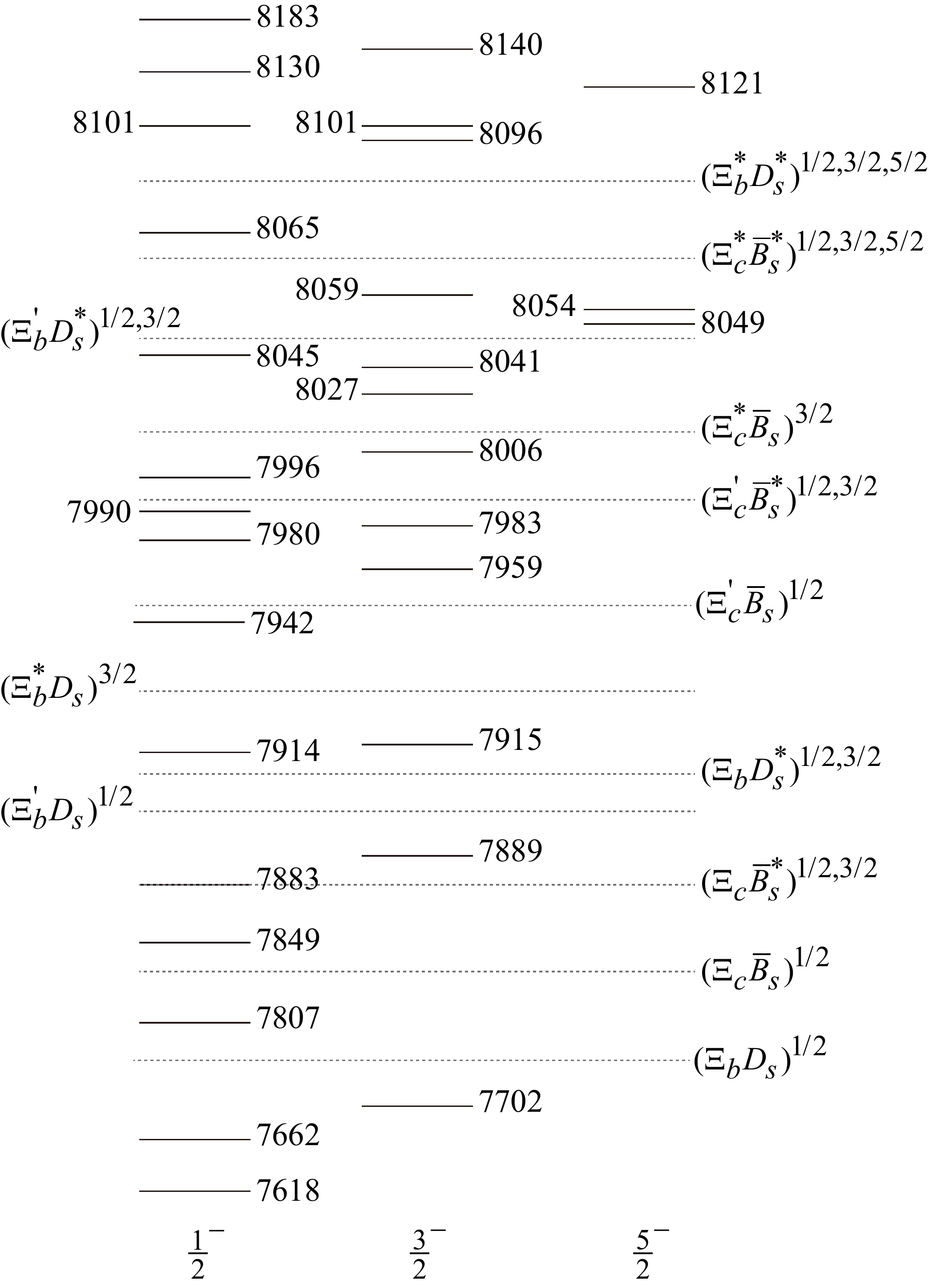}\\
(a) $I=0$ and $I=1$ (solid) $bcns\bar{n}$ states &&(b) $I=\frac{1}{2}$ (solid) $bcns\bar{s}$ states\\
\end{tabular}
\caption{Relative positions (units: MeV) for the $bcns\bar{q}$
pentaquark states labeled with solid lines. The dotted lines
indicate various baryon-meson thresholds. The $I=0$ and $I=1$
$bcns\bar{n}$ states have the same mass spectrum and are shown in
the diagram (a). When the spin of an initial pentaquark state is
equal to a number in the superscript of a baryon-meson state, its
decay into that baryon-meson channel through $S$- or $D$-wave is allowed
by the angular momentum conservation. {We have adopted
the masses estimated with the reference thresholds of (a) $\Xi_{b}^{'}D$ and (b) $\Xi_{b}^{'}D_{s}$.}}\label{fig-bcnsqbar}
\end{figure}
\end{widetext}

\section{Discussions and summary}\label{sec5}

Up to now, some candidates of the tetraquark states have been
confirmed by different experiments. The observation of the
$P_{c}(4380)$ and $P_{c}(4450)$ at LHCb gave us significant evidence
for the existence of pentaquak states and opened a new door for
studying hidden-charm exotic states. More possible pentaquarks have
been predicted in various theoretical calculations and await further
confirmation. In this paper, motivated by the $P_{c}(4380)$ and
$P_{c}(4450)$ and the observation of the $\Xi_{cc}$ at LHCb, we have
discussed the doubly heavy $QQqq\bar{q}$ pentaquark states in a CMI
model and shown their possible rearrangement decay patterns.
Although the model we adopt is simple and is not a dynamical model,
it may give us some qualitative properties with which the
experimentalists may be used to search for such exotic baryons. In
the early stage studies on the multiquark properties, chromomagnetic
effects were also intensively considered as the primary contribution
in an attempt to explain the narrow hadronic resonances
\cite{Montanet:1980te}. In recent years, this model as a widely used
method was adopted to study the multiquark states, such as the
investigations in Refs. \cite{Hogaasen:2005jv,Luo:2017eub,Cui:2006mp,Buccella:2006fn,Zhao:2014qva,Wu:2016vtq,Chen:2016ont,Yuan:2012wz}.

In the estimation of the rough masses, we have used two approaches
for comparison: one with the quark masses and the other with a
reference threshold. The results obtained with the former approach
are larger and can be treated as theoretical upper limits. In the
estimation with the latter approach, we mainly adopt the (heavy
baryon)-(heavy meson) type thresholds. Although no enough experimental data
for the doubly heavy $3q$ baryons are available, we may employ the
masses calculated in the quark model \cite{Lu:2017meb,Weng:2018mmf}. For the
investigated systems, we find that stable pentaquarks with $I=0$ are
possible in the $bcnn\bar{s}$ case. The lowest threshold of the
rearrangement decay product is for the $\Xi_{bc}K$ state while
the lowest pentaquarks can be below such thresholds. The typical
examples are the two lowest $I=0$ $bcnn\bar{s}$ states in Fig.
\ref{fig-bcnnqbar-bcssqbar} (b) and the $I=0$ $ccnn\bar{s}$ state in
Fig. \ref{fig-ccnnqbar-bbnnqbar-ccssqbar-bbssqbar} (b). In the
$Q_1Q_2nn\bar{n}$ and $Q_1Q_2ns\bar{n}$ cases, the lowest threshold
of the rearrangement decay product is for the $\Xi_{Q_1Q_2}\pi$ or
$\Omega_{Q_1Q_2}\pi$, but the lowest pentaquarks we obtain are hard
to be below such thresholds. Good news is that the lowest pentaquark
may be below the (heavy baryon)-(heavy meson) threshold and one may
search for such pentaquarks with the strong decay modes containing a
pion. In the $Q_1Q_2ss\bar{n}$ and $Q_1Q_2ns\bar{s}$ cases, the
lowest pentaquarks may be above the $\Omega_{Q_1Q_2}\bar{K}$ or
$\Omega_{Q_1Q_2}K$ threshold and can be discoveried with the decay
modes containing a kaon. Contrary to the above systems, the strong decay channel with lowest threshold in the $Q_1Q_2ss\bar{s}$ case may be the (heavy
baryon)-(heavy meson) type. Since the doubly heavy baryons are very
difficult to be used to reconstruct pentaquark spectra, maybe one
should notice the $Q_1Q_2ss\bar{s}$ pentaquarks experimentally.
Alternatively, the $J=5/2$ pentaquarks may be searched for first
since they have many $D$-wave decay modes but one or two $S$-wave
decay modes and probably they are not so broad. This feature is
similar to the hidden-charm pentaquarks \cite{Wu:2017weo}.

In the study of multiquark states, the number of color-spin
structures may be more than ten. The mixing or channel-coupling
effects could be important. The lowest pentaquarks we obtain get
contributions from such effects significantly. Whether there are
substructures in multiquark states and whether the configuration
mixing effects are that important need more studies. In the near
future, further experimental and theoretical studies on pentaquarks
are still important, especially with the running of LHC at 13 TeV
and the forthcoming BelleII.

In summary, we have studied preliminarily the mass spectra of doubly
heavy pentaquark states in a color-magnetic model. We find
candidates of possible narrow states. If they do exist, the
identification may be not difficult from their exotic quantum
numbers. We hope that the present study may inspire experimental
exploration to exotic states.

\section*{Acknowledgments}
YRL thanks the hospitality from Prof. M. Oka and other colleagues at
Tokyo Institute of Technology, where the draft was completed. This
project is supported by National Natural Science Foundation of China
under Grants No. 11775132, No. 11222547, No. 11175073, No.
11261130311 and 973 program. XL is also the National Program for Support of Top-notch Young Professionals.

\end{document}